\documentclass[english,prb,twocolumn,groupedaddress,reprint]{revtex4}
\usepackage[T1]{fontenc}
\usepackage[latin9]{inputenc}
\usepackage{color}
\usepackage{amsmath}
\usepackage{amssymb}
\usepackage{graphicx}
\usepackage{esint}

\makeatletter

\@ifundefined{textcolor}{}
{%
 \definecolor{BLACK}{gray}{0}
 \definecolor{WHITE}{gray}{1}
 \definecolor{RED}{rgb}{1,0,0}
 \definecolor{GREEN}{rgb}{0,1,0}
 \definecolor{BLUE}{rgb}{0,0,1}
 \definecolor{CYAN}{cmyk}{1,0,0,0}
 \definecolor{MAGENTA}{cmyk}{0,1,0,0}
 \definecolor{YELLOW}{cmyk}{0,0,1,0}
}

\usepackage{bbold}
\usepackage{tikz}
\newcommand*\circled[1]{\tikz[baseline=(char.base)]{
            \node[shape=circle,draw,inner sep=2pt] (char) {#1};}}

\makeatother

\usepackage{babel}
\usepackage{minitoc}

\begin{document}

\title{Nuclear Spin Dynamics in Double Quantum Dots: Multi-Stability, Dynamical
Polarization, Criticality and Entanglement }

\author{M. J. A. Schuetz,$^{1}$ E. M. Kessler,$^{2,3}$ L. M. K. Vandersypen,$^{4}$
J. I. Cirac,$^{1}$ and G. Giedke$^{1}$ }

\affiliation{$^{1}$Max-Planck-Institut für Quantenoptik, Hans-Kopfermann-Str.
1, 85748 Garching, Germany}

\affiliation{$^{2}$Physics Department, Harvard University, Cambridge, MA 02318,
USA}

\affiliation{$^{3}$ITAMP, Harvard-Smithsonian Center for Astrophysics, Cambridge,
MA 02318, USA}

\affiliation{$^{4}$Kavli Institute of NanoScience, TU Delft, P.O. Box 5046, 2600
GA, Delft, The Netherlands}

\date{\today}
\begin{abstract}
We theoretically study the nuclear spin dynamics driven by electron
transport and hyperfine interaction in an electrically defined double
quantum dot in the Pauli-blockade regime. We derive a master-equation-based
framework and show that the coupled electron-nuclear system displays
an instability towards the buildup of large nuclear spin polarization
gradients in the two quantum dots. In the presence of such inhomogeneous
magnetic fields, a quantum interference effect in the collective hyperfine
coupling results in sizable nuclear spin entanglement between the
two quantum dots in the steady state of the evolution. We investigate
this effect using analytical and numerical techniques, and demonstrate
its robustness under various types of imperfections.
\end{abstract}
\maketitle

\section{\textcolor{black}{Introduction}}

The prospect of building devices capable of quantum information processing
(QIP) has fueled an impressive race to implement well-controlled two-level
quantum systems (qubits) in a variety of physical settings.\cite{zoller05}
For any such system, generating and maintaining entanglement---one
of the most important primitives of QIP---is a hallmark achievement.
It serves as a benchmark of experimental capabilities and enables
essential information processing tasks such as the implementation
of quantum gates and the transmission of quantum information.\cite{nielsen00} 

In the solid state, electron spins confined in electrically defined
semiconductor quantum dots have emerged as a promising platform for
QIP: \cite{hanson07,chekhovich13,awschalom02,loss98} Essential ingredients
such as initialization, single-shot readout, universal quantum gates
and, quite recently, entanglement have been demonstrated experimentally.\cite{nowack11,petta05,shulman12,koppens06,koppens05,nowack07}
In this context, nuclear spins in the surrounding semiconductor host
environment have attracted considerable theoretical\cite{khaetskii02,erlingsson01,schliemann03,coish04,coish08,cywinski09,merkulov02}
and experimental\cite{vink09,bluhm10,foletti09,johnson05,ono04,ono02}
attention, as they have been identified as the main source of electron
spin decoherence due to the relatively strong hyperfine (HF) interaction
between the electronic spin and $N\sim10^{6}$ nuclei.\cite{chekhovich13}
However, it has also been noted that the nuclear spin bath itself,
with nuclear spin coherence times ranging from hundreds of microseconds
to a millisecond,\cite{takahashi11,chekhovich13} could be turned
into an asset, for example, as a resource for quantum memories or
quantum computation. \cite{taylor03,taylor04,witzel07,cappellaro09,kurucz09}
Since these applications require yet unachieved control of the nuclear
spins, novel ways of understanding and manipulating the dynamics of
the nuclei are called for. The ability to control and manipulate the
nuclei will open up new possibilities for nuclear spin-based information
storage and processing, but also directly improve electron spin decoherence
timescales.\cite{rudner07,rudner07b,rudner11a} 

Dissipation has recently been identified as a novel approach to control
a quantum system, create entangled states or perform quantum computing
tasks.\cite{verstraete09,diehl08,sanchez13,tomadin12,poyatos96} This
is done by properly engineering the continuous interaction of the
system with its environment. In this way, dissipation---previously
often viewed as a vice from a QIP perspective---can turn into a virtue
and become the driving force behind the emergence of coherent quantum
phenomena. The idea of actively using dissipation rather than relying
on coherent evolution extends the traditional DiVincenzo criteria\cite{divincenzo00}
to settings in which no unitary gates are available; also, it comes
with potentially significant practical advantages, as dissipative
methods are inherently robust against weak random perturbations, allowing,
in principle, to stabilize entanglement for arbitrary times. Recently,
these concepts have been put into practice experimentally in different
QIP architectures, namely atomic ensembles,\cite{muschik11} trapped
ions\cite{lin13,barreiro11} and superconducting qubits.\cite{shankar13}

\begin{figure}[b]
\includegraphics[width=0.95\columnwidth]{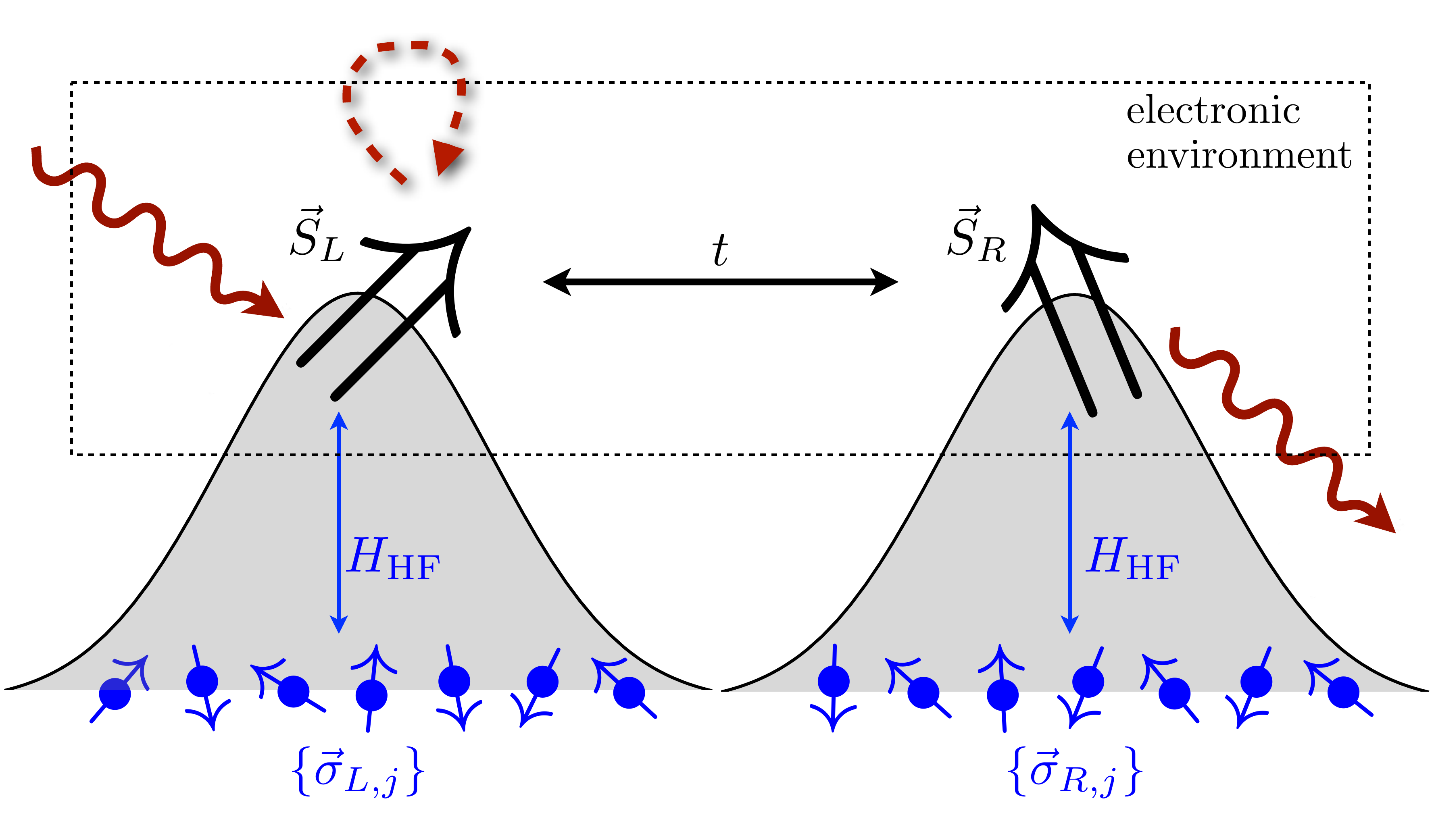}

\caption{\label{fig:entanglement-transport-sketch}(color online). Schematic
illustration of the DQD system under study. Two nuclear spin ensembles
are hyperfine-coupled to the electronic subsytem; due to various \textit{fast}
coherent (double-arrow) and incoherent processes (arrows) the latter
settles to a quasisteady state on a timescale shorter than the nuclear
dynamics. }
\end{figure}

Here, we apply these ideas to a quantum dot system and investigate
a scheme for the deterministic generation of steady-state entanglement
between the two spatially separated nuclear spin ensembles in an electrically
defined double quantum dot (DQD), operated in the Pauli-blockade regime.\cite{ono02,hanson07}
Expanding upon our proposal presented in Ref.\cite{schuetz13}, we
develop in detail the underlying theoretical framework, and discuss
in greater depth the coherent phenomena emerging from the hyperfine
coupled electron and nuclear dynamics in a DQD in spin blockade regime.
The analysis is based on the fact that the electron spins evolve rapidly
on typical timescales of the nuclear spin dynamics. This allows us
to derive a coarse-grained quantum master equation for the nuclear
spins only, disclosing the nuclei as the quantum system coupled to
an electronic environment with an exceptional degree of tunability;
see Fig.~~\ref{fig:entanglement-transport-sketch} for a schematic illustration.
This approach provides valuable insights by building up a straightforward
analogy between mesoscopic solid-state physics and a generic setting
in quantum optics (compare, for example, Ref.\cite{muschik11}): The
nuclear spin ensemble can be identified with an atomic ensemble, with
individual nuclear spins corresponding to the internal levels of a
single atom and electrons playing the role of photons.\cite{schuetz12} 

Our theoretical analysis goes beyond this simple analogy by incorporating
nonlinear, feedback-driven effects resulting from a backaction of
the effective magnetic field generated by the nuclei (Overhauser shift)
on the electron energy levels. In accordance with previous theoretical\cite{rudner07,rudner11a,rudner11b,jouravlev96,lopez-moniz11,economou13,danon09b,lunde13}
and experimental\cite{baugh07,petta08,kobayashi11,koppens05,ono04,barthel12}
observations, this feedback mechanism is shown to lead to a rich set
of phenomena such as multistability, criticality, and dynamic nuclear
polarization (DNP). In our model, we study the nuclear dynamics in
a systematic expansion of the master equation governing the evolution
of the combined electron-nuclear system, which allows us efficiently
trace out the electronic degrees of freedom yielding a compact dynamical
equation for the nuclear system alone. This mathematical description
can be understood in terms of the so-called slaving principle: The
electronic subsystem settles to a \textit{quasisteady} state on a
timescale much faster than the nuclear dynamics, and creates an effective
environment with tunable properties for the nuclear spins. Consequently,
we analyze the nuclear dynamics subject to this artificial environment.
Feedback effects kick in as the generated nuclear spin polarization
acts back on the electronic subsystem via the Overhauser shift changing
the electronic quasisteady state. We derive explicit expressions for
the nuclear steady state which allows us to fully assess the nuclear properties
in dependence on the external control parameters. In particular, we
find that, depending on the parameter regime, the polarization of
the nuclear ensemble can show two distinct behaviors: The nuclear
spins either saturate in a dark state without any nuclear polarization
or, upon surpassing a certain threshold gradient, turn self-polarizing
and build up sizable Overhauser field differences. Notably, the high-polarization
stationary states feature steady-state entanglement between the two
nuclear spin ensembles, even though the electronic quasisteady state
is separable, underlining the very robustness of our scheme against
electronic noise. 

To analyze the nuclear spin dynamics in detail, we employ different
analytical approaches, namely a semiclassical calculation and a fully
quantum mechanical treatment. This is based on a hierarchy of timescales:
While the nuclear polarization process occurs on a typical timescale
of $\tau_{\mathrm{pol}}\gtrsim1\mathrm{s}$, the timescale for building
up quantum correlations $\tau_{\mathrm{gap}}$ is collectively\cite{schuetz12}
enhanced by a factor $N\sim10^{5}-10^{6}$; i.e., $\tau_{\mathrm{gap}}\approx\left(3-30\right)\mu\mathrm{s}$.
Since nuclear spins dephase due to internal dipole-dipole interactions
on a timescale of $\tau_{\mathrm{dec}}\approx\left(0.1-1\right)\mathrm{ms}$,\cite{gullans10,takahashi11,chekhovich13}
our system exhibits the following separation of typical timescales:
$\tau_{\mathrm{pol}}\gg\tau_{\mathrm{dec}}\gg\tau_{\mathrm{gap}}$.
While the first inequality allows us to study the (slow) dynamics
of the macroscopic semiclassical part of the nuclear fields in a mean-field
treatment (which essentially disregards quantum correlations) on long
timescales, based on the second inequality we investigate the generation
of (comparatively small) quantum correlations on a much faster timescale
where we neglect decohering processes due to internal dynamics among
the nuclei. Lastly, numerical results complement our analytical findings
and we discuss in detail detrimental effects typically encountered
in experiments. 

This paper is organized as follows. Section \ref{sec:The-System}
introduces the master-equation-based theoretical framework. Based
on a simplified model, in Sec. \ref{sec:Effective-Nuclear-Dynamics}
we study the coupled electron nuclear dynamics. Using adiabatic elimination
techniques, we can identify two different regimes as possible fixed
points of the nuclear evolution which differ remarkably in their nuclear
polarization and entanglement properties. Subsequently, in Sec. \ref{sec:Polarization-Dynamics}
the underlying multi-stability of the nuclear system is revealed within
a semiclassical model. Based on a self-consistent Holstein-Primakoff
approximation, in Sec. \ref{sec:Steady-State-Entanglement} we study
in great detail the nuclear dynamics in the vicinity of a high-polarization
fixed point. This analysis puts forward the main result of our work,
the steady-state generation of entanglement between the two nuclear
spin ensembles in a DQD. Within the framework of the Holstein-Primakoff
analysis, Sec. \ref{sec:Criticality} highlights the presence of a
dissipative phase transition in the nuclear spin dynamics. Generalizations
of our findings to inhomogeneous hyperfine coupling and other weak
undesired effects are covered in Sec. \ref{sec:Implementation}. Finally,
in Sec. \ref{sec:Conclusion-and-Outlook} we draw conclusions and
give an outlook on possible future directions of research.

\section{The System \label{sec:The-System}}

This section presents a detailed description of the system under study,
a gate-defined double quantum dot (DQD) in the Pauli-blockade regime.
To model the dynamics of this system, we employ a master equation
formalism.\cite{schuetz12} This allows us to study the irreversible
dynamics of the DQD coupled to source and drain electron reservoirs.
By tracing out the unobserved degrees of freedom of the leads, we
show that---under appropriate conditions to be specified below---the
dynamical evolution of the reduced density matrix of the system $\rho$
can formally be written as 
\begin{equation}
\dot{\rho}=\underset{\circled1}{\underbrace{-i\left[H_{\mathrm{el}},\rho\right]+\mathcal{V}\rho+\mathcal{L}_{\Gamma}\rho}}+\underset{\circled2}{\underbrace{\mathcal{L}_{\pm}\rho+\mathcal{L}_{\mathrm{deph}}\rho}},\label{eq:effective-QME-full-model}
\end{equation}
Here, $H_{\mathrm{el}}$ describes the electronic degrees of freedom
of the DQD in the relevant two-electron regime, $\mathcal{V}$ refers
to the coherent hyperfine coupling between electronic and nuclear
spins and $\mathcal{L}_{\Gamma}$ is a Liouvillian of Lindblad form
describing electron transport in the spin-blockade regime. The last
two terms labeled by $\circled2$ account for different physical mechanisms
such as cotunneling, spin-exchange with the leads or spin-orbital
coupling in terms of effective dissipative terms in the electronic
subspace.

\subsection{Microscopic Model}

We consider an electrically defined DQD in the Pauli-blockade regime.\cite{hanson07,ono02}
Microscopically, our analysis is based on a two-site Anderson Hamiltonian:
Due to strong confinement, both the left and right dot are assumed
to support a single orbital level $\epsilon_{i}\left(i=L,R\right)$
only which can be Zeeman split in the presence of a magnetic field
and occupied by up to two electrons forming a localized spin singlet.
For now, excited states, forming on-site triplets that could lift
spin-blockade, are disregarded, since they are energetically well
separated by the singlet-triplet splitting $\Delta_{\mathrm{st}}\gtrsim400\mu\mathrm{eV}$.\cite{hanson07}
Cotunneling effects due to energetically higher lying localized triplet
states will be addressed separately below. 

Formally, the Hamiltonian for the global system $\mathcal{H}$ can
be decomposed as 
\begin{equation}
\mathcal{H}=H_{\mathrm{DQD}}+H_{B}+H_{T},
\end{equation}
where $H_{B}$ refers to two independent reservoirs of non-interacting
electrons, the left $\left(L\right)$ and right $\left(R\right)$
lead, respectively, 
\begin{equation}
H_{B}=\sum_{i,k,\sigma}\epsilon_{ik}c_{ik\sigma}^{\dagger}c_{ik\sigma},
\end{equation}
with $i=L,R$, $\sigma=\uparrow,\downarrow$ and $H_{T}$ models the
coupling of the DQD to the leads in terms of the tunnel Hamiltonian
\begin{equation}
H_{T}=\sum_{i,k,\sigma}T_{i}d_{i\sigma}^{\dagger}c_{ik\sigma}+\mathrm{h.c.}.\label{eq:tunnel-Hamiltonian-leads}
\end{equation}
The tunnel matrix element $T_{i}$, specifying the transfer coupling
between the leads and the system, is assumed to be independent of
momentum $k$ and spin $\sigma$ of the electron. The fermionic operator
$c_{ik\sigma}^{\dagger}\left(c_{ik\sigma}\right)$ creates (annihilates)
an electron in lead $i=L,R$ with wavevector $k$ and spin $\sigma=\uparrow,\downarrow$.
Similarly, $d_{i\sigma}^{\dagger}$ creates an electron with spin
$\sigma$ inside the dot in the orbital $i=L,R$. Accordingly, the
localized electron spin operators are 
\begin{equation}
\vec{S}_{i}=\frac{1}{2}\sum_{\sigma,\sigma'}d_{i\sigma}^{\dagger}\vec{\sigma}_{\sigma\sigma'}d_{i\sigma'},
\end{equation}
where $\vec{\sigma}$ refers to the vector of Pauli matrices. Lastly,
\begin{equation}
H_{\mathrm{DQD}}=H_{S}+H_{t}+V_{\mathrm{HF}}
\end{equation}
describes the coherent electron-nuclear dynamics inside the DQD. In
the following, $H_{S}$, $H_{t}$ and $V_{\mathrm{HF}}$ are presented.
First, $H_{S}$ accounts for the bare electronic energy levels in
the DQD and Coulomb interaction terms 
\begin{equation}
H_{S}=\sum_{i\sigma}\epsilon_{i\sigma}n_{i\sigma}+\sum_{i}U_{i}n_{i\uparrow}n_{i\downarrow}+U_{LR}n_{L}n_{R},\label{eq:Anderson-Hamiltonian}
\end{equation}
where $U_{i}$ and $U_{LR}$ refer to the on-site and interdot Coulomb
repulsion; $n_{i\sigma}=d_{i\sigma}^{\dagger}d_{i\sigma}$ and $n_{i}=n_{i\uparrow}+n_{i\downarrow}$
are the spin-resolved and total electron number operators, respectively.
Typical values are $U_{i}\approx1-4m\mathrm{eV}$ and $U_{LR}\approx200\mu\mathrm{eV}$.\cite{hayashi03,hanson07,ono02}
Coherent, spin-preserving interdot tunneling is described by 
\begin{equation}
H_{t}=t\sum_{\sigma}d_{L\sigma}^{\dagger}d_{\mathrm{R\sigma}}+\mathrm{h.c.}
\end{equation}

\textit{Spin-blockade regime}.---By appropriately tuning the chemical
potentials of the leads $\mu_{i}$, one can ensure that at maximum
two conduction electrons reside in the DQD.\cite{hanson07,sanchez13}
Moreover, for $\epsilon_{R\sigma}<\mu_{R}$ the right dot always stays
occupied. In what follows, we consider a transport setting where an
applied bias between the two dots approximately compensates the Coulomb
energy of two electrons occupying the right dot, that is $\epsilon_{L}\approx\epsilon_{R}+U_{R}-U_{LR}$.
Then, a source drain bias across the DQD device induces electron transport
via the cycle $\left(0,1\right)\rightarrow\left(1,1\right)\rightarrow\left(0,2\right)$.
Here, $\left(m,n\right)$ refers to a configuration with $m\left(n\right)$
electrons in the left (right) dot, respectively. In our Anderson model,
the only energetically accessible $\left(0,2\right)$ state is the
localized singlet, referred to as $\left|S_{02}\right\rangle =d_{R\uparrow}^{\dagger}d_{R\downarrow}^{\dagger}\left|0\right\rangle $.
As a result of the Pauli principle, the interdot charge transition
$\left(1,1\right)\rightarrow\left(0,2\right)$ is allowed only for
the $\left(1,1\right)$ spin singlet $\left|S_{11}\right\rangle =\left(\left|\Uparrow\Downarrow\right\rangle -\left|\Downarrow\Uparrow\right\rangle \right)/\sqrt{2}$,
while the spin triplets $\left|T_{\pm}\right\rangle $ and $\left|T_{0}\right\rangle =\left(\left|\Uparrow\Downarrow\right\rangle +\left|\Downarrow\Uparrow\right\rangle \right)/\sqrt{2}$
are Pauli blocked. Here, $\left|T_{+}\right\rangle =\left|\Uparrow\Uparrow\right\rangle $,
$\left|T_{-}\right\rangle =\left|\Downarrow\Downarrow\right\rangle $,
and $\left|\sigma\sigma'\right\rangle =d_{L\sigma}^{\dagger}d_{R\sigma'}^{\dagger}\left|0\right\rangle $.
For further details on how to realize this regime we refer to Appendix
\ref{sec:Two-Electron-Regime}.

\textit{Hyperfine interaction}.---The electronic spins $\vec{S}_{i}$
confined in either of the two dots $\left(i=L,R\right)$ interact
with two different sets of nuclear spins $\left\{ \sigma_{i,j}^{\alpha}\right\} $
in the semiconductor host environment via hyperfine (HF) interaction.
It is dominated by the isotropic Fermi contact term \cite{schliemann03}
given by 
\begin{equation}
H_{\mathrm{HF}}=\frac{a_{\mathrm{hf}}}{2}\sum_{i=L,R}\left(S_{i}^{+}A_{i}^{-}+S_{i}^{-}A_{i}^{+}\right)+a_{\mathrm{hf}}\sum_{i=L,R}S_{i}^{z}A_{i}^{z}.
\end{equation}
Here, $S_{i}^{\alpha}$ and $A_{i}^{\alpha}=\sum_{j}a_{i,j}\sigma_{i,j}^{\alpha}$
for $\alpha=\pm,z$ denote electron and collective nuclear spin operators.
The coupling coefficients $a_{i,j}$ are proportional to the weight
of the electron wavefunction at the $j$th lattice site and define
the individual unitless HF coupling constant between the electron
spin in dot $i$ and the $j$th nucleus. They are normalized such
that $\sum_{j=1}^{N_{i}}a_{i,j}=N$, where $N=\left(N_{1}+N_{2}\right)/2\sim10^{6}$;
$a_{\mathrm{hf}}$ is related to the total HF coupling strength $A_{\mathrm{HF}}\approx100\mu\mathrm{eV}$
via $a_{\mathrm{hf}}=A_{\mathrm{HF}}/N$ and $g_{\mathrm{hf}}=A_{\mathrm{HF}}/\sqrt{N}\approx0.1\text{\ensuremath{\mu\mathrm{eV}}}$
quantifies the typical HF interaction strength. The individual nuclear
spin operators $\sigma_{i,j}^{\alpha}$ are assumed to be spin-$\frac{1}{2}$
for simplicity. We neglect the nuclear Zeeman and dipole-dipole terms
which will be slow compared to the system's dynamics\cite{schliemann03};
these simplifications will be addressed in more detail in Sec. \ref{sec:Implementation}.

The effect of the hyperfine interaction can be split up into a perpendicular
component 
\begin{equation}
H_{\mathrm{ff}}=\frac{a_{\mathrm{hf}}}{2}\sum_{i=L,R}\left(S_{i}^{+}A_{i}^{-}+S_{i}^{-}A_{i}^{+}\right),
\end{equation}
which exchanges excitations between the electronic and nuclear spins,
and a parallel component, referred to as Overhauser (OH) field, 
\begin{equation}
H_{\mathrm{OH}}=a_{\mathrm{hf}}\sum_{i=L,R}S_{i}^{z}A_{i}^{z}.
\end{equation}
The latter can be recast into the following form 
\begin{equation}
H_{\mathrm{OH}}=H_{\mathrm{sc}}+H_{\mathrm{zz}},
\end{equation}
where 
\begin{equation}
H_{\mathrm{sc}}=\bar{\omega}_{\mathrm{OH}}\left(S_{L}^{z}+S_{R}^{z}\right)+\Delta_{\mathrm{OH}}\left(S_{R}^{z}-S_{L}^{z}\right)
\end{equation}
describes a (time-dependent) semiclassical OH field which comprises
a homogeneous $\bar{\omega}_{\mathrm{OH}}$ and inhomogeneous $\Delta_{\mathrm{OH}}$
component, respectively, 
\begin{eqnarray}
\bar{\omega}_{\mathrm{OH}} & = & \frac{a_{\mathrm{hf}}}{2}\left(\left\langle A_{L}^{z}\right\rangle _{t}+\left\langle A_{R}^{z}\right\rangle _{t}\right),\\
\Delta_{\mathrm{OH}} & = & \frac{a_{\mathrm{hf}}}{2}\left(\left\langle A_{R}^{z}\right\rangle _{t}-\left\langle A_{L}^{z}\right\rangle _{t}\right),
\end{eqnarray}
and 
\begin{equation}
H_{\mathrm{zz}}=a_{\mathrm{hf}}\sum_{i=L,R}S_{i}^{z}\delta A_{i}^{z},
\end{equation}
with $\delta A_{i}^{z}=A_{i}^{z}-\left\langle A_{i}^{z}\right\rangle _{t}$,
refers to residual quantum fluctuations due to deviations of the Overhauser
field from its expectation value.\cite{schuetz12} The semiclassical
part $H_{\mathrm{sc}}$ only acts on the electronic degrees of freedom
and can therefore be absorbed into $H_{S}$. Then, the coupling between
electronic and nuclear degrees of freedom is governed by the operator
\begin{equation}
V_{\mathrm{HF}}=H_{\mathrm{ff}}+H_{\mathrm{zz}}.
\end{equation}

\subsection{Master Equation}

To model the dynamical evolution of the DQD system, we use a master
equation approach. Starting from the full von Neumann equation for
the global density matrix $\varrho$ 
\begin{equation}
\dot{\varrho}=-i\left[\mathcal{H},\varrho\right],\label{eq:von-Neumann}
\end{equation}
we employ a Born-Markov treatment, trace out the reservoir degrees
of freedom, apply the so-called approximation of independent rates
of variation \cite{cohen-tannoudji92}, and assume fast recharging
of the DQD which allows us to eliminate the single-electron levels;\cite{petersen13,giavaras13}
for details, see Appendix \ref{sec:Quantum-Master-Equation}. Then,
we arrive at the following master equation for the system's density
matrix $\rho=\mathsf{Tr}_{\mathsf{B}}\left[\varrho\right]$ 
\begin{equation}
\dot{\rho}=-i\left[H_{\mathrm{el}},\rho\right]+\mathcal{L}_{\Gamma}\rho+\mathcal{V}\rho,\label{eq:QME-Markov-without-cotunneling}
\end{equation}
where $\mathsf{Tr}_{\mathsf{B}}\left[\dots\right]$ denotes the trace
over the bath degrees of freedom in the leads. 
In the following, the Hamiltonian $H_{\mathrm{el}}$ and the superoperators
$\mathcal{L}_{\Gamma}$, $\mathcal{V}$ will be discussed in detail [cf. 
Eqs.(\ref{eq:H-el}), (\ref{eq:Electron-transport-dissipator-dressed}) and  (\ref{eq:Hyperfine-superoperator}), respectively].

\begin{figure}
\includegraphics[width=1\columnwidth]{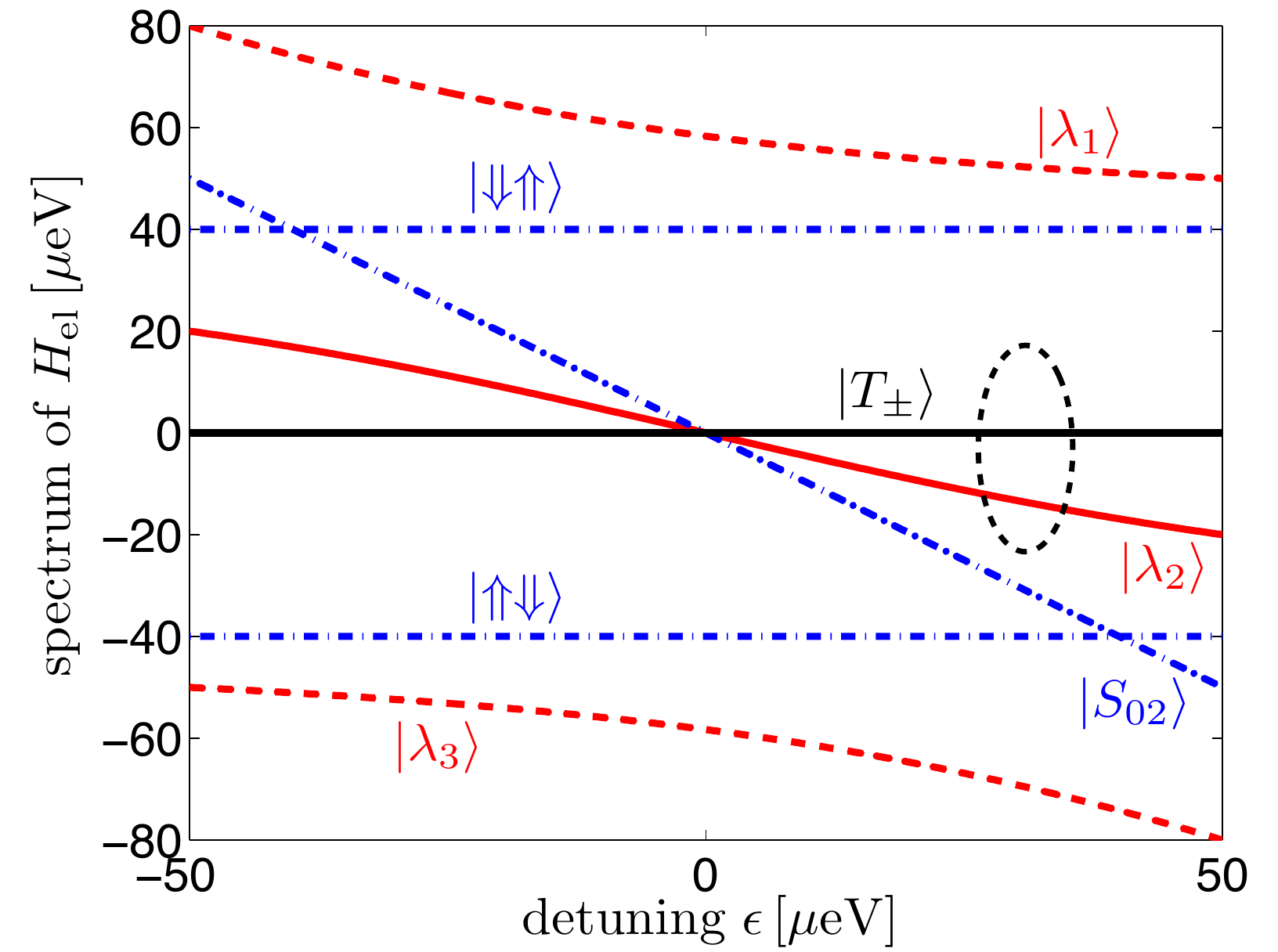}

\caption{\label{fig:spectrum_Hel}(color online). Spectrum of $H_{\mathrm{el}}$
in the relevant two-electron regime for $\Delta=40\mu\mathrm{eV}$
and $t=30\mu\mathrm{eV}$; shown here as a function of the interdot
detuning parameter $\epsilon$. The three hybridized electronic eigenstates
$\left|\lambda_{k}\right\rangle $ within the $S_{\mathrm{tot}}^{z}=0$
subspace are displayed in red, while the bare electronic states are
shown in blue (dash-dotted lines). The homogeneous Zeeman splitting
$\omega_{0}$ has been set to zero, so that the Pauli-blocked triplets
$\left|T_{\pm}\right\rangle $ are degenerate. In this setting, the
levels $\left|\lambda_{1,3}\right\rangle $ are far detuned from $\left|T_{\pm}\right\rangle $.
Therefore, the spin-blockade is lifted pre-dominantly via the non-local
electronic level $\left|\lambda_{2}\right\rangle $. The black dashed
ellipse refers to a potential operational area of our scheme. }
\end{figure}

\textit{Electronic Hamiltonian}.---In Eq.(\ref{eq:QME-Markov-without-cotunneling}),
$H_{\mathrm{el}}$ describes the electronic degrees of freedom of
the DQD within the relevant two-electron subspace. It can be written
as $\left(\hbar=1\right)$ 
\begin{eqnarray}
H_{\mathrm{el}} & = & \omega_{0}\left(S_{L}^{z}+S_{R}^{z}\right)+\Delta\left(S_{R}^{z}-S_{L}^{z}\right)-\epsilon\left|S_{02}\right\rangle \left\langle S_{02}\right|\nonumber \\
 &  & +t\left(\left|\Uparrow\Downarrow\right\rangle \left\langle S_{02}\right|-\left|\Downarrow\Uparrow\right\rangle \left\langle S_{02}\right|+\mathrm{h.c.}\right), \label{eq:H-el}
\end{eqnarray}
where the nuclear-polarization-dependent 'mean-field' quantities $\bar{\omega}_{\mathrm{OH}}$
and $\Delta_{\mathrm{OH}}$ have been absorbed into the definitions
of $\omega_{0}$ and $\Delta$ as $\omega_{0}=\omega_{\mathrm{ext}}+\bar{\omega}_{\mathrm{OH}}$
and $\Delta=\Delta_{\mathrm{ext}}+\Delta_{\mathrm{OH}}$, respectively.
In previous theoretical work, this feedback of the Overhauser shift
on the electronic energy levels has been identified as a means for
controlling the nuclear spins via instabilities towards self-polarization;
compare for example Ref.\cite{rudner07}. Apart from the OH contributions,
$\omega_{\mathrm{ext}}$ and $\Delta_{\mathrm{ext}}$ denote the Zeeman
splitting due to the homogeneous and inhomogeneous component of a
potential external magnetic field, respectively. Furthermore, $\epsilon$
refers to the relative interdot energy detuning between the left and
right dot. The interdot tunneling with coupling strength $t$ occurs
exclusively in the singlet subspace due to Pauli spin-blockade. It
is instructive to diagonalize the effective five-dimensional electronic
Hamiltonian $H_{\mathrm{el}}$. The eigenstates of $H_{\mathrm{el}}$
within the $S_{\mathrm{tot}}^{z}=S_{L}^{z}+S_{R}^{z}=0$ subspace
can be expressed as 
\begin{equation}
\left|\lambda_{k}\right\rangle =\mu_{k}\left|\Uparrow\Downarrow\right\rangle +\nu_{k}\left|\Downarrow\Uparrow\right\rangle +\kappa_{k}\left|S_{02}\right\rangle ,
\end{equation}
for $k=1,2,3$ with corresponding eigenenergies $\epsilon_{k}$; compare
Fig.~\ref{fig:spectrum_Hel}.\cite{amplitudes} Note that, throughout
this work, the hybridized level $\left|\lambda_{2}\right\rangle $
plays a crucial role for the dynamics of the DQD system: Since the
levels $\left|\lambda_{1,3}\right\rangle $ are energetically separated
from all other electronic levels (for $t\gg\omega_{0},g_{\mathrm{hf}}$),
$\left|\lambda_{2}\right\rangle $ represents the dominant channel
for lifting of the Pauli-blockade; compare Fig.~\ref{fig:spectrum_Hel}.

\textit{Electron transport}.---After tracing out the reservoir degrees
of freedom, electron transport induces dissipation in the electronic
subspace: The Liouvillian
\begin{equation}
\mathcal{L}_{\Gamma}\rho=\sum_{k,\nu=\pm}\Gamma_{k}\mathcal{D}\left[\left|T_{\nu}\right\rangle \left\langle \lambda_{k}\right|\right]\rho, 
\label{eq:Electron-transport-dissipator-dressed}
\end{equation}
with the short-hand notation for the Lindblad form $\mathcal{D}\left[c\right]\rho=c\rho c^{\dagger}-\frac{1}{2}\left\{ c^{\dagger}c,\rho\right\} $,
effectively models electron transport through the DQD; here, we have
applied a rotating-wave approximation by neglecting terms rotating
at a frequency of $\epsilon_{k}-\epsilon_{l}$ for $k\neq l$ (see
Appendix \ref{sec:Quantum-Master-Equation} for details). Accordingly,
the hybridized electronic levels $\left|\lambda_{k}\right\rangle \left(k=1,2,3\right)$
acquire a finite lifetime \cite{rudner11b} and decay with a rate
\begin{equation}
\Gamma_{k}=\left|\left<\lambda_{k}|S_{02}\right>\right|^{2}\Gamma=\kappa_{k}^{2}\Gamma,
\end{equation}
determined by their overlap with the localized singlet $\left|S_{02}\right\rangle $,
back into the Pauli-blocked triplet subspace $\left\{ \left|T_{\pm}\right\rangle \right\} $.
Here, $\Gamma=\Gamma_{R}/2$, where $\Gamma_{R}$ is the sequential
tunneling rate to the right lead.

\textit{Hyperfine interaction}.---After splitting off the semiclassical
quantities $\bar{\omega}_{\mathrm{OH}}$ and $\Delta_{\mathrm{OH}}$,
the superoperator
\begin{equation}
\mathcal{V}\rho=-i\left[V_{\mathrm{HF}},\rho\right], \label{eq:Hyperfine-superoperator}
\end{equation}
captures the remaining effects due to the HF coupling between electronic
and nuclear spins. Within the eigenbasis of $H_{\mathrm{el}}$, the
hyperfine flip-flop dynamics $H_{\mathrm{ff}}$, accounting for the
exchange of excitations between the electronic and nuclear subsystem,
takes on the form
\begin{equation}
H_{\mathrm{ff}}=\frac{a_{\mathrm{hf}}}{2}\sum_{k}\left[\left|\lambda_{k}\right\rangle \left\langle T_{+}\right|\otimes L_{k}+\left|\lambda_{k}\right\rangle \left\langle T_{-}\right|\otimes\mathbb{L}_{k}+\mathrm{h.c.}\right],
\end{equation}
where the \textit{non-local} nuclear jump operators 
\begin{eqnarray}
L_{k} & = & \nu_{k}A_{L}^{+}+\mu_{k}A_{R}^{+},\\
\mathbb{L}_{k} & = & \mu_{k}A_{L}^{-}+\nu_{k}A_{R}^{-},
\end{eqnarray}
are associated with lifting the spin-blockade from $\left|T_{+}\right\rangle $
and $\left|T_{-}\right\rangle $ via $\left|\lambda_{k}\right\rangle $,
respectively. These operators characterize the effective coupling
between the nuclear system and its electronic environment; they can
be controlled externally via gate voltages as the parameters $t$
and $\epsilon$ define the amplitudes $\mu_{k}$ and $\nu_{k}$. Since
generically $\mu_{k}\neq\nu_{k}$, the non-uniform electron spin density
of the hybridized eigenstates $\left|\lambda_{k}\right\rangle $ introduces
an asymmetry to flip a nuclear spin on the first or second dot.\cite{rudner11b} 

\begin{figure}
\includegraphics[width=1\columnwidth]{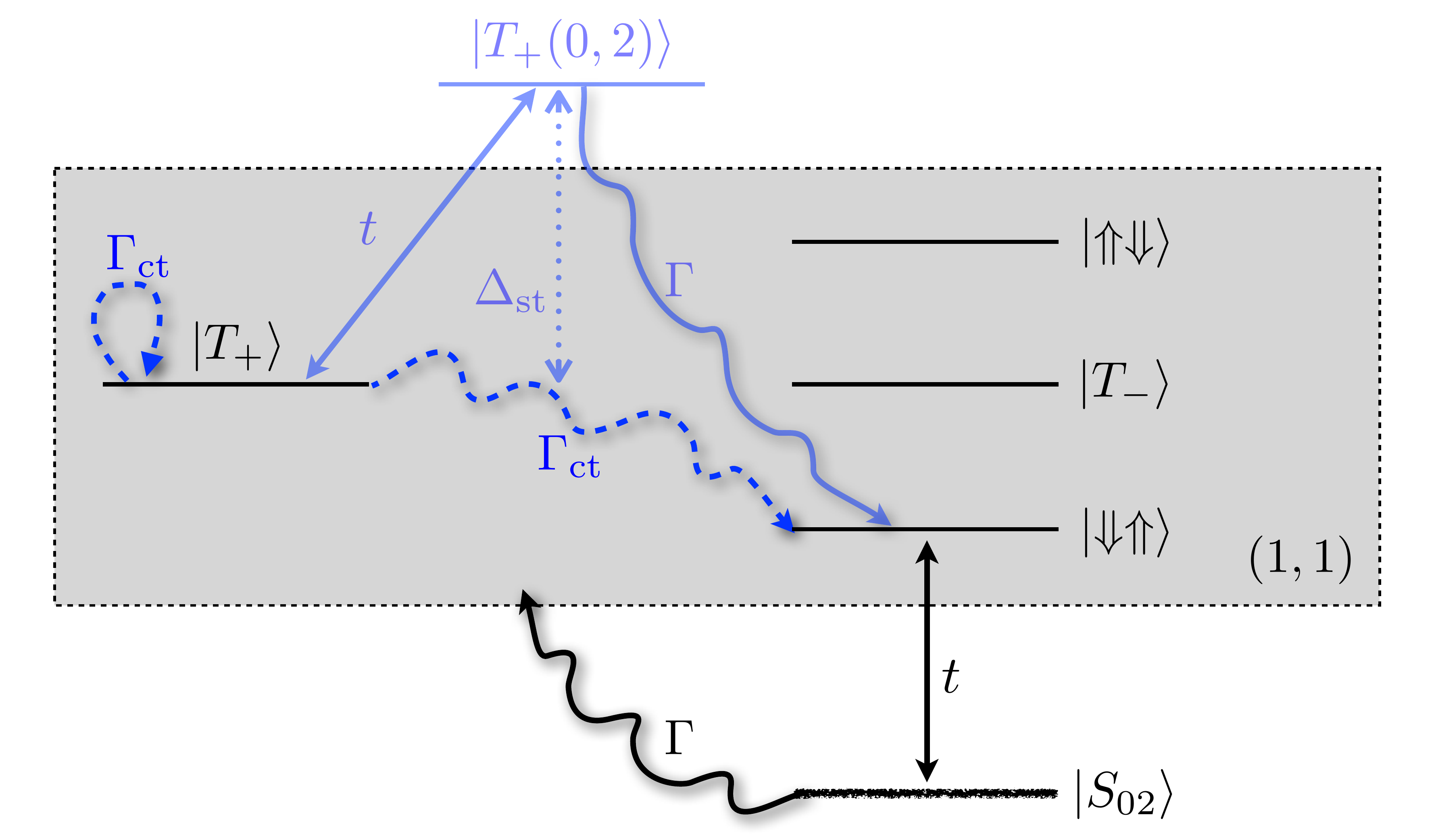}

\caption{\label{fig:scheme-cotunneling-spin-exchange}(color online). Scheme
for the phenomenological cotunneling analysis. The spin-blocked triplet
$\left|T_{+}\right\rangle $ is tunnel-coupled to the (virtually occupied)
triplet $\left|T_{+}\left(0,2\right)\right\rangle $, localized on
the right dot. Due to Pauli exclusion, this level is energetically
well separated by the singlet-triplet splitting $\Delta_{\mathrm{st}}\gtrsim400\mu\mathrm{eV}$.
It has a finite lifetime $\Gamma^{-1}$ and may decay back (via a
singly occupied level on the right dot) to $\left|T_{+}\right\rangle $
or via a series of fast coherent and incoherent intermediate processes
end up in any level within the $(1,1)$ charge sector (shaded box),
since $\left|S_{02}\right\rangle $ decays with a rate $\Gamma$ to
all four $\left(1,1\right)$ states. 
The overall effectiveness of the
process is set by the effective rate 
$\Gamma_{\mathrm{ct}}\approx\left(t/\Delta_{\mathrm{st}}\right)^{2}\Gamma$, depicted by dashed arrows.}
\end{figure}

\begin{figure*}
\includegraphics[width=1.95\columnwidth]{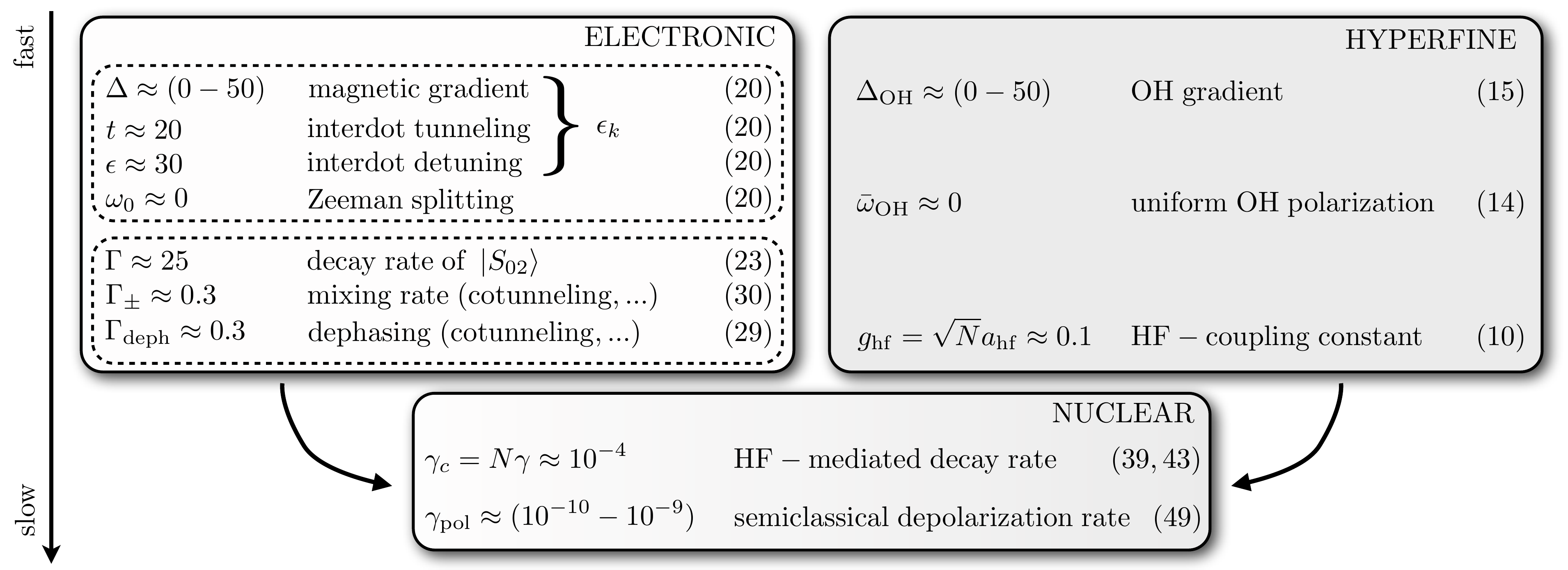}

\caption{\label{fig:overview-symbols}(color online). Schematic overview of the most important parameters in our model, 
grouped into electronic, hyperfine and HF-mediated nuclear quantities. Within the electronic quantities, we can differentiate
between coherent and incoherent processes (compare dashed boxes). 
Typical numbers are given in $\mu\mathrm{eV}$, while the numbers in parentheses $(\cdot)$ refer to the corresponding equations in the text.}
\end{figure*}

\textit{Electronic spin-blockade lifting}.---Apart from the hyperfine
mechanism described above, the Pauli blockade may also be lifted by
other, purely electronic processes such as
(i) cotunneling, (ii) spin-exchange with the leads, or (iii) spin-orbit coupling.\cite{rudner13}
Although they do not exchange excitations with the nuclear spin bath,
these processes have previously been shown to be essential to describe
the nuclear spin dynamics in the Pauli blockade regime.\cite{rudner11b,rudner07,qassemi09}
In our analysis, it is crucial to include them as they affect the
average electronic quasisteady state seen by the nuclei,  
 while the exact, microscopic nature of the electronic decoherence processes 
does not play an important role for our proposal. 
Therefore, for concreteness, here we only describe \textit{exemplarily} virtual tunneling processes via the doubly occupied
triplet state labeled as $\left|T_{+}\left(0,2\right)\right\rangle $, 
while spin-exchange with the leads or spin-orbital effects are discussed in detail 
in Appendix \ref{sec:Electronic-Lifting-of-Pauli-Blockade}.
Cotunneling via $\left|T_{-}\left(0,2\right)\right\rangle $ or $\left|T_{0}\left(0,2\right)\right\rangle $
$ $can be analyzed along the same lines. As schematically depicted
in Fig.~\ref{fig:scheme-cotunneling-spin-exchange}, the triplet $\left|T_{+}\right\rangle $
with $\left(1,1\right)$ charge configuration is coherently coupled
to $\left|T_{+}\left(0,2\right)\right\rangle $ by the interdot tunnel-coupling
$t$. This transition is strongly detuned by the singlet-triplet splitting
$\Delta_{\mathrm{st}}$. Once, the energetically high lying level
$\left|T_{+}\left(0,2\right)\right\rangle $ is populated, it quickly
decays with rate $\Gamma$ either back to $\left|T_{+}\right\rangle $
giving rise to a pure dephasing process within the low-energy subspace
or to $\left\{ \left|T_{-}\right\rangle ,\left|\lambda_{k}\right\rangle \right\} $
via some fast intermediate steps, 
mediated by fast discharging and recharging of the DQD with the rate $\Gamma$.\cite{cotunneling-basis} In our
theoretical model (see below), the former is captured by the pure
dephasing rate $\Gamma_{\mathrm{deph}}$, while the latter can be
absorbed into the dissipative mixing rate $\Gamma_{\pm}$; compare
Fig.~\ref{fig:three-level-justification} for a schematic illustration
of $\Gamma_{\pm}$ and $\Gamma_{\mathrm{deph}}$, respectively. Since
the singlet-triplet splitting is the largest energy scale in this
process $\left(t,\Gamma\ll\Delta_{\mathrm{st}}\right)$, the effective
rate for this virtual cotunneling mechanism can be estimated as 
\begin{equation}
\Gamma_{\mathrm{ct}}\approx\left(t/\Delta_{\mathrm{st}}\right)^{2}\Gamma.\label{eq:cotunneling-rate}
\end{equation}
Equation (\ref{eq:cotunneling-rate}) describes a virtually assisted process
by which $t$ couples $\left|T_{+}\right\rangle $ to a virtual level,
which can then escape via sequential tunneling $\sim\Gamma$; thus,
it can be made relatively fast compared to typical nuclear timescales
by working in a regime of efficient electron exchange with the leads
$\sim\Gamma$.\cite{cotunneling-asymmetry} For example, taking $t\approx30\mu\mathrm{eV}$,
$\Delta_{\mathrm{st}}\approx400\mu\mathrm{eV}$ and $\Gamma\approx50\mu\mathrm{eV}$,
we estimate $\Gamma_{\mathrm{ct}}\approx0.3\mu\mathrm{eV}$, 
which is fast compared to typical nuclear timescales. 
Note that, for more conventional, \textit{slower} electronic parameters ($t\approx5\mu\mathrm{eV}$, $\Gamma\approx0.5\mu\mathrm{eV}$), 
indirect tunneling becomes negligibly small, $\Gamma_{\mathrm{ct}}\approx5\times10^{-5}\mu\mathrm{eV}\approx5\times10^4\mathrm{s}^{-1}$, 
in agreement with values given in Ref.\cite{rudner11b}. 
Our analysis, however, is restricted to the regime, where indirect tunneling is fast compared to the nuclear dynamics;  
this regime of motional averaging has previously been shown to be beneficial for e.g. nuclear spin squeezing. \cite{rudner11a,rudner07}
Alternatively, spin-blockade may be lifted via spin-exchange with the leads. The corresponding rate 
$\Gamma_{\mathrm{se}}$ scales as $\Gamma_{\mathrm{se}} \sim \Gamma^2$, as compared to 
$\Gamma_{\mathrm{ct}} \sim t^2\Gamma$. Moreover, $\Gamma_{\mathrm{se}}$ depends strongly on the detuning 
of the $(1,1)$ levels from the Fermi levels of the leads. 
If this detuning is $\sim500\mu\mathrm{eV}$ and for $\Gamma\approx100\mu\mathrm{eV}$, 
we estimate $\Gamma_{\mathrm{se}}\approx0.25\mu\mathrm{eV}$, 
which is commensurate with the desired motional averaging regime,
whereas, for less efficient transport ($\Gamma\approx1\mu\mathrm{eV}$) and stronger detuning $\sim1\mathrm{meV}$, 
one obtains a negligibly small rate, 
$\Gamma_{\mathrm{se}}\approx 6\times10^{-6}\mu\mathrm{eV}\approx 6\times10^{3}\mathrm{s}^{-1}$.
Again, this is in line with Ref.\cite{rudner11b}.  
As discussed in more detail in Appendix \ref{sec:Electronic-Lifting-of-Pauli-Blockade}, these spin-exchange
processes as well as spin-orbital effects can be treated on a similar
footing as the interdot cotunneling processes discussed here. Therefore, 
to describe the net effect of various non-hyperfine
mechanisms and to complete our theoretical description of electron
transport in the spin-blockade regime, we add the following phenomenological
Lindblad terms to our model 
\begin{eqnarray}
\mathcal{L}_{\mathrm{deph}}\rho & = & \frac{\Gamma_{\mathrm{deph}}}{2}\mathcal{D}\left[\left|T_{+}\right\rangle \left\langle T_{+}\right|-\left|T_{-}\right\rangle \left\langle T_{-}\right|\right]\rho,\label{eq:Liouvillian-dephasing}\\
\mathcal{L}_{\pm}\rho & = & \Gamma_{\pm}\sum_{\nu=\pm}\mathcal{D}\left[\left|T_{\bar{\nu}}\right\rangle \left\langle T_{\nu}\right|\right]\rho\label{eq:Liouvillian-mixing}\\
 &  & +\Gamma_{\pm}\sum_{k,\nu}\mathcal{D}\left[\left|T_{\nu}\right\rangle \left\langle \lambda_{k}\right|\right]\rho+\mathcal{D}\left[\left|\lambda_{k}\right\rangle \left\langle T_{\nu}\right|\right]\rho.\nonumber 
\end{eqnarray}

\textit{Summary}.---Before concluding the description of the system
under study, let us quickly reiterate the ingredients of the master
equation as stated in Eq.(\ref{eq:effective-QME-full-model}): It
accounts for (i) the unitary dynamics within the DQD governed by $-i\left[H_{\mathrm{el}}+V_{\mathrm{HF}},\rho\right]$,
(ii) electron-transport-mediated dissipation via $\mathcal{L}_{\Gamma}$
and (iii) dissipative mixing and dephasing processes described by
$\mathcal{L}_{\pm}$ and $\mathcal{L}_{\mathrm{deph}}$, respectively. 
Finally, the most important parameters of our model are summarized in Fig.~\ref{fig:overview-symbols}.

\section{Effective Nuclear Dynamics \label{sec:Effective-Nuclear-Dynamics}}

\begin{figure*}
\includegraphics[width=1.95\columnwidth]{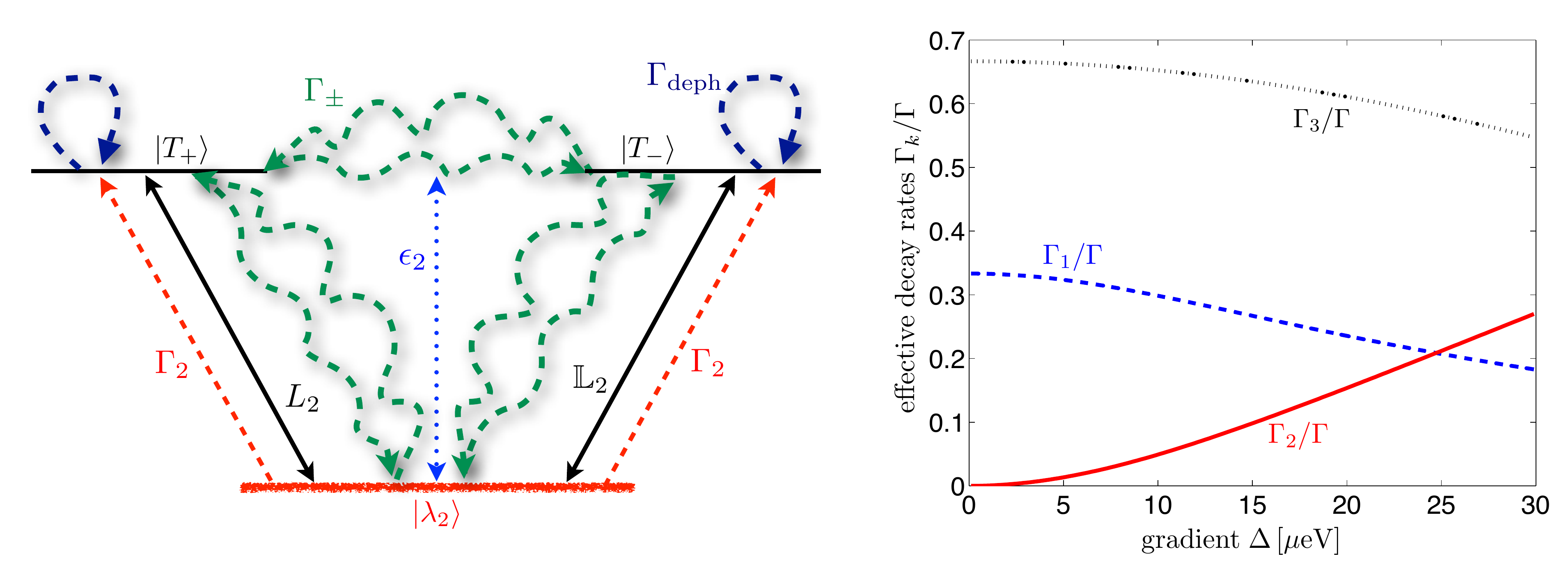}

\caption{\label{fig:three-level-justification}(color online). Left plot: Schematic
illustration of coherent and incoherent processes within the effective
three-level system $\left\{ \left|T_{\pm}\right\rangle ,\left|\lambda_{2}\right\rangle \right\} $:
The level $\left|\lambda_{2}\right\rangle $ is detuned from $\left|T_{\pm}\right\rangle $
by $\epsilon_{2}$ and decays according to its overlap with the
localized singlet with the rate $\Gamma_{2}=\kappa_{2}^{2}\Gamma$.
Moreover, it is coherently coupled to the triplets $\left|T_{+}\right\rangle $
and $\left|T_{-}\right\rangle $ via the \textit{nonlocal} nuclear
operators $L_{2}$ and $\mathbb{L}_{2}$, respectively. Purely electronic
spin-blockade lifting mechanisms such as cotunneling or spin-orbital
effects result in effective dephasing and dissipative mixing rates,
labeled as $\Gamma_{\mathrm{deph}}$ and $\Gamma_{\pm}$, respectively.
$ $The latter do not affect the nuclei directly, but lead to an unbiased
population transfer within the electronic three-level system. 
In particular, mixing between $\left|T_{\pm}\right\rangle$ can arise 
from virtual occupation of $\left|\lambda_{1,3}\right\rangle$ or spin-orbit coupling.
Right plot: Effective decay rates $\Gamma_{k}=\kappa_{k}^{2}\Gamma$, shown
here for $\epsilon=t=30\mu\mathrm{eV}$. For small gradients, $\left|\lambda_{2}\right\rangle \approx\left|T_{0}\right\rangle $
and therefore it does not decay due to Pauli-blockade. }
\end{figure*}

In this section we develop the general theoretical framework of our
analysis which is built upon the fact that, generically, the nuclear
spins evolve slowly on typical electronic timescales. Due to this
separation of electronic and nuclear timescales, the system is subject
to the slaving principle \cite{yamamoto99} implying that the electronic
subsystem settles to a quasisteady state on a timescale much shorter
than the nuclear dynamics. This allows us to adiabatically eliminate
the electronic coordinates yielding an effective master equation on
a coarse-grained timescale. Furthermore, the electronic quasisteady
state is shown to depend on the state of the nuclei resulting in feedback
mechanisms between the electronic and nuclear degrees of freedom.
Specifically, here we analyze the dynamics of the nuclei coupled to
the electronic three-level subspace spanned by the levels $\left|T_{\pm}\right\rangle $
and $\left|\lambda_{2}\right\rangle $. This simplification is justified
for $t\gg\omega_{0},g_{\mathrm{hf}}$, since in this parameter regime
the electronic levels $\left|\lambda_{1,3}\right\rangle $ are strongly
detuned from the manifold $\left\{ \left|T_{\pm}\right\rangle ,\left|\lambda_{2}\right\rangle \right\} $;
compare Fig.~\ref{fig:spectrum_Hel}. Effects due to the presence of
$\left|\lambda_{1,3}\right\rangle $ will be discussed separately
in Secs. \ref{sec:Steady-State-Entanglement} and \ref{sec:Criticality}.
Here, due to their fast decay with a rate $\Gamma_{1,3}$, 
they have already been eliminated adiabatically from the dynamics, leading to a dissipative mixing between the blocked triplet 
states $\left|T_{\pm}\right\rangle$ with rate $\Gamma_{\pm}$; alternatively, this mixing could come from spin-orbit coupling 
(see Appendix  \ref{sec:Electronic-Lifting-of-Pauli-Blockade} for details).
Moreover, for simplicity, we assume $\omega_{0}=0$ and neglect nuclear fluctuations
arising from $H_{\mathrm{zz}}$. This approximation is in line with
the semiclassical approach used below in order to study the nuclear
polarization dynamics; for details we refer to Appendix \ref{sec:Effective-Nuclear-Dynamics:OH-fluctuations}.
In summary, all relevant coherent and incoherent processes within
the effective three-level system $\left\{ \left|T_{\pm}\right\rangle ,\left|\lambda_{2}\right\rangle \right\} $
are schematically depicted in Fig.~\ref{fig:three-level-justification}.

\textit{Intuitive picture}.---The main results of this section can
be understood from the fact that the level $\left|\lambda_{2}\right\rangle $
decays according to its overlap with the localized singlet, that is
with a rate 
\begin{equation}
\Gamma_{2}=\left|\left<\lambda_{2}|S_{02}\right>\right|^{2}\Gamma\overset{\Delta\rightarrow0}{\longrightarrow}0
\end{equation}
which in the low-gradient regime $\Delta\approx0$ tends to zero,
since then $\left|\lambda_{2}\right\rangle $ approaches the triplet
$\left|T_{0}\right\rangle $ which is dark with respect to tunneling
and therefore does not allow for electron transport; see Fig.~\ref{fig:three-level-justification}.
In other words, in the limit $\Delta\rightarrow0$, the electronic
level $\left|\lambda_{2}\right\rangle \rightarrow\left|T_{0}\right\rangle $
gets stabilized by Pauli-blockade. In this regime, we expect the nuclear
spins to undergo some form of random diffusion process since the dynamics
lack any \textit{directionality}: the operators $L_{2}\left(\mathbb{L}_{2}\right)$
and their respective adjoints $L_{2}^{\dagger}(\mathbb{L}_{2}^{\dagger})$
act with equal strength on the nuclear system. In contrast, in the
high-gradient regime, $ $$\left|\lambda_{2}\right\rangle $ exhibits
a significant singlet character and therefore gets depleted very quickly.
Thus, $ $$\left|\lambda_{2}\right\rangle $ can be eliminated adiabatically
from the dynamics, the electronic subsystem settles to a maximally
mixed state in the Pauli-blocked $\left|T_{\pm}\right\rangle $ subspace
and the nuclear dynamics acquire a certain directionality in that
now the nuclear spins experience dominantly the action of the non-local
operators $L_{2}$ and $\mathbb{L}_{2}$, respectively. As will be
shown below, this directionality features both the build-up of an
Overhauser field gradient and entanglement generation between the
two nuclear spin ensembles.

\subsection{Adiabatic Elimination of Electronic Degrees of Freedom}

Having separated the macroscopic semiclassical part of the nuclear
Overhauser fields, the problem at hand features a hierarchy in the
typical energy scales since the typical HF interaction strength is
slow compared to all relevant electronic timescales. This allows for
a perturbative approach to second order in $\mathcal{V}$ to derive
an effective master equation for the nuclear subsystem.\cite{schuetz12,kessler12}
To stress the perturbative treatment, the full quantum master equation
can formally be decomposed as 
\begin{equation}
\dot{\rho}=\left[\mathcal{L}_{0}+\mathcal{V}\right]\rho,
\end{equation}
where the superoperator $\mathcal{L}_{0}$ acts on the electron degrees
of freedom only and the HF interaction represents a perturbation.
Thus, in zeroth order the electronic and nuclear dynamics are decoupled.
In what follows, we will determine the effective nuclear evolution
in the submanifold of the electronic quasisteady states of $\mathcal{L}_{0}$.
The electronic Liouvillian $\mathcal{L}_{0}$ features a \textit{unique}
steady state \cite{sanchez13}, that is $\mathcal{L}_{0}\rho_{\mathrm{ss}}^{\mathrm{el}}=0$
for
\begin{equation}
\rho_{\mathrm{ss}}^{\mathrm{el}}=p\left(\left|T_{+}\right\rangle \left\langle T_{+}\right|+\left|T_{-}\right\rangle \left\langle T_{-}\right|\right)+\left(1-2p\right)\left|\lambda_{2}\right\rangle \left\langle \lambda_{2}\right|,\label{eq:quasi-steady-state-general}
\end{equation}
where 
\begin{equation}
p=\frac{\Gamma_{\pm}+\Gamma_{2}}{3\Gamma_{\pm}+2\Gamma_{2}},\label{eq:parameter-p}
\end{equation}
completely defines the electronic quasisteady state. It captures the
competition between undirected population transfer within the the
manifold $\left\{ \left|T_{\pm}\right\rangle ,\left|\lambda_{2}\right\rangle \right\} $
due to $\Gamma_{\pm}$ and a unidirectional, electron-transport-mediated
decay of $\left|\lambda_{2}\right\rangle $. Moreover, it describes
feedback between the electronic and nuclear degrees of freedom as
the rate $\Gamma_{2}$ depends on the gradient $\Delta$ which incorporates
the nuclear-polarization-dependent Overhauser gradient $\Delta_{\mathrm{OH}}$.
We can immediately identify two important limits which will be analyzed
in greater detail below: For $\Gamma_{\pm}\gg\Gamma_{2}$ we get $p=1/3$,
whereas $\Gamma_{\pm}\ll\Gamma_{2}$ results in $p=1/2$, that is
a maximally mixed state in the $\left|T_{\pm}\right\rangle $ subspace,
since a fast decay rate $\Gamma_{2}$ leads to a complete depletion
of $\left|\lambda_{2}\right\rangle $. 

Since $\rho_{\mathrm{ss}}^{\mathrm{el}}$ is unique, the projector
$\mathcal{P}$ on the subspace of zero eigenvalues of $\mathcal{L}_{0}$,
i.e., the zeroth order steady states, is given by 
\begin{equation}
\mathcal{P}\rho=\mathsf{Tr}_{\mathrm{el}}\left[\rho\right]\otimes\rho_{\mathrm{ss}}^{\mathrm{el}}=\sigma\otimes\rho_{\mathrm{ss}}^{\mathrm{el}}.
\end{equation}
By definition, we have $\mathcal{P}\mathcal{L}_{0}=\mathcal{L}_{0}\mathcal{P}=0$
and $\mathcal{P}^{2}=\mathcal{P}$. The complement of $\mathcal{P}$
is $\mathcal{Q}=\mathbb{1}-\mathcal{P}$. Projection of the master
equation on the $\mathcal{P}$ subspace gives in second-order perturbation
theory
\begin{equation}
\frac{d}{dt}\mathcal{P}\rho=\left[\mathcal{P}\mathcal{V}\mathcal{P}-\mathcal{P}\mathcal{V}\mathcal{Q}\mathcal{L}_{0}^{-1}\mathcal{Q}\mathcal{V}\mathcal{P}\right]\rho,
\end{equation}
from which we can deduce the required equation of motion $\dot{\sigma}=\mathcal{L}_{\mathrm{eff}}\left[\sigma\right]$
for the reduced density operator of the nuclear subsystem $\sigma=\mathsf{Tr}_{\mathrm{el}}\left[\mathcal{P}\rho\right]$
as 
\begin{eqnarray}
\dot{\sigma} & = & \mathsf{Tr}_{\mathrm{el}}\left[\mathcal{P}\mathcal{V}\mathcal{P}\rho-\mathcal{P}\mathcal{V}\mathcal{Q}\mathcal{L}_{0}^{-1}\mathcal{Q}\mathcal{V}\mathcal{P}\rho\right].\label{eq:effective-nuclear-QME-2nd-order-general}
\end{eqnarray}
The subsequent, full calculation follows the general framework developed
in Ref.\cite{kessler12b} and is presented in detail in Appendices
\ref{sec:Effective-Nuclear-Master-Equation-Adiabatic-Elimination}
and \ref{sec:Appendix-Effective-Nuclear-QME-High-Gradient-Regime}.
We then arrive at the following effective master equation for nuclear
spins

\begin{eqnarray}
\dot{\sigma} & = & \gamma\left\{ p\left[\mathcal{D}\left[L_{2}\right]\sigma+\mathcal{D}\left[\mathbb{L}_{2}\right]\sigma\right]\right.\label{eq:effective-nuclear-QME-electronic-general}\\
 &  & \left.+\left(1-2p\right)\left[\mathcal{D}\left[L_{2}^{\dagger}\right]+\mathcal{D}\left[\mathbb{L}_{2}^{\dagger}\right]\sigma\right]\right\} \nonumber \\
 &  & +i\delta\left\{ p\left(\left[L_{2}^{\dagger}L_{2},\sigma\right]+\left[\mathbb{L}_{2}^{\dagger}\mathbb{L}_{2},\sigma\right]\right)\right.\nonumber \\
 &  & \left.-\left(1-2p\right)\left(\left[L_{2}L_{2}^{\dagger},\sigma\right]+\left[\mathbb{L}_{2}\mathbb{L}_{2}^{\dagger},\sigma\right]\right)\right\} .\nonumber 
\end{eqnarray}
Here, we have introduced the effective quantities 
\begin{eqnarray}
\gamma & = & \frac{a_{\mathrm{hf}}^{2}\tilde{\Gamma}}{2\left[\tilde{\Gamma}^{2}+\text{\ensuremath{\epsilon}}_{2}^{2}\right]},\label{eq:nuc-diss-rate-gamma}\\
\delta & = & \frac{a_{\mathrm{hf}}^{2}\epsilon_{2}}{4\left[\tilde{\Gamma}^{2}+\text{\ensuremath{\epsilon}}_{2}^{2}\right]},
\end{eqnarray}
and 
\begin{equation}
\tilde{\Gamma}=\Gamma_{2}+2\Gamma_{\pm}+\frac{\Gamma_{\mathrm{deph}}}{4}.
\end{equation}
The master equation in Eq.(\ref{eq:effective-nuclear-QME-electronic-general})
is our first main result. It is of Lindblad form and incorporates
electron-transport-mediated jump terms as well as Stark shifts. The
two main features of Eq.(\ref{eq:effective-nuclear-QME-electronic-general})
are: (i) The dissipative nuclear jump terms are governed by the \textit{nonlocal}
jump operators $L_{2}$ and $\mathbb{L}_{2}$, respectively. (ii)
The effective dissipative rates $\sim p\gamma$ incorporate intrinsic
electron-nuclear feedback effects as they depend on the macroscopic
state of the nuclei via the parameter $p$ and the decay rate $\Gamma_{2}$.
Because of this feedback mechanism, we can distinguish two very different
fixed points for the coupled electron-nuclear evolution. This is discussed
below.

\subsection{Low-Gradient Regime: Random Nuclear Diffusion}

\begin{figure}
\includegraphics[width=1\columnwidth]{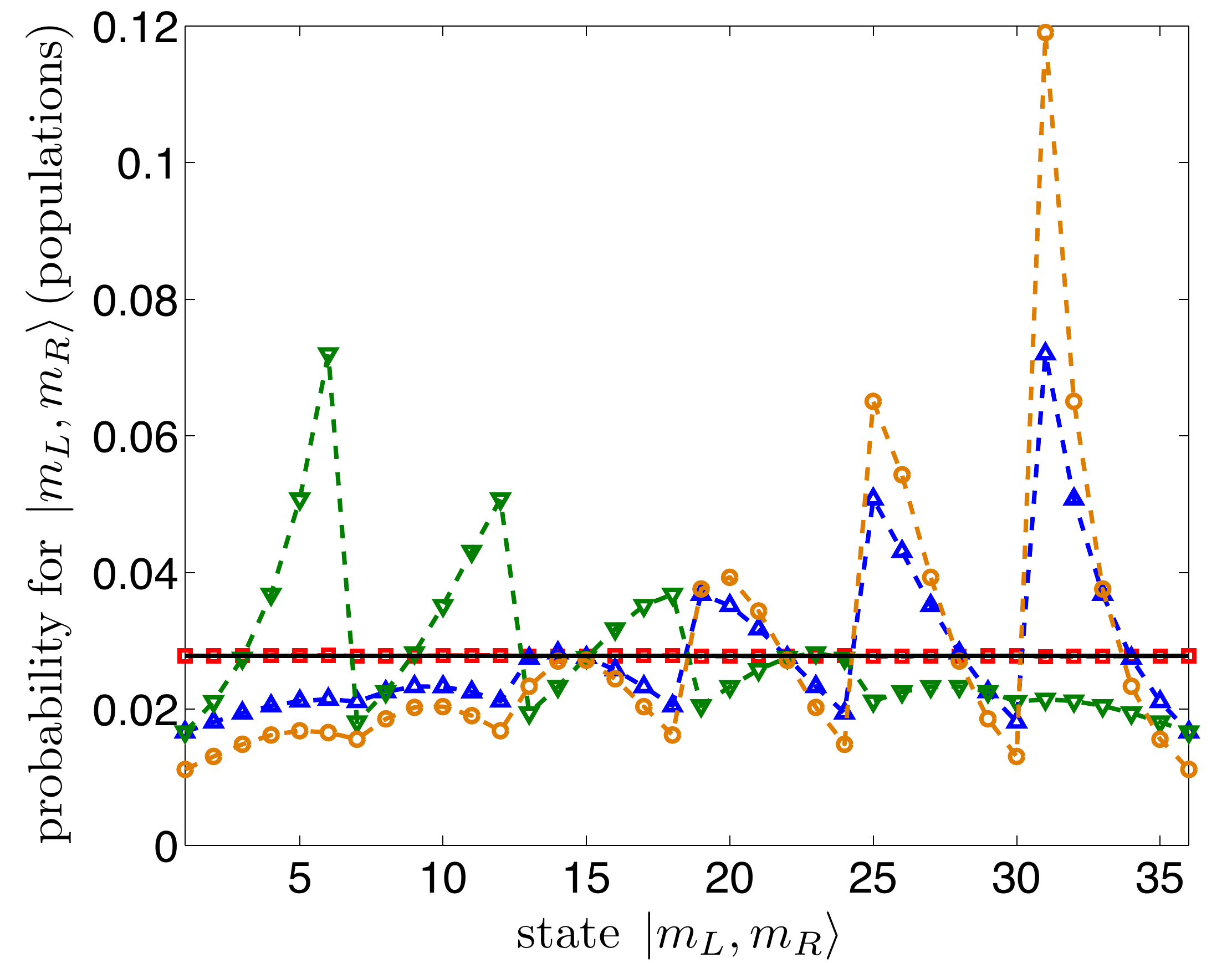}

\caption{\label{fig:external-gradient-N10}(color online). Exact steady-state
as a function of the inhomogeneous splitting $\Delta$; results for
$10$ nuclear spins, five in each quantum dot. We plot the diagonal
elements of the nuclear steady-state density matrix $\sigma_{\mathrm{ss}}$
(i.e. the nuclear populations); its dimension is $\left(2J_{L}+1\right)\left(2J_{R}+1\right)=36$.
For a small external gradient of the order of natural fluctuations
of the Overhauser field (red squares) the nuclear system settles into
the fully mixed state, as evidenced by the uniform populations of
the nuclear levels. However, as we increase the gradient $\Delta$,
the nuclear steady state starts to display a structure different from
the fully mixed state, showing a dominant peak in the occupation of
the nuclear level with maximum gradient, that is $\left|-J_{L},J_{R}\right\rangle $
and $\left|J_{L},-J_{R}\right\rangle $ for $\Delta>0$ and $\Delta<0$,
respectively. The upward triangles, downward triangles and circles
refer to $\Delta=5\mu\mathrm{eV}$, $\Delta=-5\mu\mathrm{eV}$ and
$\Delta=10\mu\mathrm{eV}$, respectively. Other numerical parameters
are: $\Gamma=10\mu\mathrm{eV}$, $\Gamma_{\pm}=0.3\mu\mathrm{eV}$,
$\Gamma_{\mathrm{deph}}=3\mu\mathrm{eV}$, $\omega_{0}=0$, $t=\mathrm{20}\mu\mathrm{eV}$
and $\epsilon=30\mu\mathrm{eV}$. }
\end{figure}

As argued qualitatively above, in the low-gradient regime where $\left|\lambda_{2}\right\rangle \approx\left|T_{0}\right\rangle $,
the nuclear master equation given in Eq.(\ref{eq:effective-nuclear-QME-electronic-general})
lacks any directionality. Accordingly, the resulting dynamics may
be viewed as a random nuclear diffusion process. Indeed, in the limit
$\Gamma_{2}\rightarrow0$, it is easy to check that $p=1/3$ and $\sigma_{\mathrm{ss}}\propto\mathbb{1}$
is a steady-state solution. Therefore, both the electronic and the
nuclear subsystem settle into the fully mixed state with no preferred
direction nor any peculiar polarization characteristics. 

This analytical argument is corroborated by exact numerical simulations
(i.e., without having eliminated the electronic degrees of freedom)
for the full five-level electronic system coupled to ten $(N_{L}=N_{R}=5)$
nuclear spins. Here, we assume homogeneous HF coupling (effects due
to non-uniform HF couplings are discussed in Section \ref{sec:Implementation}):
Then, the total spins $J_{i}$ are conserved and it is convenient
to describe the nuclear spin system in terms of Dicke states $\left|J_{i},m_{i}\right\rangle $
with total spin quantum number $J_{i}$ and spin projection $m_{i}=-J_{i},\dots,J_{i}$.
Fixing the (conserved) total spin quantum numbers $J_{i}=N_{i}/2$,
we write in short $\left|J_{L},m_{L}\right\rangle \otimes\left|J_{R},m_{R}\right\rangle =\left|m_{L},m_{R}\right\rangle $.
In order to realistically mimic the perturbative treatment of the
HF coupling in an experimentally relevant situation where $N\approx10^{6}$,
here the HF coupling constant $g_{\mathrm{hf}}=A_{\mathrm{HF}}/\sqrt{N}$
is scaled down to a constant value of $g_{\mathrm{hf}}=0.1\mu\mathrm{eV}$.
Moreover, let us for the moment neglect the nuclear fluctuations due
to $H_{\mathrm{zz}}$, in order to restrict the following analysis
to the semiclassical part of the nuclear dynamics; compare also previous
theoretical studies.\cite{rudner07,rudner11b,vink09} In later setions,
this part of the dynamics will be taken into account again. In particular,
we compute the steady state and analyze its dependence on the gradient
$\Delta$: Experimentally, $\Delta$ could be induced intrinsically
via a nuclear Overhauser gradient $\Delta_{\mathrm{OH}}$ or extrinsically
via a nano- or micro-magnet.\cite{petersen13,pioro-ladriere08} The
results are displayed in Fig.~\ref{fig:external-gradient-N10}: Indeed,
in the low-gradient regime the nuclear subsystem settles into the
fully mixed state. However, outside of the low-gradient regime, the
nuclear subsystem is clearly driven away from the fully mixed state
and shows a tendency towards the build-up of a nuclear Overhauser
gradient. For $\Delta>0$, we find numerically an increasing population
(in descending order) of the levels $ $$\left|-J_{L},J_{R}\right\rangle ,\left|-J_{L}+1,J_{R}-1\right\rangle $
etc., whereas for $\Delta<0$ strong weights are found at $\left|J_{L},-J_{R}\right\rangle ,\left|J_{L}-1,-J_{R}+1\right\rangle ,\dots$
which effectively increases $\Delta$ such that the nuclear spins
actually tend to self-polarize. This trend towards self-polarization
and the peculiar structure of the nuclear steady state $\sigma_{\mathrm{ss}}$
displayed in Fig.~\ref{fig:external-gradient-N10} is in very good
agreement with the ideal nuclear two-mode squeezedlike steady-state
that we are to construct analytically in the next subsection.

\subsection{High-Gradient Regime: Entanglement Generation}

\begin{figure}
\includegraphics[width=1\columnwidth]{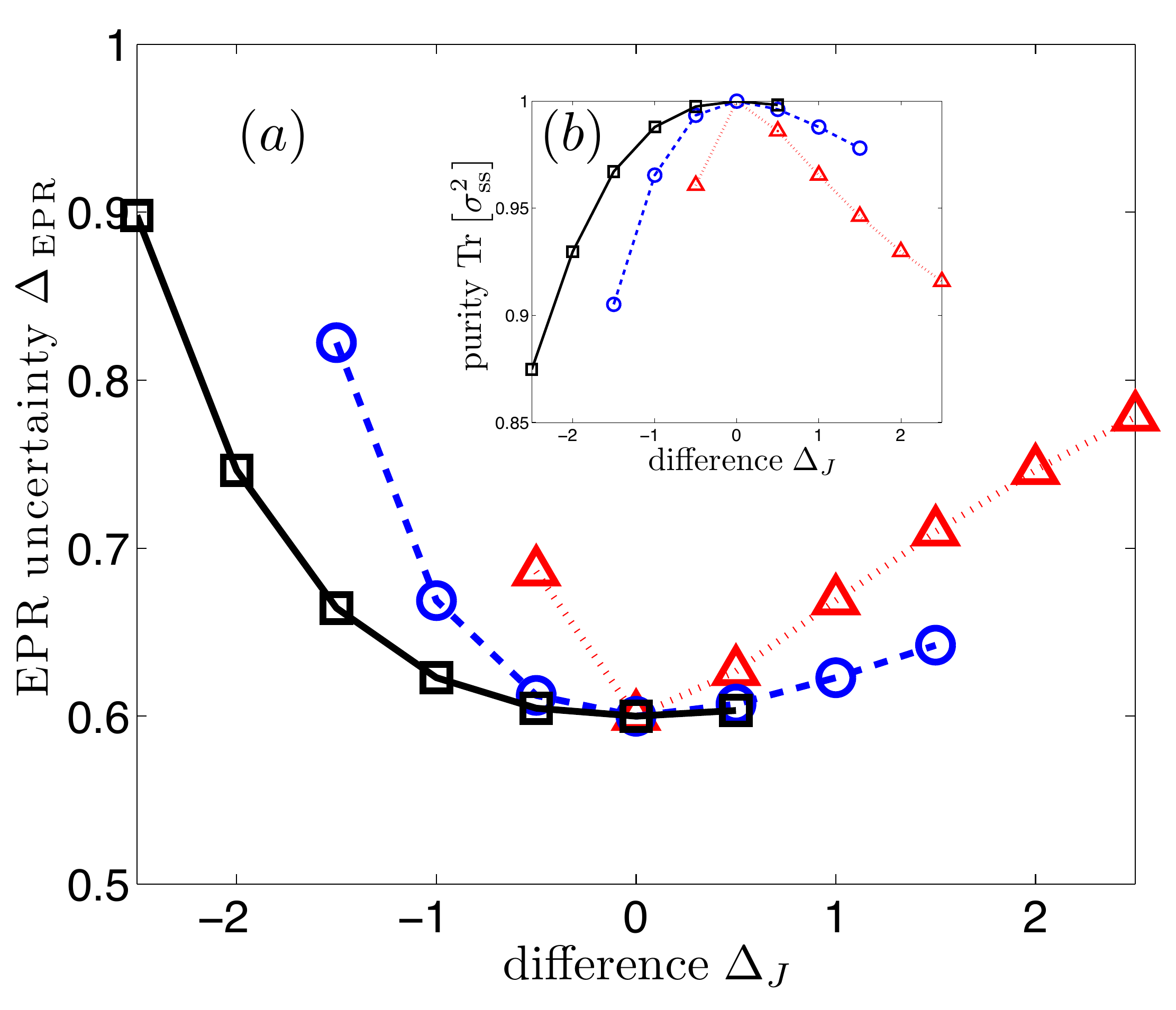}

\caption{\label{fig:ideal-target-state-uniform-HF-coupling}(color online).
EPR-uncertainty $\Delta_{\mathrm{EPR}}$ and purity of the (exact)
nuclear dark states fulfilling $ $$\mathcal{D}\left[L_{2}\right]\sigma_{\mathrm{ss}}+\mathcal{D}\left[\mathbb{L}_{2}\right]\sigma_{\mathrm{ss}}=0$
for small system sizes $J_{i}$.\cite{simiplification-delta=00003D0}
We fix $J_{L}$ to $J_{L}=1$ (triangles), $J_{L}=2$ (circles) and
$J_{L}=3$ (squares) and compute $\sigma_{\mathrm{ss}}$ for different
values of $\Delta_{J}$; $J_{R}$ runs from $0.5$ up to $3.5$. In
the symmetric scenario $\Delta_{J}=0$, $\sigma_{\mathrm{ss}}$ is
pure and given by the two-mode squeezed like state $\sigma_{\mathrm{ss}}=\left|\xi_{\mathrm{ss}}\right\rangle \left\langle \xi_{\mathrm{ss}}\right|$.
For $\Delta_{J}\neq0$, $\sigma_{\mathrm{ss}}$ is mixed; however,
the purity $\mathsf{Tr}\left[\sigma_{\mathrm{ss}}^{2}\right]$ (inset)
as well as $\Delta_{\mathrm{EPR}}$ increase with the system size
$J_{L}+J_{R}$. In all cases, $\sigma_{\mathrm{ss}}$ was found to
be unique. Here, we have set $\left|\xi\right|=0.25$. }
\end{figure}

In the high-gradient regime the electronic level $\left|\lambda_{2}\right\rangle $
overlaps significantly with the localized singlet $\left|S_{02}\right\rangle $.
For $\Gamma_{2}\gg\Gamma_{\pm}$ it decays sufficiently fast such
that it can be eliminated adiabatically from the dynamics. As can
be seen from Eqs.(\ref{eq:quasi-steady-state-general}) and (\ref{eq:parameter-p}),
on typical nuclear timescales, the electronic subsystem then quickly
settles into the quasisteady state given by $\rho_{\mathrm{ss}}^{\mathrm{el}}=\left(\left|T_{+}\right\rangle \left\langle T_{+}\right|+\left|T_{-}\right\rangle \left\langle T_{-}\right|\right)/2$
and the effective master equation for the nuclear spin density matrix
$\sigma$ simplifies to
\begin{equation}
\dot{\sigma}=\frac{\gamma}{2}\left[\mathcal{D}\left[L_{2}\right]\sigma+\mathcal{D}\left[\mathbb{L}_{2}\right]\sigma\right]+i\frac{\delta}{2}\left(\left[L_{2}^{\dagger}L_{2},\sigma\right]+\left[\mathbb{L}_{2}^{\dagger}\mathbb{L}_{2},\sigma\right]\right).\label{eq:effective-QME-nuclear-spins-high-gradient-simple}
\end{equation}
For later reference, the typical timescale of this dissipative dynamics
is set by the rate 
\begin{equation}
\gamma_{c}=N\gamma=\frac{g_{\mathrm{hf}}^{2}\tilde{\Gamma}}{2\left[\tilde{\Gamma}^{2}+\text{\ensuremath{\epsilon}}_{2}^{2}\right]},
\end{equation}
which is \textit{collectively} enhanced by a factor of $N\approx10^{6}$
to account for the norm of the collective nuclear spin operators $A_{i}^{\pm}$.
This results in the typical HF-mediated interaction strength of $g_{\mathrm{hf}}=\sqrt{N}a_{\mathrm{hf}}$,\cite{schuetz12}
and for typical parameter values we estimate $\gamma_{c}\approx10^{-4}\mu\mathrm{eV}$. 

This evolution gives rise to the desired, entangling nuclear squeezing
dynamics: It is easy to check that all \textit{pure} stationary solutions
$\left|\xi_{\mathrm{ss}}\right\rangle $ of this Lindblad evolution
can be found via the dark-state condition $L_{2}\left|\xi_{\mathrm{ss}}\right\rangle =\mathbb{L}_{2}\left|\xi_{\mathrm{ss}}\right\rangle =0$.
Next, we explicitly construct $\left|\xi_{\mathrm{ss}}\right\rangle $
in the limit of equal dot sizes $\left(N_{L}=N_{R}\right)$ and uniform
HF coupling $\left(a_{i,j}=N/N_{i}\right)$, and generalize our results
later. In this regime, again it is convenient to describe the nuclear
system in terms of Dicke states $\left|J_{i},k_{i}\right\rangle $,
where $k_{i}=0,\dots,2J_{i}$. For the symmetric scenario $J_{L}=J_{R}=J$,
one can readily verify that the dark state condition is satisfied
by the (unnormalized) pure state
\begin{equation}
\left|\xi_{\mathrm{ss}}\right\rangle =\sum_{k=0}^{2J}\xi^{k}\left|J,k\right\rangle _{L}\otimes\left|J,2J-k\right\rangle _{R}.\label{eq:target-squeezed-state-J-subspaces-uniform}
\end{equation}
This nuclear state may be viewed as an extension of the two-mode squeezed
state familiar from quantum optics \cite{muschik11} to finite dimensional
Hilbert spaces. The parameter $\xi=-\nu_{2}/\mu_{2}$ quantifies the
entanglement and polarization of the nuclear system. Note that unlike
in the bosonic case (discussed in detail in Section \ref{sec:Steady-State-Entanglement}),
the modulus of $\xi$ is unconfined. Both $\left|\xi\right|<1$ and
$\left|\xi\right|>1$ are allowed and correspond to states of large
positive (negative) OH field gradients, respectively, and the system
is invariant under the corresponding symmetry transformation ($\mu_{2}\leftrightarrow\nu_{2}$,
$A_{L,R}^{z}\rightarrow-A_{L,R}^{z}$). As we discuss in detail in
Section \ref{sec:Polarization-Dynamics}, this symmetry gives rise
to a bistability in the steady state, as for every solution with positive
OH field gradient ($\Delta_{\mathrm{OH}}>0$), we find a second one
with negative gradient ($\Delta_{\mathrm{OH}}<0$). As a first indication
for this bistability, also compare the green and blue curve in Fig.~\ref{fig:external-gradient-N10}:
For $\Delta\gg0$, the dominant weight of the nuclear steady state
is found in the level $\left|-J_{L},J_{R}\right\rangle $, that is
the Dicke state with maximum \textit{positive} Overhauser gradient,
whereas for $\Delta\ll0$, the weight of the nuclear stationary state
is peaked symmetrically at $\left|J_{L},-J_{R}\right\rangle $, corresponding
to the Dicke state with maximum \textit{negative} Overhauser gradient. 

In the asymmetric scenario $J_{L}\neq J_{R}$, one can readily show
that a \textit{pure} dark-state solution does not exist. Thus, we
resort to exact numerical solutions for small system sizes $J_{i}\approx3$
to compute the nuclear steady state-solution $\sigma_{\mathrm{ss}}$.
To verify the creation of steady-state entanglement between the two
nuclear spin ensembles, we take the EPR uncertainty as a figure of
merit. It is defined via 
\begin{equation}
\Delta_{\mathrm{EPR}}=\frac{\mathrm{var}\left(I_{L}^{x}+I_{R}^{x}\right)+\mathrm{var}\left(I_{L}^{y}+I_{R}^{y}\right)}{\left|\left\langle I_{L}^{z}\right\rangle \right|+\left|\left\langle I_{R}^{z}\right\rangle \right|},\label{eq:EPR-collective-spins}
\end{equation}
and measures the degree of nonlocal correlations. For an arbitrary
state, $\Delta_{\mathrm{EPR}}<1$ implies the existence of such non-local
correlations, whereas $ $$\Delta_{\mathrm{EPR}}\geq1$ for separable
states.\cite{muschik11} The results are displayed in Fig.~\ref{fig:ideal-target-state-uniform-HF-coupling}.
First of all, the numerical solutions confirm the analytical result
in the symmetric limit where the asymmetry parameter $\Delta_{J}=J_{R}-J_{L}$
is zero. In the asymmetric setting, where $J_{L}\neq J_{R}$, the
steady state $\sigma_{\mathrm{ss}}$ is indeed found to be mixed,
that is $\mathsf{Tr}\left[\sigma_{\mathrm{ss}}^{2}\right]<1$. However,
both the amount of generated entanglement as well as the purity of
$\sigma_{\mathrm{ss}}$ tend to increase, as we increase the system
size $J_{L}+J_{R}$ for a fixed value of $\Delta_{J}$. For fixed
$J_{i}$, we have also numerically verified that the steady-state
solution is \textit{unique}. 

In practical experimental situations one deals with a mixture of different
$J_{i}$ subspaces. The width of the nuclear spin distribution is
typically $\Delta_{J}\sim\sqrt{N}$, but may even be narrowed further
actively; see for example Refs.\cite{rudner07,vink09}. The numerical
results displayed above suggest that the amount of entanglement and
purity of the nuclear steady state increases for smaller absolute
values of the relative asymmetry $\Delta_{J}/J=\left(J_{R}-J_{L}\right)/\left(J_{L}+J_{R}\right)$.
In Fig.~\ref{fig:ideal-target-state-uniform-HF-coupling}, $\Delta_{\mathrm{EPR}}<1$
is still observed even for $\left|\Delta_{J}\right|/J=2.5/3.5\approx0.7$.
Thus, experimentally one might still obtain entanglement in a mixture
of different \textit{large} $J_{i}$ subspaces for which the relative
width is comparatively small, $\Delta_{J}/J\approx\sqrt{N}/N\approx10^{-3}\ll1$.
Intuitively, the idea is that for every pair $\left\{ J_{L},J_{R}\right\} $
with $J_{L}\approx J_{R}$ the system is driven towards a state similar
to the ideal two-mode squeezedlike state given in Eq.(\ref{eq:target-squeezed-state-J-subspaces-uniform}).
This will also be discussed in more detail in Section \ref{sec:Steady-State-Entanglement}.

\section{Dynamic Nuclear Polarization \label{sec:Polarization-Dynamics}}

In the previous section we have identified a low-gradient regime,
where the nuclear spins settle into a fully mixed state, and a high-gradient
regime, where the ideal nuclear steady state was found to be a highly
polarized, entangled two-mode squeezedlike state. Now, we provide
a thorough analysis which reveals the multi-stability of the nuclear
subsystem and determines the connection between these two very different
regimes. It is shown that, beyond a critical polarization, the nuclear
spin system becomes self-polarizing and is driven towards a highly
polarized OH gradient. 

\begin{figure}
\includegraphics[width=0.9\columnwidth]{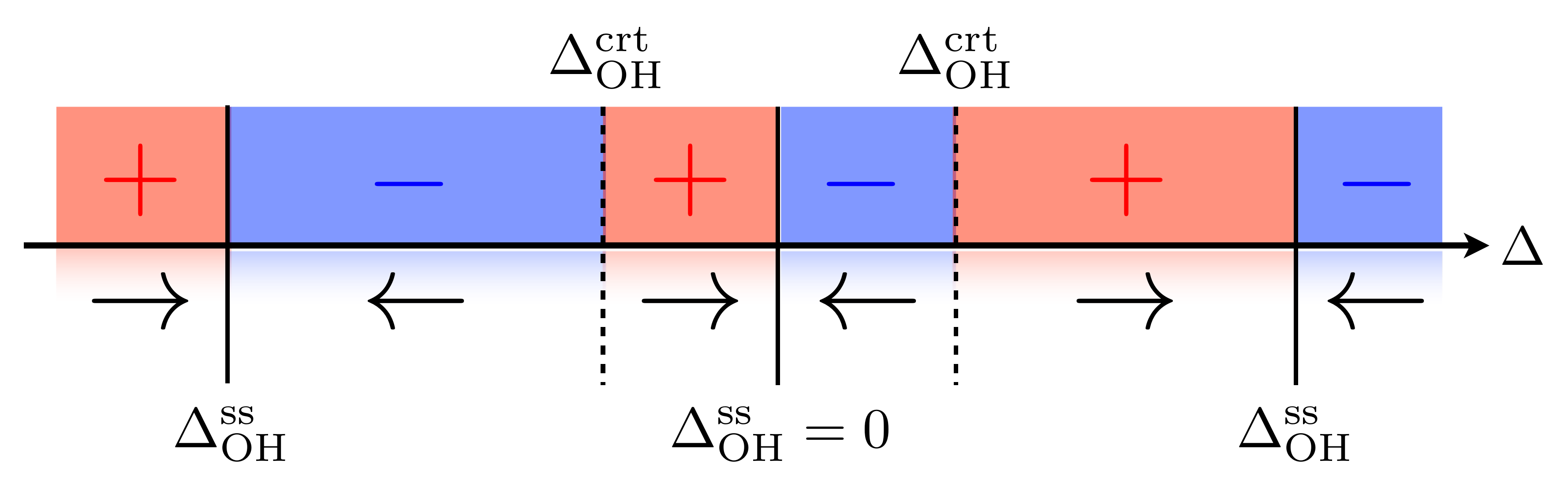}

\caption{\label{fig:Schematic-tristability}(color online). Schematic representation of the
multistability of the nuclear dynamics. For initial nuclear gradients
smaller than $\Delta_{\mathrm{OH}}^{\mathrm{crt}}$ the nuclear system
is attracted towards the trivial zero-polarization solution ($\Delta_{\mathrm{OH}}^{\mathrm{ss}}=0$).
Upon surpassing $\Delta_{\mathrm{OH}}^{\mathrm{crt}}$, however, the
system enters into an electron-nuclear feedback loop and the nuclear
dynamics turn self-polarizing such that large OH gradients can be
reached in the steady state. This is schematically denoted by $\pm$
referring to the sign of $\dot{\Delta}_{I^{z}}$ which determines
the stable fixed point the nuclear system is attracted to in the steady
state (see arrows). }

\end{figure}

To this end, we analyze the nuclear spin evolution within a semiclassical
approximation which neglects coherences among different nuclei. This
approach has been well studied in the context of central spin systems
(see for example Ref.\cite{gullans10} and references therein) and
is appropriate on timescales longer than nuclear dephasing times.\cite{christ07}
This approximation will be justified self-consistently. The analysis
is based on the effective QME given in Eq.(\ref{eq:effective-nuclear-QME-electronic-general}).
First, assuming homogeneous HF coupling and equal dot sizes $\left(N_{L}=N_{R}=N\right)$,
we construct dynamical equations for the expectation values of the
collective nuclear spins $\left\langle I_{i}^{z}\right\rangle _{t}$,
$i=L,R$, where $I_{i}^{\nu}=\sum_{j}\sigma_{i,j}^{\nu}$ for $\nu=\pm,z$.
To close the corresponding differential equations we use a semiclassical
factorization scheme resulting in two equations of motion for the
two nuclear dynamical variables $\left\langle I_{L}^{z}\right\rangle _{t}$
and $\left\langle I_{R}^{z}\right\rangle _{t}$, respectively. This
extends previous works on spin dynamics in double quantum dots, where
a single dynamical variable for the nuclear polarization was used
to explain the feedback mechanism in this system; see for example
Refs.\cite{rudner07,lopez-moniz11}. The corresponding nonlinear differential
equations are then shown to yield nonlinear equations for the equilibrium
polarizations. Generically, the nuclear polarization is found to be
multi-stable (compare also Refs.\cite{rudner11b,danon09b}) and, depending
on the system's parameters, we find up to three stable steady state
solutions for the OH gradient $\Delta_{\mathrm{OH}}^{\mathrm{ss}}$,
two of which are highly polarized in opposite directions and one is
unpolarized; compare Fig.~\ref{fig:Schematic-tristability} for a schematic
illustration. 

At this point, some short remarks are in order: First, the analytical
results obtained within the semiclassical approach are confirmed by
exact numerical results for small sets of nuclei; see Appendix \ref{sec:Numerical-Results-for-DNP}.
Second, by virtue of the semiclassical decoupling scheme used here,
our results can be generalized to the case of inhomogeneous HF coupling
in a straightforward way with the conclusions remaining essentially
unchanged. Third, for simplicity here we assume the symmetric scenario
of vanishing external fields $\omega_{\mathrm{ext}}=\Delta_{\mathrm{ext}}=0$;
therefore, $\Delta=\Delta_{\mathrm{OH}}$. However, as shown in Section
\ref{sec:Implementation} and Appendix \ref{sec:Numerical-Results-for-DNP}
one may generalize our results to finite external fields: This opens
up another experimental knob to tune the desired steady-state properties
of the nuclei. 

\textit{Intuitive picture}.---Before going through the calculation,
let us sketch an intuitive picture that can explain the instability
of the nuclear spins towards self-polarization and the corresponding
build-up of a macroscopic nuclear OH gradient: In the high-gradient
regime, the nuclear spins predominantly experience the action of the
nonlocal jump operators $L_{2}=\nu_{2}A_{L}^{+}+\mu_{2}A_{R}^{+}$
and $\mathbb{L}_{2}=\mu_{2}A_{L}^{-}+\nu_{2}A_{R}^{-}$, respectively,
both of them acting with the same rate $\gamma$ on the nuclear spin
ensembles. For example, for $\Delta>0$ and $\epsilon>0$, where $\mu_{2}>\nu_{2}$,
the first nuclear ensemble gets exposed more strongly to the action
of the collective lowering operator $A_{L}^{-}$, whereas the second
ensemble preferentially experiences the action of the raising operator
$A_{R}^{+}$; therefore, the two nuclear ensembles are driven towards
polarizations of opposite sign. The second steady solution featuring
a large OH gradient with opposite sign is found along the same lines
for $\mu_{2}<\nu_{2}$. Therefore, our scheme provides a good dynamic
nuclear polarization (DNP) protocol for $\mu_{2}\gg\nu_{2}$ $\left(\left|\xi\right|\ll1\right)$,
or vice-versa for $\mu_{2}\ll\nu_{2}$ $\left(\left|\xi\right|\gg1\right)$. 

\textit{Semiclassical analysis}.---Using the usual angular momentum
commutation relations $\left[I^{z},I^{\pm}\right]=\pm I^{\pm}$ and
$\left[I^{+},I^{-}\right]=2I^{z}$, Eq.(\ref{eq:effective-nuclear-QME-electronic-general})
readily yields two rate equations for the nuclear polarizations $\left\langle I_{i}^{z}\right\rangle _{t}$,
$i=L,R$. We then employ a semiclassical approach by neglecting correlations
among different nuclear spins, that is 
\begin{equation}
\left\langle \sigma_{i}^{+}\sigma_{j}^{-}\right\rangle =\begin{cases}
0 & ,i\neq j\\
\left\langle \sigma_{i}^{z}\right\rangle +\frac{1}{2} & ,i=j
\end{cases}\label{eq:semiclassical-factorization-1}
\end{equation}
which allows us to close the equations of motion for the nuclear polarizations
$\left\langle I_{i}^{z}\right\rangle $. This leads to the two following
nonlinear equations of motion, 
\begin{eqnarray}
\frac{d}{dt}\left\langle I_{L}^{z}\right\rangle _{t} & = & -\gamma_{\mathrm{pol}}\left[\left\langle I_{L}^{z}\right\rangle _{t}+\frac{N}{2}\frac{\chi}{\gamma_{\mathrm{pol}}}\right],\label{eq:EOM-I1z-semiclassical-closed}\\
\frac{d}{dt}\left\langle I_{R}^{z}\right\rangle _{t} & = & -\gamma_{\mathrm{pol}}\left[\left\langle I_{R}^{z}\right\rangle _{t}-\frac{N}{2}\frac{\chi}{\gamma_{\mathrm{pol}}}\right],\label{eq:EOM-I2z-semiclassical-closed}
\end{eqnarray}
where we have introduced the effective HF-mediated depolarization
rate $\gamma_{\mathrm{pol}}$ and pumping rate $\chi$ as 
\begin{eqnarray}
\gamma_{\mathrm{pol}} & = & \gamma\left(\mu_{2}^{2}+\nu_{2}^{2}\right)\left(1-p\right),\label{eq:semiclassical-depolarization-rate-gamma-eff}\\
\chi & = & \gamma\left(\mu_{2}^{2}-\nu_{2}^{2}\right)\left(3p-1\right),\label{eq:semiclassical-pumping-rate-chi}
\end{eqnarray}
with the rate $\gamma$ given in Eq.(\ref{eq:nuc-diss-rate-gamma}).
Clearly, Eqs.(\ref{eq:EOM-I1z-semiclassical-closed}) and (\ref{eq:EOM-I2z-semiclassical-closed})
already suggest that the two nuclear ensembles are driven towards
opposite polarizations. The nonlinearity is due to the fact that both
$\chi$ and $\gamma_{\mathrm{pol}}$ depend on the gradient $\Delta$
which itself depends on the nuclear polarizations $\left\langle I_{i}^{z}\right\rangle _{t}$;
at this stage of the anlysis, however, $\Delta$ simply enters as
a parameter of the underlying effective Hamiltonian. Equivalently,
the macroscopic dynamical evolution of the nuclear system may be expressed
in terms of the total net polarization $P\left(t\right)=\left\langle I_{L}^{z}\right\rangle _{t}+\left\langle I_{R}^{z}\right\rangle _{t}$
and the polarization gradient $\Delta_{I^{z}}=\left\langle I_{R}^{z}\right\rangle _{t}-\left\langle I_{L}^{z}\right\rangle _{t}$
as 
\begin{eqnarray}
\dot{P}\left(t\right) & = & -\gamma_{\mathrm{pol}}P\left(t\right),\label{eq:EOM-P-semiclassical-closed}\\
\frac{d}{dt}\Delta_{I^{z}} & = & -\gamma_{\mathrm{pol}}\left[\Delta_{I^{z}}-N\frac{\chi}{\gamma_{\mathrm{pol}}}\right].\label{eq:EOM-Delta-Iz-semiclassical-closed}
\end{eqnarray}

\textit{Fixed-point analysis}.---In what follows, we examine the fixed
points of the semiclassical equations derived above. First of all,
since $\gamma_{\mathrm{pol}}>0\,\forall P,\Delta_{I^{z}}$, Eq.(\ref{eq:EOM-P-semiclassical-closed})
simply predicts that in our system no homogeneous nuclear net polarization
$P$ will be produced. In contrast, any potential initial net polarization
is exponentially damped to zero in the long-time limit, since in the
steady state $\lim_{t\rightarrow\infty}P\left(t\right)=0$. This finding
is in agreement with previous theoretical results showing that, due
to angular momentum conservation, a net nuclear polarization cannot
be pumped in a system where the HF-mediated relaxation rate for the
blocked triplet levels $\left|T_{+}\right\rangle $ and $\left|T_{-}\right\rangle $,
respectively, is the same; see, e.g., Ref.\cite{rudner07} and references
therein. 

The dynamical equation for $\Delta_{I^{z}}$, however, is more involved:
The effective rates $\gamma_{\mathrm{pol}}=\gamma_{\mathrm{pol}}\left(\Delta\right)$
and $\chi=\chi\left(\Delta\right)$ in Eq.(\ref{eq:EOM-Delta-Iz-semiclassical-closed})
depend on the nuclear-polarization dependent parameter $\Delta$.
This nonlinearity opens up the possibility for multiple steady-state
solutions. From Eqs.(\ref{eq:EOM-I1z-semiclassical-closed}) and (\ref{eq:EOM-I2z-semiclassical-closed})
we can immediately identify the fixed points $\left\langle I_{i}^{z}\right\rangle _{\mathrm{ss}}$
of the nuclear polarization dynamics as $\pm\left(N/2\right)\chi/\gamma_{\mathrm{pol}}$.
Consequently, the two nuclear ensembles tend to be polarized along
opposite directions, that is $\left\langle I_{L}^{z}\right\rangle _{\mathrm{ss}}=-\left\langle I_{R}^{z}\right\rangle _{\mathrm{ss}}$.
The corresponding steady-state nuclear polarization gradient $\Delta_{I^{z}}^{\mathrm{ss}}$,
scaled in terms of its maximum value $N$, is given by 
\begin{equation}
\frac{\Delta_{I^{z}}^{\mathrm{ss}}}{N}=\mathcal{R}\left(\Delta\right)=\Lambda\frac{3p-1}{1-p}.\label{eq:Delta-Iz-gradient-ss}
\end{equation}
Here, we have introduced the nonlinear function $\mathcal{R}\left(\Delta\right)$
which depends on the purely electronic quantity 
\begin{equation}
\Lambda=\Lambda\left(\Delta\right)=\frac{\mu_{2}^{2}-\nu_{2}^{2}}{\mu_{2}^{2}+\nu_{2}^{2}}=\frac{1-\xi^{2}}{1+\xi^{2}}.
\end{equation}
According to Eq.(\ref{eq:Delta-Iz-gradient-ss}), the function $\mathcal{R}\left(\Delta\right)$
determines the nuclear steady-state polarization. While the functional
dependence of $\Lambda$ on the gradient $\Delta$ can give rise to
two highly polarized steady-state solutions with opposite nuclear
spin polarization, for $\left|\mu_{2}\right|\gg\left|\nu_{2}\right|$
and $\left|\mu_{2}\right|\ll\left|\nu_{2}\right|$, respectively,
the second factor in Eq.(\ref{eq:Delta-Iz-gradient-ss}) may prevent
the system from reaching these highly polarized fixed points. Based
on Eq.(\ref{eq:Delta-Iz-gradient-ss}), we can identify the two important
limits discussed previously: For $\Gamma_{2}\ll\Gamma_{\pm}$, the
electronic subsystem settles into the steady-state solution $p=1/3$
and the nuclear system is unpolarized, as the second factor in Eq.(\ref{eq:Delta-Iz-gradient-ss})
vanishes. This is what we identified above as the nuclear diffusion
regime in which the nuclear subsystem settles into the unpolarized
fully mixed state. In the opposite limit, where $\Gamma_{2}\gg\Gamma_{\pm}$,
the electronic subsystem settles into $p\approx1/2$. In this limit,
the second factor in Eq.(\ref{eq:Delta-Iz-gradient-ss}) becomes $1$
and the functional dependence of $\Lambda\left(\Delta\right)$ dominates
the behavior of $\mathcal{R}\left(\Delta\right)$ such that large
nuclear OH gradients can be achieved in the steady state. The electron-nuclear
feedback loop can then be closed self-consistently via $\Delta_{\mathrm{OH}}^{\mathrm{ss}}/\Delta_{\mathrm{OH}}^{\mathrm{max}}=\mathcal{R}\left(\Delta_{\mathrm{OH}}^{\mathrm{ss}}\right),$
where, in analogy to Eq.(\ref{eq:Delta-Iz-gradient-ss}), $\Delta_{\mathrm{OH}}^{\mathrm{ss}}$
has been scaled in units of its maximum value $\Delta_{\mathrm{OH}}^{\mathrm{max}}=A_{\mathrm{HF}}/2$.
Points fulfilling this condition can be found at intersections of
$\mathcal{R}\left(\Delta\right)$ with $\Delta_{\mathrm{OH}}^{\mathrm{ss}}/\Delta_{\mathrm{OH}}^{\mathrm{max}}$
. This is elaborated below. 

\begin{figure}
\includegraphics[width=1\columnwidth]{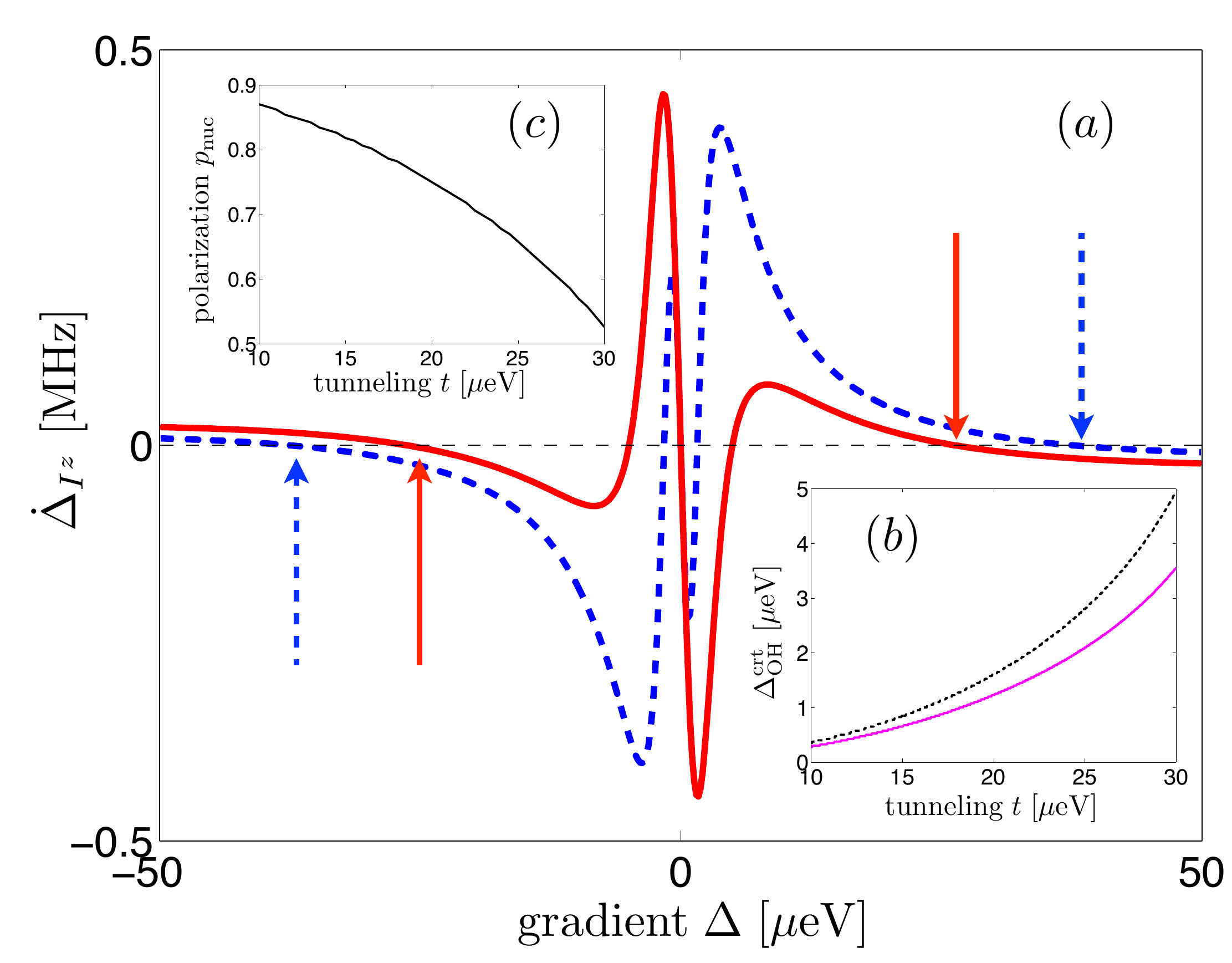}

\caption{\label{fig:ss-OH-gradient}(color online). Semiclassical solution
to the nuclear polarization dynamics: tristability of the nuclear
steady state. (a) Instantaneous nuclear polarization rate $\dot{\Delta}_{I^{z}}$
for $t=20\mu\mathrm{eV}$ (dashed) and $t=30\mu\mathrm{eV}$ (solid),
respectively. Stable fixed points are found at $\dot{\Delta}_{I^{z}}=0$
and $d\dot{\Delta}_{I^{z}}/d\Delta<0$. The nuclear system is driven
towards one of the highly polarized fixed points (indicated by arrows),
if the initial gradient $\Delta$ exceeds a critical threshold $\left|\Delta_{\mathrm{OH}}^{\mathrm{crt}}\right|$,
shown in (b) for $\Gamma_{\pm}=0.1\mu\mathrm{eV}$ (dashed) and $\Gamma_{\pm}=0.05\mu\mathrm{eV}$
(solid), respectively. (c) By tuning $t$, one can achieve $\left|\xi\right|\ll1$
leading to a nuclear polarization of $\lesssim90\%$. Other numerical
parameters in $\mu\mathrm{eV}$: $\Gamma=25$, $\epsilon=30$, $\Gamma_{\pm}=0.1$
(except for the solid line in (b) where $\Gamma_{\pm}=0.05$) and
$\Gamma_{\mathrm{deph}}=0.1$. }
\end{figure}

To gain further insights into the nuclear polarization dynamics, we
evaluate $\dot{\Delta}_{I^{z}}$ as given in Eq.(\ref{eq:EOM-Delta-Iz-semiclassical-closed}).
The results are displayed in Fig.~\ref{fig:ss-OH-gradient}. \textit{Stable}
fixed points of the dynamics are determined by $\dot{\Delta}_{I^{z}}=0$
and $d\dot{\Delta}_{I^{z}}/d\Delta<0$ as opposed to unstable fixed
points where $d\dot{\Delta}_{I^{z}}/d\Delta>0$. In this way it is
ensured that fluctuations of $\Delta_{I^{z}}$ away from a stable
fixed point are corrected by a restoring intrinsic pump effect.\cite{vink09,danon09b,bluhm10b}
We can identify parameter regimes in which the nuclear system features
three stable fixed points. As schematically depicted in Fig.~\ref{fig:Schematic-tristability},
they are interspersed by two unstable points referred to as $\Delta_{\mathrm{OH}}^{\mathrm{crt}}$.
Therefore, in general, the nuclear steady-state polarization is found
to be tri-stable: Two of the stable fixed points are high-polarization
solutions of opposite sign, supporting a macroscopic OH gradient,
while one is the trivial zero-polarization solution. The unstable
points $\Delta_{\mathrm{OH}}^{\mathrm{crt}}$ represent critical values
for the initial OH gradient marking the boundaries of a critical region.
If the initial gradient lies outside of this critical region, the
OH gradient runs into one of the highly polarized steady states. Otherwise,
the nuclear system gets stuck in the zero-polarization steady state.
Note that $\Delta_{\mathrm{OH}}^{\mathrm{crt}}$ is tunable: To surpass
the critical region one needs $\Gamma_{2}\gg\Gamma_{\pm}$; thus,
the critical region can be destabilized by making $\Gamma_{\pm}$
smaller {[}compare Fig.~\ref{fig:ss-OH-gradient}(b){]} which is lower
bounded by $\Gamma_{\pm}\gg\gamma_{c}$ in order to justify the elimination
of the electronic degrees of freedom. For typical parameters we thus
estimate $\Delta_{\mathrm{OH}}^{\mathrm{crt}}\approx(3-5)\mu\mathrm{eV}$
which sets the required initial $\Delta$ in order to kick-start the
nuclear self-polarization process. Experimentally, this could be realized
either via an initial nuclear polarization of $p_{\mathrm{nuc}}\approx(5-10)\%$
or an on-chip nanomagnet.\cite{petersen13,pioro-ladriere08} 

\textit{Timescales}.---In order to reach a highly polarized steady
state, approximately $\sim10^{5}$ nuclear spin flips are required.
We estimate $\dot{\Delta}_{I^{z}}\approx0.1\mathrm{MHz}$ and, thus,
the total time for the polarization process is therefore approximately
$\sim10^{5}/0.1\mathrm{MHz}\approx1\mathrm{s}$. This order of magnitude
estimate is in very good agreement with typical timescales observed
in nuclear polarization experiments.\cite{takahashi11} Moreover,
$\gamma_{\mathrm{pol}}^{-1}\approx1\mathrm{s}$ is compatible with
our semiclassical approach, since nuclear spins typically dephase
at a rate of $\sim\mathrm{kHz}$.\cite{takahashi11,gullans10} Finally,
in any experimental situation, the nuclear spins are subject to relaxation
and diffusion processes which prohibit complete polarization of the
nuclear spins. Therefore, in order to capture other depolarizing processes
that go beyond our current analysis, one could add an additional phenomenological
nuclear depolarization rate $\gamma_{\mathrm{dp}}$ by simply making
the replacement $\gamma_{\mathrm{pol}}\left(\Delta\right)\rightarrow\gamma_{\mathrm{pol}}\left(\Delta\right)+\gamma_{\mathrm{dp}}$.
Since typically $\gamma_{\mathrm{dp}}^{-1}\approx15\mathrm{s}$,\cite{danon09b}
however, these additional processes are slow in comparison to the
intrinsic rate $\gamma_{\mathrm{pol}}$ and should not lead to any
qualitative changes of our results.

\section{\textcolor{black}{Steady-State Entanglement Generation} \label{sec:Steady-State-Entanglement}}

In Section \ref{sec:Effective-Nuclear-Dynamics} we have identified
a high-gradient regime which---after adiabatically eliminating all
electronic coordinates---supports a rather simple description of the
nuclear dynamics on a coarse-grained timescale. Now, we extend our
previous analysis and provide a detailed analysis of the nuclear dynamics
in the high-gradient regime. In particular, this includes perturbative
effects due to the presence of the so far neglected levels $\left|\lambda_{1,3}\right\rangle $.
To this end, we apply a self-consistent Holstein-Primakoff approximation,
which reexpresses nuclear fluctuations around the semiclassical state
in terms of bosonic modes. This enables us to approximately solve
the nuclear dynamics analytically, to directly relate the ideal nuclear
steady state to a two-mode squeezed state familiar from quantum optics
and to efficiently compute several entanglement measures.

\subsection{Extended Nuclear Master Equation in the High-Gradient Regime}

In the high-gradient regime the electronic system settles to a quasisteady
state $\rho_{\mathrm{ss}}^{\mathrm{el}}=\rho_{\mathrm{target}}^{\mathrm{el}}=\left(\left|T_{+}\right\rangle \left\langle T_{+}\right|+\left|T_{-}\right\rangle \left\langle T_{-}\right|\right)/2$
{[}compare Eqs.(\ref{eq:quasi-steady-state-general}) and (\ref{eq:parameter-p}){]}
on a timescale short compared to the nuclear dynamics; deviations
due to (small) populations of the hybridized levels are discussed
in Appendix \ref{sec:Nonidealities-In-Electronic-Quasisteady-State}.
We then follow the general adiabatic elimination procedure discussed
in Section \ref{sec:Effective-Nuclear-Dynamics} to obtain an effective
master equation for the nuclear spins in the submanifold of the electronic
quasisteady state $\rho_{\mathrm{target}}^{\mathrm{el}}$. The full
calculation is presented in detail in Appendix \ref{sec:Appendix-Effective-Nuclear-QME-High-Gradient-Regime}.
In summary, the generalized effective master equation reads
\begin{eqnarray}
\dot{\sigma} & = & \sum_{k}\left[\frac{\gamma_{k}^{+}}{2}\mathcal{D}\left[L_{k}\right]\sigma+\frac{\gamma_{k}^{-}}{2}\mathcal{D}\left[\mathbb{L}_{k}\right]\sigma\right]+i\left[H_{\mathrm{Stark}},\sigma\right]\nonumber \\
 &  & +\gamma_{\mathrm{zz}}\sum_{i,j}\left[\delta A_{i}^{z}\sigma\delta A_{j}^{z}-\frac{1}{2}\left\{ \delta A_{j}^{z}\delta A_{i}^{z},\sigma\right\} \right].\label{eq:effective-QME-uniqueSS-explicit}
\end{eqnarray}
Here, we have introduced the effective HF-mediated decay rates 
\begin{eqnarray}
\gamma_{k}^{+} & = & \frac{a_{\mathrm{hf}}^{2}\tilde{\Gamma}_{k}}{2\left[\Delta_{k}^{2}+\tilde{\Gamma}_{k}^{2}\right]},\\
\gamma_{k}^{-} & = & \frac{a_{\mathrm{hf}}^{2}\tilde{\Gamma}_{k}}{2\left[\delta_{k}^{2}+\tilde{\Gamma}_{k}^{2}\right]},
\end{eqnarray}
where $\tilde{\Gamma}_{k}=\Gamma_{k}+3\Gamma_{\pm}+\Gamma_{\mathrm{deph}}/4$
and the detuning parameters 
\begin{eqnarray}
\Delta_{k} & = & \epsilon_{k}-\omega_{0},\\
\delta_{k} & = & \epsilon_{k}+\omega_{0},
\end{eqnarray}
specify the splitting between the electronic eigenstate $\left|\lambda_{k}\right\rangle $
and the Pauli-blocked triplet states $\left|T_{+}\right\rangle $
and $\left|T_{-}\right\rangle $, respectively. The effective nuclear
Hamiltonian 
\begin{equation}
H_{\mathrm{Stark}}=\sum_{k}\frac{\Delta_{k}^{+}}{2}L_{k}^{\dagger}L_{k}+\frac{\Delta_{k}^{-}}{2}\mathbb{L}_{k}^{\dagger}\mathbb{L}_{k}\label{eq:nuclear-Stark-shift-Hamiltonian}
\end{equation}
is given in terms of the second-order Stark shifts 
\begin{eqnarray}
\Delta_{k}^{+} & = & \frac{a_{\mathrm{hf}}^{2}\Delta_{k}}{4\left[\Delta_{k}^{2}+\tilde{\Gamma}_{k}^{2}\right]},\\
\Delta_{k}^{-} & = & \frac{a_{\mathrm{hf}}^{2}\delta_{k}}{4\left[\delta_{k}^{2}+\tilde{\Gamma}_{k}^{2}\right]}.
\end{eqnarray}
Lastly, in Eq.(\ref{eq:effective-QME-uniqueSS-explicit}) we have
set $\gamma_{\mathrm{zz}}=a_{\mathrm{hf}}^{2}/(5\Gamma_{\pm})$. For
$\omega_{0}=0$, we have $\gamma_{k}^{+}=\gamma_{k}^{-}$ and $\Delta_{k}^{+}=\Delta_{k}^{-}$.
When disregarding effects due to $H_{\mathrm{zz}}$ and neglecting
the levels $\left|\lambda_{1,3}\right\rangle $, i.e., only keeping
$k=2$ in Eq.(\ref{eq:effective-QME-uniqueSS-explicit}), indeed,
we recover the result of the Section \ref{sec:Effective-Nuclear-Dynamics};
see Eq.(\ref{eq:effective-QME-nuclear-spins-high-gradient-simple}).
As shown in Appendix \ref{sec:Diagonalization-of-Nuclear-Disspator},
the nuclear HF-mediated jump terms in Eq.(\ref{eq:effective-QME-uniqueSS-explicit})
can be brought into diagonal form which features a clear hierarchy
due to the predominant coupling to $\left|\lambda_{2}\right\rangle $.
To stress this hierarchy in the effective nuclear dynamics $\dot{\sigma}=\mathcal{L}_{\mathrm{eff}}\sigma$,
we write 
\begin{equation}
\dot{\sigma}=\mathcal{L}_{\mathrm{id}}\sigma+\mathcal{L}_{\mathrm{nid}}\sigma,
\end{equation}
where the first term captures the dominant coupling to the electronic
level $\left|\lambda_{2}\right\rangle $ only and is given as 
\begin{eqnarray}
\mathcal{L}_{\mathrm{id}}\sigma & = & \frac{\gamma_{2}^{+}}{2}\mathcal{D}\left[L_{2}\right]\sigma+\frac{\gamma_{2}^{-}}{2}\mathcal{D}\left[\mathbb{L}_{2}\right]\sigma\nonumber \\
 &  & +i\frac{\Delta_{2}^{+}}{2}\left[L_{2}^{\dagger}L_{2},\sigma\right]+i\frac{\Delta_{2}^{-}}{2}\left[\mathbb{L}_{2}^{\dagger}\mathbb{L}_{2},\sigma\right],\label{eq:ideal-nuclear-Liouvillian}
\end{eqnarray}
whereas the remaining non-ideal part $\mathcal{L}_{\mathrm{nid}}$
captures all remaining effects due to the coupling to the far-detuned
levels $\left|\lambda_{1,3}\right\rangle $ and the OH fluctuations
described by $H_{\mathrm{zz}}$.

\subsection{Holstein-Primakoff Approximation and Bosonic Formalism}

To obtain further insights into the nuclear spin dynamics in the high-gradient
regime, we now restrict ourselves to uniform hyperfine coupling $\left(a_{i,j}=N/N_{i}\right)$
and apply a Holstein-Primakoff (HP) transformation to the collective
nuclear spin operators $I_{i}^{\alpha}=\sum_{j}\sigma_{i,j}^{\alpha}$
for $\alpha=\pm,z$; generalizations to non-uniform coupling will
be discussed separately below in Section \ref{sec:Implementation}.
This treatment of the nuclear spins has proven valuable already in
previous theoretical studies.\cite{kessler12} In the present case,
it allows for a detailed study of the nuclear dynamics including perturbative
effects arising from $\mathcal{L}_{\mathrm{nid}}$. 

The (exact) Holstein-Primakoff (HP) transformation expresses the truncation
of the collective nuclear spin operators to a total spin $J_{i}$
subspace in terms of a bosonic mode.\cite{kessler12} Note that for
uniform HF coupling the total nuclear spin quantum numbers $J_{i}$
are conserved quantities. Here, we consider two nuclear spin ensembles
that are polarized in opposite directions of the quantization axis
$\hat{z}$. Then, the HP transformation can explicitly be written
as 
\begin{eqnarray}
I_{L}^{-} & = & \sqrt{2J_{L}}\sqrt{1-\frac{b_{L}^{\dagger}b_{L}}{2J_{L}}}b_{L},\label{eq:Holstein-Primakoff-trafo}\\
I_{L}^{z} & = & b_{L}^{\dagger}b_{L}-J_{L},\nonumber 
\end{eqnarray}
for the first ensemble, and similarly for the second ensemble 
\begin{eqnarray}
I_{R}^{+} & = & \sqrt{2J_{R}}\sqrt{1-\frac{b_{R}^{\dagger}b_{R}}{2J_{R}}}b_{R},\label{eq:Holstein-Primakoff-trafo-2}\\
I_{R}^{z} & = & J_{R}-b_{R}^{\dagger}b_{R}.\nonumber 
\end{eqnarray}
Here, $b_{i}$ denotes the annihilation operator of the bosonic mode
$i=L,R$. Next, we expand the operators of Eqs.(\ref{eq:Holstein-Primakoff-trafo})
and (\ref{eq:Holstein-Primakoff-trafo-2}) in orders of $\varepsilon_{i}=1/\sqrt{J_{i}}$
which can be identified as a perturbative parameter.\cite{kessler12}
This expansion can be justified self-consistently provided that the
occupation numbers of the bosonic modes $b_{i}$ are small compared
to $2J_{i}$. Thus, here we consider the subspace with large collective
spin quantum numbers, that is $J_{i}\sim\mathcal{O}\left(N/2\right)$.
Accordingly, up to second order in $\varepsilon_{L}\approx\varepsilon_{R}$,
the hyperfine Hamiltonian can be rewritten as
\begin{equation}
H_{\mathrm{HF}}=H_{\mathrm{sc}}+H_{\mathrm{ff}}+H_{\mathrm{zz}},
\end{equation}
where the semiclassical part $H_{\mathrm{sc}}$ reads 
\begin{eqnarray}
H_{\mathrm{sc}} & = & a_{R}J_{R}S_{R}^{z}-a_{L}J_{L}S_{L}^{z}\\
 & = & \text{\ensuremath{\bar{\omega}}}_{\mathrm{OH}}\left(S_{L}^{z}+S_{R}^{z}\right)+\Delta_{\mathrm{OH}}\left(S_{R}^{z}-S_{L}^{z}\right).
\end{eqnarray}
Here, we have introduced the individual HF coupling constants $a_{i}=A_{\mathrm{HF}}/N_{i}$
and 
\begin{eqnarray}
\text{\ensuremath{\bar{\omega}}}_{\mathrm{OH}} & = & \Delta_{\mathrm{OH}}^{\mathrm{max}}\left(\mathrm{p}_{R}-\mathrm{p}_{L}\right)/2,\\
\text{\ensuremath{\bar{\Delta}}}_{\mathrm{OH}} & = & \Delta_{\mathrm{OH}}^{\mathrm{max}}\left(\mathrm{p}_{L}+\mathrm{p}_{R}\right)/2,
\end{eqnarray}
with $\mathrm{p}_{i}=J_{i}/J_{i}^{\mathrm{max}}=2J_{i}/N_{i}$ denoting
the degree of polarization in dot $i=L,R$ and $ $$\Delta_{\mathrm{OH}}^{\mathrm{max}}=A_{\mathrm{HF}}/2\approx50\mu\mathrm{eV}$.
Within the HP approximation, the hyperfine dynamics read 
\begin{equation}
H_{\mathrm{ff}}=\frac{a_{L}}{2}\sqrt{2J_{L}}S_{L}^{+}b_{L}+\frac{a_{R}}{2}\sqrt{2J_{R}}S_{R}^{+}b_{R}^{\dagger}+\mathrm{h.c.},
\end{equation}
and 
\begin{equation}
H_{\mathrm{zz}}=a_{L}S_{L}^{z}b_{L}^{\dagger}b_{L}-a_{R}S_{R}^{z}b_{R}^{\dagger}b_{R}.
\end{equation}
Note that, due to the different polarizations in the two dots, the
collective nuclear operators $I_{i}^{+}$ map onto bosonic annihilation
(creation) operators in the left (right) dot, respectively. The expansion
given above implies a clear hierarchy in the Liouvillian $\mathcal{L}_{0}+\mathcal{V}$
allowing for a perturbative treatment of the leading orders and adiabatic
elimination of the electron degrees of freedom whose evolution is
governed by the fastest timescale of the problem: while the semiclassical
part $H_{\mathrm{sc}}/J_{i}\sim\mathcal{O}\left(1\right)$, the HF
interaction terms scales as $H_{\mathrm{ff}}/J_{i}\sim\mathcal{O}\left(\varepsilon\right)$
and $H_{\mathrm{zz}}/J_{i}\sim\mathcal{O}\left(\varepsilon^{2}\right)$;
also compare Ref.\cite{kessler12}. To make connection with the analysis
of the previous subsection, we give the following explicit mapping
\begin{eqnarray}
A_{L}^{+} & \approx & \eta_{L}b_{L}^{\dagger},\qquad\delta A_{L}^{z}=\zeta_{L}b_{L}^{\dagger}b_{L},\label{eq:HP-mapping-imperfection-parameter}\\
A_{R}^{+} & \approx & \eta_{R}b_{R},\qquad\delta A_{R}^{z}=-\zeta_{R}b_{R}^{\dagger}b_{R}.\nonumber 
\end{eqnarray}
Here, the parameters $\zeta_{i}=N/N_{i}$ and $\eta_{i}=\zeta_{i}\sqrt{2J_{i}}$
capture imperfections due to either different dot sizes $\left(N_{L}\neq N_{R}\right)$
and/or different total spin manifolds $\left(J_{L}\neq J_{R}\right)$.
Moreover, within the HP treatment $\mathcal{V}$ can be split up into
a first $\left(\mathcal{L}_{\mathrm{ff}}\right)$ and a second-order
effect $\left(\mathcal{L}_{\mathrm{zz}}\right)$; therefore,
in second-order perturbation theory, the effective nuclear dynamics
simplify to {[}compare Eq.(\ref{eq:effective-nuclear-QME-2nd-order-general}){]}
\begin{equation}
\dot{\sigma}=\mathsf{Tr}_{\mathrm{el}}\left[\mathcal{P}\mathcal{L}_{\mathrm{ff}}\mathcal{P}\rho+\mathcal{P}\mathcal{L}_{\mathrm{zz}}\mathcal{P}\rho-\mathcal{P}\mathcal{L}_{\mathrm{ff}}\mathcal{Q}\mathcal{L}_{0}^{-1}\mathcal{Q}\mathcal{L}_{\mathrm{ff}}\mathcal{P}\rho\right],
\end{equation}
since higher-order effects due to $\mathcal{L}_{\mathrm{zz}}$ can
be neglected self-consistently to second order. 

\textit{Ideal nuclear target state}.---Within the HP approximation
and for the symmetric setting $\eta_{1}=\eta_{2}=\eta$, the dominant
nuclear jump operators $L_{2}$ and $\mathbb{L}_{2}$, describing
the lifting of the spin blockade via the electronic level $\left|\lambda_{2}\right\rangle $,
can be expressed in terms of \textit{nonlocal} bosonic modes as 
\begin{eqnarray}
L_{2} & = & \eta\sqrt{\mu_{2}^{2}-\nu_{2}^{2}}a,\label{eq:L2-nonlocal-bosonic-mode-a}\\
\mathbb{L}_{2} & = & \eta\sqrt{\mu_{2}^{2}-\nu_{2}^{2}}\tilde{a},\label{eq:LL2-nonlocal-bosonic-mode-a}
\end{eqnarray}
where $a=\nu b_{L}^{\dagger}+\mu b_{R}$ and $\tilde{a}=\mu b_{L}+\nu b_{R}^{\dagger}$.
Here, $\mu=\mu_{2}/\sqrt{\mu_{2}^{2}-\nu_{2}^{2}}$ and $\nu=\nu_{2}/\sqrt{\mu_{2}^{2}-\nu_{2}^{2}}$,
such that $\mu^{2}-\nu^{2}=1$. Therefore, due to $\left[a,a^{\dagger}\right]=1=\left[\tilde{a},\tilde{a}^{\dagger}\right]$
and $\left[a,\tilde{a}^{\dagger}\right]=0=\left[a,\tilde{a}\right]$,
the operators $a$ and $\tilde{a}$ refer to two independent, properly
normalized nonlocal bosonic modes. In this picture, the (unique) ideal
nuclear steady state belonging to the dissipative evolution $\mathcal{L}_{\mathrm{id}}\sigma$
in Eq.(\ref{eq:ideal-nuclear-Liouvillian}) is well known to be a
two-mode squeezed state $ $
\begin{equation}
\left|\Psi_{\mathrm{TMS}}\right\rangle =\mu^{-1}\sum_{n}\xi^{n}\left|n\right\rangle _{L}\otimes\left|n\right\rangle _{R}
\end{equation}
with $\xi=-\nu/\mu$:\cite{muschik11} $\left|\Psi_{\mathrm{TMS}}\right\rangle $
is the common vacuum of the non-local bosonic modes $a$ and $\tilde{a}$,
$a\left|\Psi_{\mathrm{TMS}}\right\rangle =\tilde{a}\left|\Psi_{\mathrm{TMS}}\right\rangle =0$.
It features entanglement between the number of excitations $n$ in
the first and second dot. Going back to collective nuclear spins,
this translates to perfect correlations between the degree of polarization
in the two nuclear ensembles. Note that $\left|\Psi_{\mathrm{TMS}}\right\rangle $
represents the dark state $\left|\xi_{\mathrm{ss}}\right\rangle $
given in Eq.(\ref{eq:target-squeezed-state-J-subspaces-uniform})
in the zeroth-order HP limit where the truncation of the collective
spins to $J_{i}$ subspaces becomes irrelevant. 

\textit{Bosonic steady-state solution}.---Within the HP approximation,
the nuclear dynamics generated by the full effective Liouvillian $\dot{\sigma}=\mathcal{L}_{\mathrm{eff}}\sigma$
are quadratic in the bosonic creation $b_{i}^{\dagger}$ and annihilation
operators $b_{i}$. Therefore, the nuclear dynamics are purely Gaussian
and an exact solution is feasible. Based on Eq.(\ref{eq:effective-QME-uniqueSS-explicit})
and Eq.(\ref{eq:HP-mapping-imperfection-parameter}), one readily
derives a closed dynamical equation for the second-order moments
\begin{equation}
\frac{d}{dt}\pmb\gamma=\mathcal{M}\pmb\gamma+\pmb C,\label{eq:closed-EOM-2ndorder-moments}
\end{equation}
where $\pmb\gamma$ is a vector comprising the second-order moments,
that is $\pmb\gamma=\left(\left\langle b_{i}^{\dagger}b_{j}\right\rangle _{t},\left\langle b_{i}^{\dagger}b_{j}^{\dagger}\right\rangle _{t},\dots\right)^{\top}$
and $\pmb C$ is a constant vector. The solution to Eq.(\ref{eq:closed-EOM-2ndorder-moments})
is given by 
\begin{equation}
\pmb\gamma\left(t\right)=e^{\mathcal{M}t}\pmb c_{0}-\mathcal{M}^{-1}\pmb C,\label{eq:solution-EOM-2ndorder-moments}
\end{equation}
where $\pmb c_{0}$ is an integration constant. Accordingly, provided
that the dynamics generated by $\mathcal{M}$ is contractive (see
section \ref{sec:Criticality} for more details), the steady-state
solution is found to be 
\begin{equation}
\pmb\gamma_{\mathrm{ss}}=-\mathcal{M}^{-1}\pmb C.
\end{equation}
Based on $\pmb\gamma_{\mathrm{ss}}$, one can construct the steady-state
covariance matrix (CM), defined as $\Gamma_{ij}^{\mathrm{CM}}=\left\langle \left\{ R_{i},R_{j}\right\} \right\rangle -2\left\langle R_{i}\right\rangle \left\langle R_{j}\right\rangle ,$
where $\left\{ R_{i},i=1,\dots,4\right\} =\left\{ X_{L},P_{L},X_{R},P_{R}\right\} $;
here, $X_{i}=(b_{i}+b_{i}^{\dagger})/\sqrt{2}$ and $P_{i}=i(b_{i}^{\dagger}-b_{i})/\sqrt{2}$
refer to the quadrature operators related to the bosonic modes $b_{i}$.
By definition, Gaussian states are fully characterized by the first
and second moments of the field operators $R_{i}$. Here, the first
order moments can be shown to vanish. The entries of the CM are real
numbers: since they constitute the variances and covariances of quantum
operators, they can be detected experimentally via nuclear spin variance
and correlation measurements.\cite{laurat05} 

We now turn to the central question of whether the steady-state entanglement
inherent to the ideal target state $\left|\Psi_{\mathrm{TMS}}\right\rangle $
is still present in the presence of the undesired terms described
by $\mathcal{L}_{\mathrm{nid}}$. In our setting, this is conveniently
done via the CM, which encodes all information about the entanglement
properties:\cite{schwager10} It allows us to compute certain entanglement
measures efficiently in order to make qualitative and quantitative
statements about the degree of entanglement.\cite{schwager10} Here,
we will consider the following quantities: For symmetric states, the
entanglement of formation $E_{F}$ can be computed easily.\cite{giedke03,wolf04}
It measures the minimum number of singlets required to prepare the
state through local operations and classical communication. For symmetric
states, this quantification of entanglement is fully equivalent to
the one provided by the logarithmic negativity $E_{\mathcal{N}}$;
the latter is determined by the smallest symplectic eigenvalues of
the CM of the partially transposed density matrix.\cite{vidal02}
Lastly, in the HP picture the EPR uncertainty defined in Eq.(\ref{eq:EPR-collective-spins})
translates to $\Delta_{\mathrm{EPR}}=\left[\mathrm{var}\left(X_{L}+X_{R}\right)+\mathrm{var}\left(P_{L}-P_{R}\right)\right]/2$.
For the ideal target state $\left|\Psi_{\mathrm{TMS}}\right\rangle $,
we find $\Delta_{\mathrm{EPR}}^{\mathrm{id}}=\left(\mu-\nu\right)^{2}=\left(1-\left|\xi\right|\right)/\left(1+\left|\xi\right|\right)<1$.
Finally, one can also compute the fidelity $\mathcal{F}\left(\sigma_{\mathrm{ss}},\sigma_{\mathrm{target}}\right)$
which measures the overlap between the steady state generated by the
full dynamics $\dot{\sigma}=\mathcal{L}_{\mathrm{eff}}\sigma$ and
the ideal target state $\sigma_{\mathrm{target}}=\left|\Psi_{\mathrm{TMS}}\right\rangle \left\langle \Psi_{\mathrm{TMS}}\right|$.
\cite{schwager10}

\begin{figure}
\includegraphics[width=1\columnwidth]{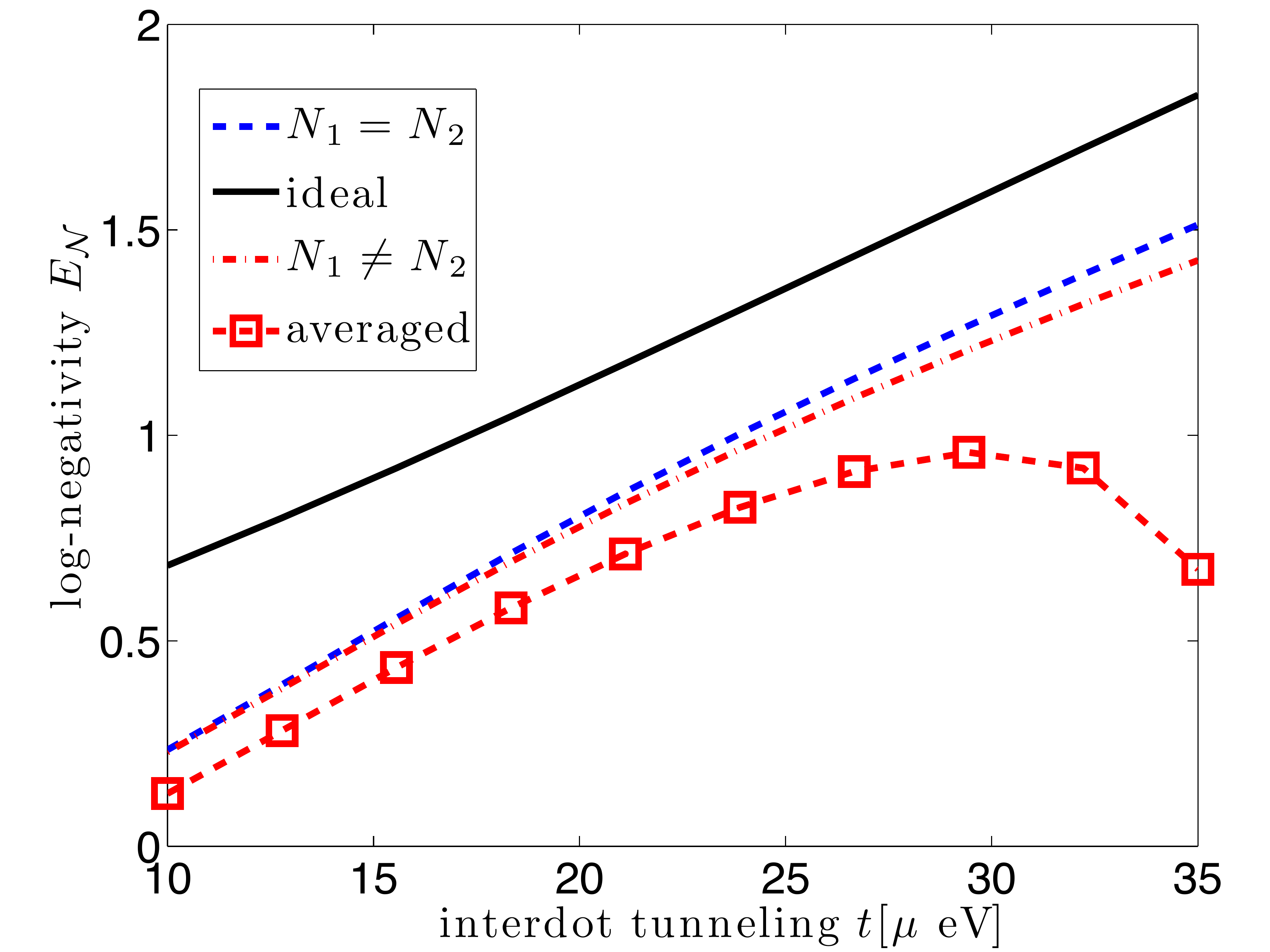}

\caption{\label{fig:steady-state-log-negativity}(color online). Steady-state
entanglement generation between the two nuclear spin ensembles. $E_{\mathcal{N}}>0$
indicates the creation of entanglement. The black solid curve refers
to the idealized, symmetric setting where the undesired HF-coupling
to $\left|\lambda_{1,3}\right>$ has been ignored and where $J_{L}=J_{R}=\mathrm{p}J_{\mathrm{max}}$;
here, the nuclear polarization $\mathrm{p}=0.8$ and $N_{L}=N_{R}=2J_{\mathrm{max}}=10^{6}$.
The blue-dashed line then also takes into account coupling to $\left|\lambda_{1,3}\right>$,
while the red (dash-dotted) curve in addition accounts for an asymmetric
dot size: $N_{R}=0.8N_{L}=8\times10^{5}$. Additionally, classical
uncertainty (red squares) in the total spin $J_{i}$ quantum numbers
leads to a reduced amount of entanglement, but does not disrupt it
completely; here, we have set the range of the (uniform) distribution
to $\Delta_{J_{i}}=50\sqrt{N_{i}}$. Other numerical parameters: $\omega_{0}=0$,
$\Gamma=25\mu\mathrm{eV}$, $\epsilon=30\mu\mathrm{eV}$, $3\Gamma_{\pm}+\Gamma_{\mathrm{deph}}/4=0.5\mu\mathrm{eV}$. }
\end{figure}

As illustrated in Fig.~\ref{fig:steady-state-log-negativity}, the
generation of steady-state entanglement persists even in presence
of the undesired noise terms described by $\mathcal{L}_{\mathrm{nid}}$,
asymmetric dot sizes $\left(N_{L}\neq N_{R}\right)$ and classical
uncertainty in total spins $J_{i}$: The maximum amount of entanglement
that we find (in the symmetric scenario $N_{L}=N_{R}$) is approximately
$E_{\mathcal{N}}\approx1.5$, corresponding to an entanglement of
formation $E_{F}\approx(1-2)\mathrm{ebit}$ and an EPR uncertainty
of $\Delta_{\mathrm{EPR}}\approx0.4$. When tuning the interdot tunneling
parameter from $t=10\mu\mathrm{eV}$ to $t=35\mu\mathrm{eV}$, the
squeezing parameter $\left|\xi\right|=\left|\nu_{2}/\mu_{2}\right|$
increases from $\sim0.2$ to $\sim0.6$, respectively; this is because
(for fixed $\Delta,\epsilon>0$) and increasing $t$, $\epsilon_{2}$
approaches 0 and the relative weight of $\nu_{2}$ as compared to
$\mu_{2}$ increases. Ideally, this implies stronger squeezing of
the steady state of $\mathcal{L}_{\mathrm{id}}$ and therefore a greater
amount of entanglement (compare the solid line in Fig.~\ref{fig:steady-state-log-negativity}),
but, at the same time, it renders the target state more susceptible
to undesired noise terms. Stronger squeezing leads to a larger occupation
of the bosonic HP modes (pictorially, the nuclear target state leaks
farther into the Dicke ladder) and eventually to a break-down of the
approximative HP description. The associated critical behavior in
the nuclear spin dynamics can be understood in terms of a dynamical
phase transition \cite{kessler12}, which will be analyzed in greater
detail in the next section.

\section{Criticality \label{sec:Criticality}}

\begin{figure}
\includegraphics[width=1\columnwidth]{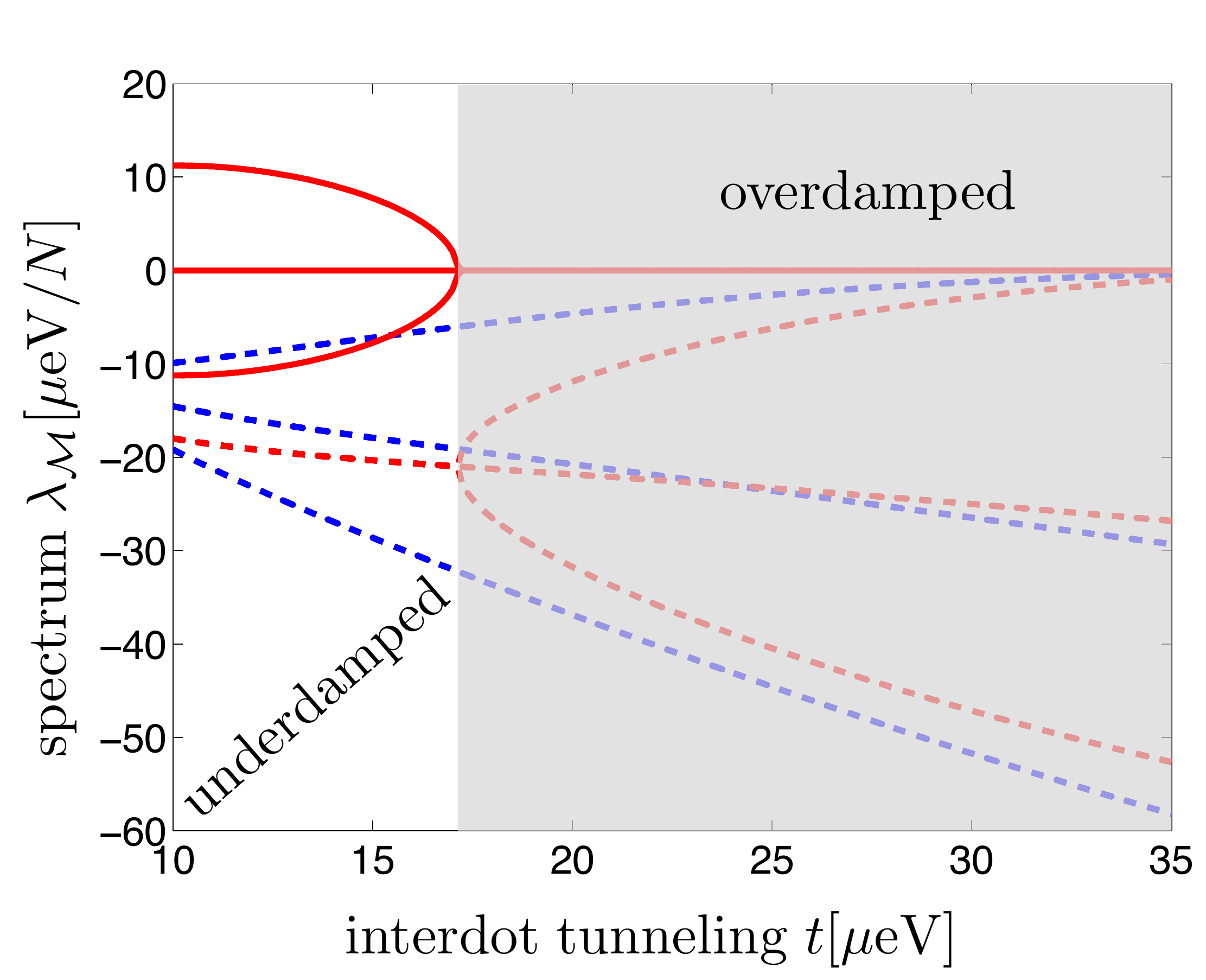}

\caption{\label{fig:dynamical-phase-transition}(color online). Spectrum of
$\mathcal{M}$. Real (dashed) and imaginary parts (solid) for $\Gamma=25\mu\mathrm{eV}$
(blue) and $\Gamma=50\mu\mathrm{eV}$ (red). A dynamical phase transition
is found at the bifurcation separating an underdamped from an overdamped
region (see gray shading for the spectrum displayed in red). The critical
point $t_{\mathrm{crt}}\approx37\mu\mathrm{eV}$ is reached where
the smallest decay rate $\left(\mathrm{ADR}\right)$ becomes zero.
Other numerical parameters as those for the dashed curve in Fig.~\ref{fig:steady-state-log-negativity}.}
\end{figure}

Based on the Holstein-Primakoff analysis outlined above, we now show
that the nuclear spin dynamics exhibit a dynamical quantum phase transition
which originates from the competition between dissipative terms and
unitary dynamics. This rather generic phenomenon in open quantum systems
results in nonanalytic behaviour in the spectrum of the nuclear spin
Liouvillian, as is well known from the paradigm example of the
Dicke model.\cite{kessler12,nagy11,dimer07,diehl10} 

The nuclear dynamics in the vicinity of the stationary state are described
by the stability matrix $\mathcal{M}$. Resulting from a systematic
expansion in the system size, the (complex) eigenvalues of $\mathcal{M}$
correspond exactly to the low-excitation spectrum of the full system
Liouvillian given in Eq.(\ref{eq:effective-QME-full-model}) in the
thermodynamic limit ($J\rightarrow\infty$). A non-analytic change
of steady state properties (indicating a steady state phase transition)
can only occur if the spectral gap of $\mathcal{M}$ closes.\cite{kessler12,hoening12}
The relevant gap in this context is determined by the eigenvalue with
the largest real part different from zero {[}from here on referred
to as the asymptotic decay rate (ADR){]}. The ADR determines the rate
by which the steady state is approached in the long time limit. 

As depicted in Fig.~\ref{fig:dynamical-phase-transition}, the system
reaches such a critical point at $t_{\mathrm{crt}}\approx37\mu\mathrm{eV}$
where the ADR (red/blue dotted lines closest to zero) becomes zero.
At this point, the dynamics generated by $\mathcal{M}$ become non-contractive
{[}compare Eq.(\ref{eq:solution-EOM-2ndorder-moments}){]} and the
nuclear fluctuations diverge, violating the self-consistency condition
of low occupation numbers in the bosonic modes $b_{i}$ and thus leading
to a break-down of the HP approximation. Consequently, the dynamics
cannot furhter be described by the dynamical matrix $\mathcal{M}$
indicating a qualitative change in the system properties and a steady
state phase transition. 

\begin{figure}
\includegraphics[width=1\columnwidth]{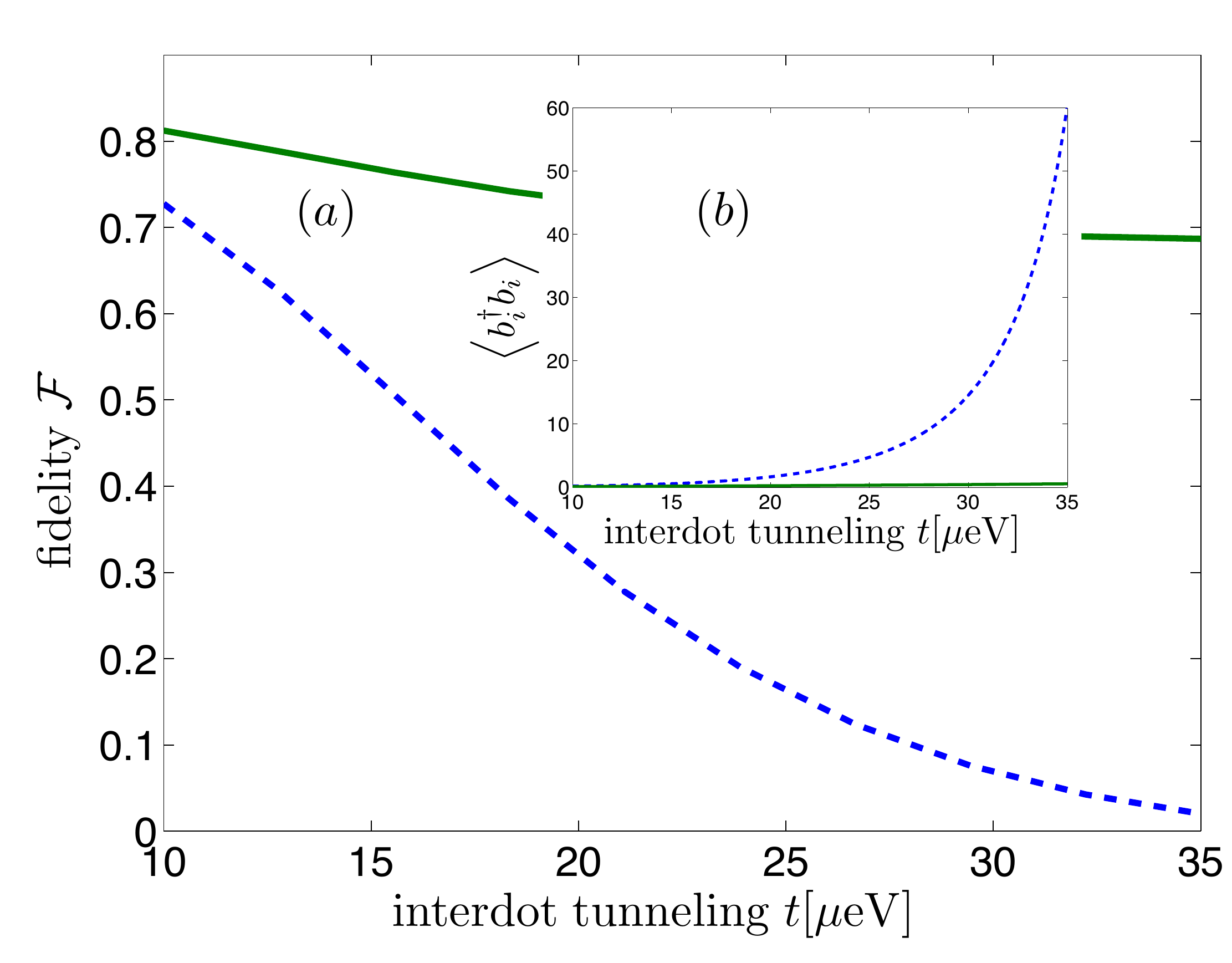}

\caption{\label{fig:steady-state-fidelity-Stark}(color online). (a) Fidelity
$\mathcal{F}$ of the nuclear steady-state with the two-mode squeezed
target state. The blue-dashed line accounts for the full nuclear Liouvillian
$\mathcal{L}_{\mathrm{eff}}$ for the symmetric setting $\left(N_{L}=N_{R}\right)$,
while the green solid line refers to the same setting in the absence
of any Stark-shift terms. Therefore, the decreasing fidelity $\mathcal{F}$
(blue dashed line) and a diverging number of HP bosons shown in (b)
is due to undesired Stark shift terms included in $\mathcal{L}_{\mathrm{eff}}$.
Here, $\eta_{L}=\eta_{R}$ and therefore $\left\langle b_{L}^{\dagger}b_{L}\right\rangle =\left\langle b_{R}^{\dagger}b_{R}\right\rangle $;
asymmetric settings where $\eta_{L}\neq\eta_{R}$ entail small asymmetries
in the number of HP bosons. For other numerical parameters compare
the dashed curve in Fig.~\ref{fig:steady-state-log-negativity}.}
\end{figure}

To obtain further insights into the cross-over of the maximum real
part of the eigenvalues $\lambda_{\mathcal{M}}$ of the matrix $\mathcal{M}$
from negative to positive values, we analyze the effect of the nuclear
Stark shift terms {[}Eq.(\ref{eq:nuclear-Stark-shift-Hamiltonian}){]}
in more detail. In the HP regime, up to irrelevant constant terms,
the Stark shift Hamiltonian $H_{\mathrm{Stark}}$ can be written as
\begin{equation}
H_{\mathrm{Stark}}=\epsilon_{L}^{\mathrm{st}}b_{L}^{\dagger}b_{L}+\epsilon_{R}^{\mathrm{st}}b_{R}^{\dagger}b_{R}+\epsilon_{LR}^{\mathrm{st}}\left[b_{L}b_{R}+b_{L}^{\dagger}b_{R}^{\dagger}\right],
\end{equation}
The relevant parameters $\epsilon_{\nu}^{\mathrm{st}}$ introduced
above are readily obtained from Eqs.(\ref{eq:nuclear-Stark-shift-Hamiltonian})
and (\ref{eq:HP-mapping-imperfection-parameter}). In the symmetric
setting $\eta_{L}=\eta_{R}$, it is instructive to re-express $H_{\mathrm{Stark}}$
in terms of the squeezed, non-local bosonic modes $a=\nu b_{L}^{\dagger}+\text{\ensuremath{\mu}}b_{R}$
and $\tilde{a}=\mu b_{L}+\nu b_{R}^{\dagger}$ {[}see Eqs. (\ref{eq:L2-nonlocal-bosonic-mode-a})
and (\ref{eq:LL2-nonlocal-bosonic-mode-a}){]} whose common vacuum
is the ideal steady state of $\mathcal{L}_{\mathrm{id}}$. Up to an
irrelevant constant term, $H_{\mathrm{Stark}}$ takes on the form
\begin{equation}
H_{\mathrm{Stark}}=\Delta_{a}a^{\dagger}a+\Delta_{\tilde{a}}\tilde{a}^{\dagger}\tilde{a}+g_{a\tilde{a}}\left(a\tilde{a}+a^{\dagger}\tilde{a}^{\dagger}\right).\label{eq:HStark-nonlocal-modes}
\end{equation}
With respect to the entanglement dynamics, the first two terms do
not play a role as the ideal steady state $\left|\Psi_{\mathrm{TMS}}\right\rangle $
is an eigenstate thereof. However, the last term is an active squeezing
term in the non-local bosonic modes: It does not preserve the excitation
number in the modes $a,\tilde{a}$ and may therefore drive the nuclear
system away from the vacuum by pumping excitations into the system.
Numerically, we find that the relative strength of $g_{a\tilde{a}}$
increases compared to the desired entangling dissipative terms when
tuning the interdot tunneling parameter $t$ towards $t_{\mathrm{crt}}$.
We therefore are confronted with two competing effects while tuning
the interdot coupling $t$. On the one hand, the dissipative dynamics
tries to pump the system into the vacuum of the modes $a$ and $\tilde{a}$
{[}see Eqs. (\ref{eq:L2-nonlocal-bosonic-mode-a}) and (\ref{eq:LL2-nonlocal-bosonic-mode-a}){]},
which become increasingly squeezed as we increase $t$. On the other
hand, an increase in $t$ leads to enhanced coherent dynamics (originating
from the nuclear Stark shift $H_{\mathrm{Stark}}$) which try to pump
excitations in the system {[}Eq.(\ref{eq:HStark-nonlocal-modes}){]}.
This competition between dissipative and coherent dynamics is known
to be at the origin of many dissipative phase transitions, and has
been extensively studied, e.g., in the context of the Dicke phase
transition.\cite{dimer07,nagy11}

As shown in Fig.~\ref{fig:steady-state-fidelity-Stark}, the observed
critical behaviour in the nuclear spin dynamics can indeed be traced
back to the presence of the nuclear Stark shift terms $H_{\mathrm{Stark}}$:
here, when tuning the system towards the critical point $t_{\mathrm{crt}}$,
the diverging number of HP bosons is shown to be associated with the
presence of $H_{\mathrm{Stark}}$. Moreover, for relatively low values
of the squeezing parameter $\left|\xi\right|$, we obtain a relatively
high fidelity $\mathcal{F}$ with the ideal two-mode squeezed state,
close to 80\%. For stronger squeezing, however, the target state becomes
more susceptible to the undesired noise terms, first leading to a
reduction of $\mathcal{F}$ and eventually to a break-down of the
HP approximation. 

Aside from this phase transition in the steady state, we find nonanalyticities
at non-zero values of the nuclear ADR, indicating a change in the
dynamical properties of the system which cannot be detected in steady-state
observables.\cite{kessler12} Rather, the system displays anomalous
behaviour approaching the stationary state: As shown Fig.~\ref{fig:dynamical-phase-transition},
we can distinguish two \textit{dynamical} phases,\cite{alvarez06, danieli07, eleuch14, lesanovsky13} an underdamped and
an overdamped one, respectively. The splitting of the real parts of
$\mathcal{M}$ coincides with vanishing imaginary parts. Thus, in
the overdamped regime, perturbing the system away from its steady
state leads to an exponential, non-oscillating return to the stationary
state. A similar underdamped region in direct vicinity of the phase
transition can be found in the dissipative Dicke phase transition.\cite{dimer07,nagy11}

\section{\textcolor{black}{Implementation} \label{sec:Implementation}}

This Section is devoted to the experimental realization of our proposal.
First, we summarize the experimental requirements of our scheme. Thereafter,
we address several effects that are typically encountered in realistic
systems, but which have been neglected so far in our analysis. This
includes non-uniform HF coupling, larger individual nuclear spins
$\left(I>1/2\right)$, external magnetic fields, different nuclear
species, internal nuclear dynamics and charge noise. 

\textit{Experimental requirements}.---Our proposal relies on the predominant
spin-blockade lifting via the electronic level $\left|\lambda_{2}\right\rangle $
and the adiabatic elimination of the electronic degrees of freedom:
First, the condition $t\gg\omega_{0},g_{\mathrm{hf}}$ ascertains
a predominant lifting of the Pauli-blockade via the hybridized, nonlocal
level $\left|\lambda_{2}\right\rangle $. To reach the regime in which
the electronic subsystem settles into the desired quasisteady state
$\rho_{\mathrm{ss}}^{\mathrm{el}}=\left(\left|T_{+}\right\rangle \left\langle T_{+}\right|+\left|T_{-}\right\rangle \left\langle T_{-}\right|\right)/2$
on a timescale much shorter than the nuclear dynamics, the condition
$\Gamma_{2}\gg\Gamma_{\pm}\gg\gamma_{c}$ must be fulfilled. Both,
$t\gg\omega_{0},g_{\mathrm{hf}}$ and $\Gamma_{\pm}\gg\gamma_{c}$
can be reached thanks to the extreme, separate, in-situ tunability
of the relevant, electronic parameters $t,\epsilon$ and $\Gamma$.\cite{hanson07}
Moreover, to kick-start the nuclear self-polarization process towards
a high-gradient stable fixed point, where the condition $\Gamma_{2}\gg\Gamma_{\pm}$
is fulfilled, an initial gradient of approximately $\sim\left(3-5\right)\mu\mathrm{eV}$,
corresponding to a nuclear polarization of $\sim\left(5-10\right)\%$,
is required; as shown in Sec.\ref{sec:Polarization-Dynamics}, this
ensures $\kappa_{2}^{2}\gg x_{\pm}$, where we estimate the suppression
factor $x_{\pm}=\Gamma_{\pm}/\Gamma\approx10^{-3}$. The required
gradient could be provided via an on-probe nanomagnet\cite{petersen13,pioro-ladriere08}
or alternative dynamic polarization schemes;\cite{gullans10,takahashi11,petta08,foletti09}
experimentally, nuclear spin polarizations of up to 50\% have been
reported for electrically defined quantum dots.\cite{petersen13,baugh07} 

\textit{Inhomogeneous HF coupling}.---Within the HP analysis presented
in Section \ref{sec:Steady-State-Entanglement}, we have restricted
ourselves to uniform HF coupling. Physically, this approximation amounts
to the assumption that the electron density is flat in the dots and
zero outside.\cite{rudner11a} In Ref.\cite{schwager10b}, it was
shown that corrections to this idealized scenario are of the order
of $1-p$ for a high nuclear polarization $p$. Thus, the HP analysis
for uniform HF coupling is correct to zeroth order in the small parameter
$1-p$. To make connection with a more realistic setting, where---according
to the electronic s-type wavefunction---the HF coupling constants
$a_{i,j}$ typically follow a Gaussian distribution, one may express
them as $a_{i,j}=\bar{a}+\delta_{i,j}$. Then, the uniform contribution
$\bar{a}$ enables an efficient description within fixed $J_{i}$
subspaces, whereas the non-uniform contribution leads to a coupling
between different $J_{i}$ subspaces on a much longer timescale. As
shown in Ref.\cite{christ07}, the latter is relevant in order to
avoid low-polarization dark states and to reach highly polarized nuclear
states. Let us stress that (for uniform HF coupling) we have found
that the generation of nuclear steady-state entanglement persists
in the presence of asymmetric $\left(N_{L}\neq N_{R}\right)$ dot
sizes which represents another source of inhomogeneity in our system. 

\begin{figure}
\includegraphics[width=1\columnwidth]{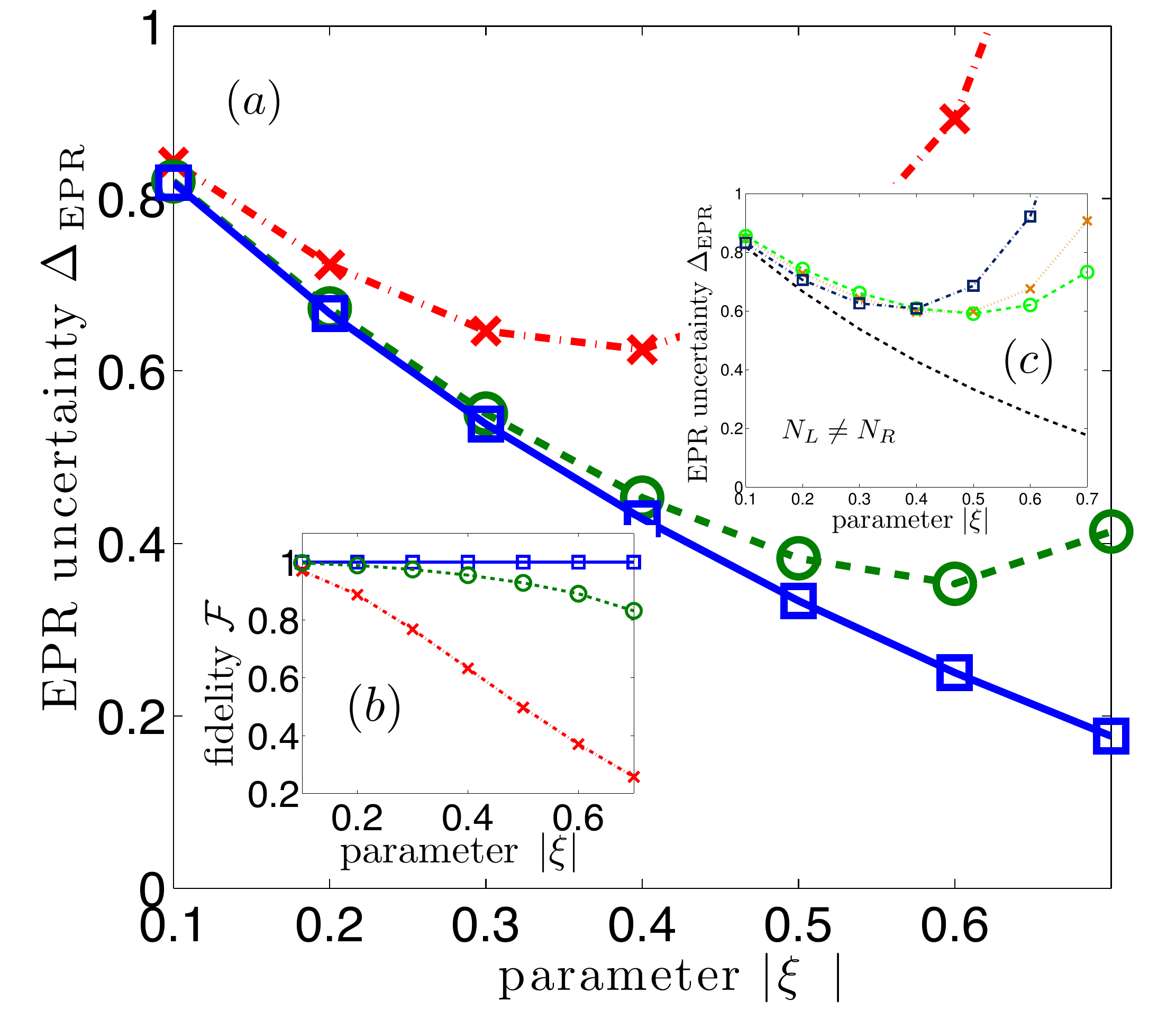}

\caption{\label{fig:EPR-inhomogeneous}(color online). EPR uncertainty (a)
and fidelity $\mathcal{F}$ with the ideal nuclear target state $\left|\xi_{\mathrm{ss}}\right\rangle $
given in Eq.(\ref{eq:steady-state-completely-inhomogenous}) (b) as
a function of the squeezing-like parameter $\left|\xi\right|$ for
$N_{L}=N_{R}=3$ \textit{inhomogeneously} coupled nuclei. The blue
curve (squares) refers to a symmetric setting where $\vec{a}_{L}=\vec{a}_{R}=\left(1.11,1.67,0.22\right)$,
whereas the green (circles) and red (crosses) solutions incorporate
asymmetries: $\vec{a}_{L}=\left(1.18,1.61,0.21\right)$, $\vec{a}_{R}=\left(1.11,1.67,0.22\right)$
and $\vec{a}_{L}=\left(1.0,1.5,0.5\right)$, $\vec{a}_{R}=\left(1.24,1.55,0.21\right)$,
respectively. (c) Exact results for the asymmetric scenario $N_{L}=2\neq3=N_{R}$.
Here, $\vec{a}_{L}=\left(1.0,1.5\right)$ was held fixed while the
green (circles), orange (crosses) and dark blue (squares) curves refer
to $\vec{a}_{R}=\left(0.98,1.47,0.05\right)$, $\vec{a}_{R}=\left(0.93,1.39,0.18\right)$
and $\vec{a}_{R}=\left(0.76,1.14,0.60\right)$, respectively; as a
benchmark, the black dashed curve refers to the ideal results in the
symmetric setting. Due to the absence of degeneracies, the steady
state solution $\sigma_{\mathrm{ss}}$ is unique in all cases considered
here. }
\end{figure}

In what follows, we show that our scheme works even in the case of
non-uniform coupling, provided that the two dots are sufficiently
similar. If the HF coupling constants are completely inhomogeneous,
that is $a_{i,j}\neq a_{i,k}$ for all $j\neq k$, but the two dots
are identical $\left(a_{1,j}=a_{2,j}\equiv a_{j}\,\forall j=1,2,\dots,N_{L}\equiv N_{R}\equiv N\right)$,
such that the nuclear spins can be grouped into pairs according to
their HF coupling constants, the two dominant nuclear jump operators
$L_{2}$ and $\mathbb{L}_{2}$ simplify to 
\begin{eqnarray}
L_{2} & = & \sum_{j}a_{j}l_{j},\,\,\,\,\,\,\,\,\,\,\mathbb{L}_{2}=\sum_{j}a_{j}\mathbb{l}_{j},
\end{eqnarray}
where the nuclear operators $l_{j}=\nu_{2}\sigma_{Lj}^{+}+\mu_{2}\sigma_{Rj}^{+}$
and $\mathbb{l}_{j}=\mu_{2}\sigma_{Lj}^{-}+\nu_{2}\sigma_{Rj}^{-}$
are nonlocal nuclear operators, comprising two nuclear spins that
belong to different nuclear ensembles, but have the same HF coupling
constant $a_{j}$. For one such pair of nuclear spins, the unique,
common nuclear dark state fulfilling 
\begin{equation}
l_{j}\left|\xi\right\rangle _{j}=\mathbb{l}_{j}\left|\xi\right\rangle _{j}=0,
\end{equation}
is easily verified to be 
\begin{equation}
\left|\xi\right\rangle _{j}=\mathcal{N}_{\xi}\left(\left|\downarrow_{j},\uparrow_{j}\right\rangle +\xi\left|\uparrow_{j},\downarrow_{j}\right\rangle \right),
\end{equation}
where $\mathcal{N}_{\xi}=1/\sqrt{1+\xi^{2}}$ for normalization. Therefore,
in the absence of degeneracies in the HF coupling constants $\left(a_{i,j}\neq a_{i,k}\,\forall j\neq k\right)$,
the \textit{pure, entangled }ideal nuclear dark state fulfilling $L_{2}\left|\xi_{\mathrm{ss}}\right\rangle =\mathbb{L}_{2}\left|\xi_{\mathrm{ss}}\right\rangle =0$
can be constructed as a tensor product of entangled pairs of nuclear
spins, 
\begin{equation}
\left|\xi_{\mathrm{ss}}\right\rangle =\otimes_{j=1}^{N}\left|\xi\right\rangle _{j}.\label{eq:steady-state-completely-inhomogenous}
\end{equation}
Again, the parameter $\xi=-\nu_{2}/\mu_{2}$ fully quantifies polarization
and entanglement properties of the nuclear stationary state; compare
Eq.(\ref{eq:target-squeezed-state-J-subspaces-uniform}): First, for
small values of the parameter $\left|\xi\right|$ the ideal nuclear
dark state $\left|\xi_{\mathrm{ss}}\right\rangle $ features an arbitrarily
high polarization gradient 
\begin{equation}
\Delta_{I^{z}}=\left\langle I_{R}^{z}\right\rangle _{\mathrm{ss}}-\left\langle I_{L}^{z}\right\rangle _{\mathrm{ss}}=N\frac{1-\xi^{2}}{1+\xi^{2}},
\end{equation}
whereas the homogeneous net polarization $P=\left\langle I_{L}^{z}\right\rangle _{\mathrm{ss}}+\left\langle I_{R}^{z}\right\rangle _{\mathrm{ss}}$
vanishes. The stationary solution for the nuclear gradient $\Delta_{I^{z}}$
is bistable as it is positive (negative) for $\left|\xi\right|<1$
$\left(\left|\xi\right|>1\right)$, respectively. Second, the amount
of entanglement inherent to the stationary solution $\left|\xi_{\mathrm{ss}}\right\rangle $
can be quantified via the EPR uncertainty ($\Delta_{\mathrm{EPR}}<1$
indicates entanglement) and is given by $\Delta_{\mathrm{EPR}}=\left(1-\left|\xi\right|\right)^{2}/\left|1-\xi^{2}\right|.$ 

Our analytical findings are verified by exact diagonalization results
for small sets of inhomogeneously coupled nuclei. Here, we compute
the exact (possibly mixed) solutions $\sigma_{\mathrm{ss}}$ to the
dark state equation $\mathcal{D}\left[L_{2}\right]\sigma_{\mathrm{ss}}+\mathcal{D}\left[\mathbb{L}_{2}\right]\sigma_{\mathrm{ss}}=0$;
compare Fig.~\ref{fig:ideal-target-state-uniform-HF-coupling} for
the special case of uniform HF coupling. As shown in Fig.~\ref{fig:EPR-inhomogeneous},
our numerical evidence indicates that small deviations from the perfect
symmetry (that is for $a_{Lj}\approx a_{Rj}$) between the QDs still
yield a (mixed) \textit{unique} entangled steady state close to $\left|\xi_{\mathrm{ss}}\right\rangle $.
In the ideal case $a_{Lj}=a_{Rj}$, we recover the pure steady state
given in Eq.(\ref{eq:steady-state-completely-inhomogenous}). Moreover,
we find that the generation of steady-state entanglement even persists
for asymmetric dot sizes, i.e. for $N_{L}\neq N_{R}$. Exact solutions
for $N_{L}=2\neq3=N_{R}$ are displayed in Fig.~\ref{fig:EPR-inhomogeneous}.
Here, we still find strong traces of the ideal dark state $\left|\xi_{\mathrm{ss}}\right\rangle $,
provided that one can approximately group the nuclear spins into pairs
of similar HF coupling strength. The interdot correlations $\left\langle \sigma_{Lj}^{+}\sigma_{Rj}^{-}\right\rangle $
are found to be close to the ideal value of $\xi/\left(1+\xi^{2}\right)$
for nuclear spins with a similar HF constant, but practically zero
otherwise. In line with this reasoning, the highest amount of entanglement
in Fig.~\ref{fig:EPR-inhomogeneous} is observed in the case where
one of the nuclear spins belonging to the bigger second ensemble is
practically uncoupled. Lastly, we note that one can \textquoteright{}continuously\textquoteright{}
go from the case of non-degenerate HF coupling constants (the case
considered in detail here) to the limit of uniform HF coupling {[}compare
Eq.(\ref{eq:target-squeezed-state-J-subspaces-uniform}){]} by grouping
spins with the same HF coupling constants to \textquoteright{}shells\textquoteright{},
which form collective nuclear spins. For degenerate couplings, however,
there are additional conserved quantities, namely the respective total
spin quantum numbers, and therefore multiple stationary states of
the above form. As argued in Section \ref{sec:Effective-Nuclear-Dynamics},
a mixture of different $J$-subspaces should still be entangled provided
that the range of $J$-subspaces involved in this mixture is small
compared to the average $J$ value. 

\textit{Larger nuclear spins}.---All natural isotopes of Ga and As
carry a nuclear spin $I=3/2$,\cite{schliemann03} whereas we have
considered $I=1/2$ for the sake of simplicity. For our purposes,
however, this effect can easily be incorporated as an individual nuclear
spin with $I=3/2$ maps onto 3 homogeneously coupled nuclear spins
with individual $I=1/2$ which are already in the fully symmetric
Dicke subspace $J=3/2$. 

\begin{figure}
\includegraphics[width=1\columnwidth]{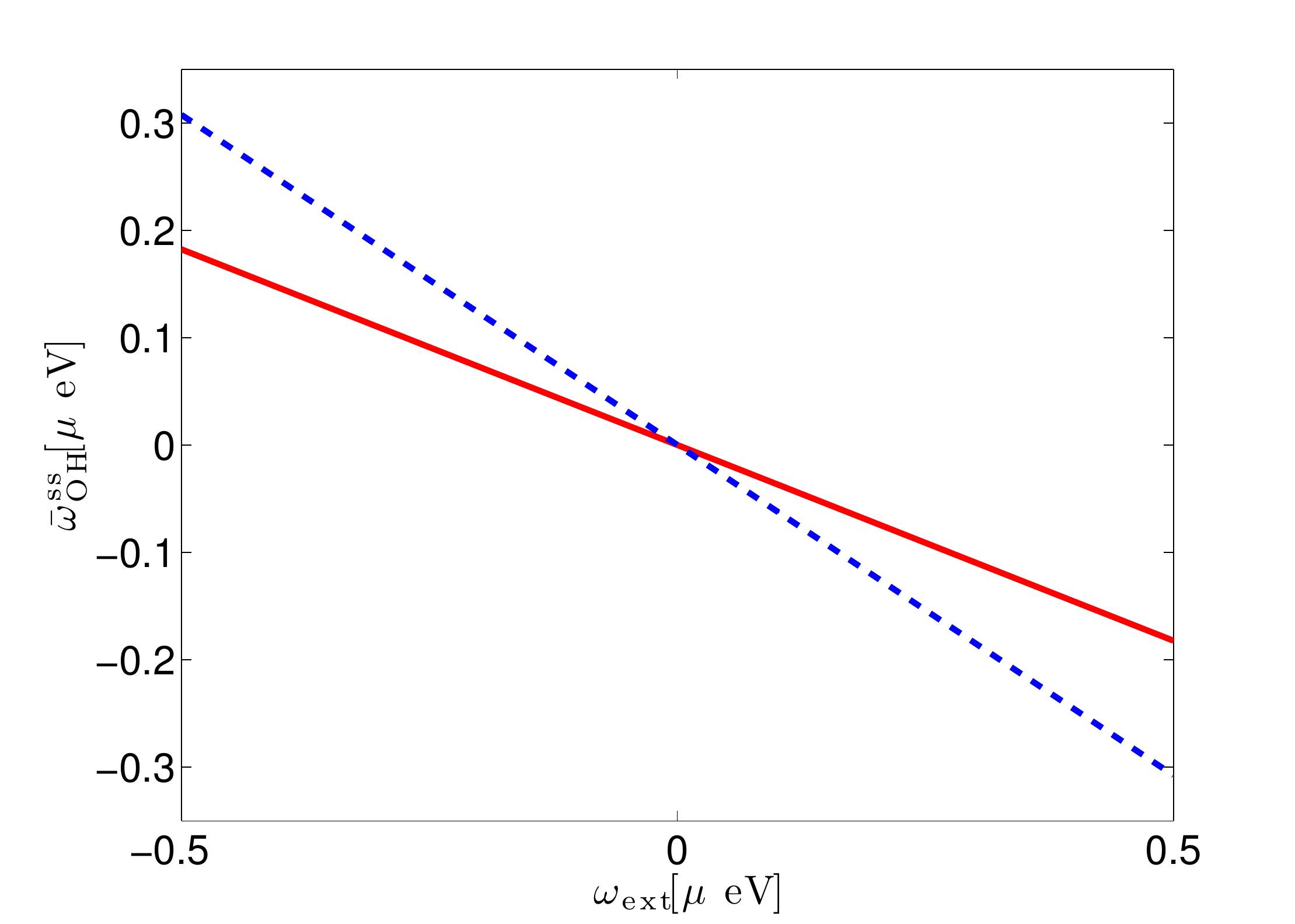}

\caption{\label{fig:omegaOH-omegaEXT}(color online). Buildup of a homogeneous
nuclear Overhauser field component $\bar{\omega}_{\mathrm{OH}}^{\mathrm{ss}}$
which partially compensates an applied external magnetic field, shown
here for $t=10\mu\mathrm{eV}$ (red solid) and $t=20\mu\mathrm{eV}$
(blue dashed). Other numerical parameters: $\Gamma=25\mu\mathrm{eV}$,
$\epsilon=30\mu\mathrm{eV}$, $\Gamma_{\pm}=\Gamma_{\mathrm{deph}}=0.1\mu\mathrm{eV}$. }
\end{figure}

\textit{External magnetic fields}.---For simplicity, our previous
analysis has focused on a symmetric setting of vanishing external
fields, $\Delta_{\mathrm{ext}}=\omega_{\mathrm{ext}}=0$. Non-vanishing
external fields, however, may be used as further experimental knobs
to tune the desired nuclear steady-state properties: First, as mentioned
above, a non-zero external gradient $\Delta_{\mathrm{ext}}$ is beneficial
for our proposal as it can provide an efficient way to destabilize
the zero-polarization solution $\left(\Delta_{\mathrm{OH}}^{\mathrm{ss}}=0\right)$
by initiating the nuclear self-polarization process. Second, non-vanishing
$\omega_{\mathrm{ext}}\neq0$ gives rise to another electron-nuclear
feedback-driven experimental knob for controlling the nuclear stationary
state. In the framework of Section \ref{sec:Polarization-Dynamics},
for $\omega_{\mathrm{ext}}\neq0$ the semiclassical dynamical equations
can be generalized to 
\begin{eqnarray}
\frac{d}{dt}\left\langle I_{L}^{z}\right\rangle _{t} & = & \alpha_{+}N_{L\downarrow}-\beta_{-}N_{L\uparrow},\\
\frac{d}{dt}\left\langle I_{R}^{z}\right\rangle _{t} & = & \beta_{+}N_{R\downarrow}-\alpha_{-}N_{R\uparrow},
\end{eqnarray}
where we have introduced the number of nuclear spin-up and spin-down
spins as $N_{i\uparrow}=N_{i}/2+\left\langle I_{i}^{z}\right\rangle $
and $N_{i\downarrow}=N_{i}/2-\left\langle I_{i}^{z}\right\rangle $,
respectively, and the generalized polarization rates 
\begin{eqnarray}
\alpha_{\pm} & = & p\gamma^{\pm}\nu_{2}^{2}+\left(1-2p\right)\gamma^{\mp}\mu_{2}^{2},\\
\beta_{\pm} & = & p\gamma^{\pm}\mu_{2}^{2}+\left(1-2p\right)\gamma^{\mp}\nu_{2}^{2}.
\end{eqnarray}
They depend on the generalized HF-mediated decay rate 
\begin{equation}
\gamma^{\pm}=\frac{a_{\mathrm{hf}}^{2}\tilde{\Gamma}}{2\left[\left(\epsilon_{2}\mp\omega_{0}\right)^{2}+\tilde{\Gamma}^{2}\right]},
\end{equation}
which accounts for different detunings for $\omega_{0}\neq0$; compare Eq.(\ref{eq:nuc-diss-rate-gamma}). 
As shown in Fig.~\ref{fig:omegaOH-omegaEXT}, in the presence of an external
magnetic splitting $\omega_{\mathrm{ext}}$, the nuclear spins build
up a homogeneous Overhauser field $\bar{\omega}_{\mathrm{OH}}$ $ $in
the steady state to partially compensate the external component. The
steady state solution then locally fulfills a detailed-balance principle,
namely $\alpha_{+}N_{L\downarrow}=\beta_{-}N_{L\uparrow}$ and $\beta_{+}N_{R\downarrow}=\alpha_{-}N_{R\uparrow}$,
which is determined by effective nuclear flip rates and the number
of spins available for a spin flip. Intuitively, this finding can
be understood as follows: For $\omega_{\mathrm{ext}}\neq0$, the degeneracy
between $\left|T_{+}\right\rangle $ and $\left|T_{-}\right\rangle $
is lifted with one of them being less detuned from $\left|\lambda_{2}\right\rangle $
than the other. This favors the build-up of a nuclear net polarization
$P$ which, however, counteracts the splitting $\omega_{\mathrm{ext}}$;
for $\omega_{\mathrm{ext}}=0$, this mechanism stabilizes $\bar{\omega}_{\mathrm{OH}}=P=0$
in the stationary state. This result has also been confirmed by numerical
results presented in Appendix \ref{sec:Numerical-Results-for-DNP}.

\textit{Species inhomogeneity}.---Nonzero external magnetic fields,
however, induce nuclear Zeeman splittings, with the nuclear magnetic
moment being about three orders of magnitude smaller than the Bohr
magneton for typical quantum dots.\cite{hanson07,schliemann03} Most
QDs consist of a few (in GaAs three) different species of nuclei with
strongly varying $g$ factors. In principle, this species-inhomogeneity
can cause dephasing between the nuclear spins. However, for a uniform
external magnetic field this dephasing mechanism only applies to nuclei
belonging to different species. In a rotating wave approximation,
this leads to few mutually decohered subsystems (in GaAs three) each
of which being driven towards a two-mode squeezedlike steady state:
note that, because of the opposite polarizations in the two dots,
the nuclear target state $\left|\xi\right\rangle _{\mathrm{ss}}$
is invariant under the application of a homogeneous magnetic field.
This argument, however, does not hold for an inhomogeneous magnetic
field which causes dephasing of $\left|\xi\right\rangle _{\mathrm{ss}}$
as the nuclear states $\left|m,-m\right\rangle $ ($m$ is the nuclear
spin projection) pick up a phase $\exp\left[2im\Delta_{\mathrm{ext}}^{\mathrm{nuc}}t\right]$,
where $\Delta_{\mathrm{ext}}^{\mathrm{nuc}}\approx10^{-3}\Delta_{\mathrm{ext}}$.
If one uses an external magnetic gradient to incite the nuclear self-polarization
process, after successful polarization one should therefore switch
off the gradient\cite{small-gradient-entanglement} to support the
generation of entanglement between the two ensembles. 

\textit{Weak nuclear interactions}.---We have neglected nuclear dipole-dipole
interactions among the nuclear spins. The strength of the effective
magnetic dipole-dipole interaction between \textit{neighboring} nuclei
in GaAs is about $g_{dd}\sim\left(100\mu\mathrm{s}\right)^{-1}$.\cite{hanson07,schliemann03}
Spin-nonconserving terms and flip-flop terms between different species
can be suppressed efficiently by applying an external magnetic field
of $B_{\mathrm{ext}}\gtrsim10\mathrm{mT}$.\cite{reilly08} As discussed
above, the corresponding (small) electron Zeeman splitting $\omega_{\mathrm{ext}}\approx0.25\mu\mathrm{eV}$
does not hamper our protocol. Then, it is sufficient to consider so-called
homonuclear flip-flop terms between nuclei of the same species only
and phase changing $zz$-terms. First, nuclear spin diffusion processes---governing
the dynamics of the spatial profile of the nuclear polarization by
changing $A_{i}^{z}$---have basically no effect within an (almost)
completely symmetric Dicke subspace. With typical timescales of $\gtrsim10\mathrm{s}$,
they are known to be very slow and therefore always negligible on
the timescale considered here.\cite{reilly08,rudner13,paget82} Second,
the interactions $\propto\sigma_{i}^{z}\sigma_{j}^{z}$ lead to dephasing
similar to the nuclear Zeeman terms discussed above: In a mean-field
treatment one can estimate the effective Zeeman splitting of a single
nuclear spin in the field of its surrounding neighbors to be a few
times $g_{dd}$.\cite{christ07} This mean field is different only
for different species and thus does not cause any homonuclear dephasing.
Still, the variance of this effective field may dephase spins of the
same species, but for a high nuclear polarization $\text{\ensuremath{\mathrm{p}}}_{\mathrm{nuc}}$
this effect is further suppressed by a factor $\sim\left(1-\mathrm{p}_{\mathrm{nuc}}^{2}\right)$
as the nuclei experience a \textit{sharp} field for a sufficiently
high nuclear polarization $\text{\ensuremath{\mathrm{p}}}_{\mathrm{nuc}}$.
Lastly, we refer to recently measured nuclear decoherence times of
$\sim1\mathrm{ms}$ in vertical double quantum dots.\cite{takahashi11}
Since this is slow compared to the dissipative gap of the nuclear
dynamics $\tau_{\mathrm{gap}}\approx\left(3-30\right)\mu\mathrm{s}$
for $N\approx10^{5}-10^{6}$, we conclude that it should be possible
to create entanglement between the two nuclear spin ensembles faster
than it gets disrupted due to dipole-dipole interactions among the
nuclear spins or other competing mechanisms.\cite{rudner11a} Moreover,
since strain is largely absent in electrically defined QDs,\cite{chekhovich13}
nuclear quadrupolar interactions have been neglected as well. For
a detailed analysis of the internal nuclear dynamics within a HP treatment,
we refer to Ref.\cite{schwager10}.

\textit{Charge noise}.---Nearly all solid-state qubits suffer from
some kind of charge noise.\cite{dial13} In a DQD device background
charge fluctuations and noise in the gate voltages may cause undesired
dephasing processes. In a recent experimental study,\cite{dial13}
voltage fluctuations in $\epsilon$ have been identified as the source
of the observed dephasing in a singlet-triplet qubit. In our setting,
however, the electronic subsystem quickly settles into the quasisteady
state $\rho_{\mathrm{ss}}^{\mathrm{el}}$ which lives solely in the
$\left(1,1\right)$ triplet subspace spanned by $\left\{ \left|T_{\pm}\right\rangle \right\} $
and is thus relatively robust against charge noise. Still, voltage
fluctuations in $\epsilon$ lead to fluctuations in the parameter
$\xi$ characterizing the nuclear two-mode target state given in Eq.(\ref{eq:target-squeezed-state-J-subspaces-uniform}).
For typical parameters $\left(t=20\mu\mathrm{eV},\,\epsilon=30\mu\mathrm{eV},\,\Delta=40\mu\mathrm{eV}\right)$,
however, $\xi$ turns out to be rather insensitive to fluctuations
in $\epsilon$, that is $\left|d\xi/d\epsilon\right|\approx10^{-2}/\mu\mathrm{eV}$.
Note that the system can be made even more robust (while keeping $\xi$
constant) by increasing both $\epsilon$ and $t$: For $t=50\mu\mathrm{eV}$,
$\epsilon=90\mu\mathrm{eV}$, the charge noise sensitivity is further
reduced to $\left|d\xi/d\epsilon\right|\approx3\times10^{-3}/\mu\mathrm{eV}$.
We can then estimate the sensitivity of the generated steady-state
entanglement via $\left|d\Delta_{\mathrm{EPR}}/d\epsilon\right|=\left|(d\xi/d\epsilon)(d\Delta_{\mathrm{EPR}}/d\xi)\right|\lesssim2\times10^{-2}/\mu\mathrm{eV}$,
where we have used $\left|d\Delta_{\mathrm{EPR}}/d\xi\right|=2/\left(1+\xi\right)^{2}<2$.
Typical fluctuations in $\epsilon$ of the order of $\sim\left(1-3\right)\mu\mathrm{eV}$
as reported in Ref.\cite{hayashi03} may then cause a reduction of
entanglement in the nuclear steady state of approximately $\sim5\%$
as compared to the optimal value of $\epsilon$. If the typical timescale
associated with charge noise $\tau_{\mathrm{noise}}$ is fast compared
to the dissipative gap of the nuclear dynamics, i.e., $\tau_{\mathrm{noise}}\ll\tau_{\mathrm{gap}}$,
the nuclear spins effectively only experience the averaged value of
$\xi$, coarse-grained over its fast fluctuations.

\section{\textcolor{black}{Conclusion and Outlook }\label{sec:Conclusion-and-Outlook}}

In summary, we have developed a theoretical master-equation-based
framework for a DQD in the Pauli-blockade regime which features coupled
dynamics of electron and nuclear spins as a result of the hyperfine
interaction. Our analysis is based on the typical separation of timescales
between (fast) electron spin evolution and (slow) nuclear spin dynamics,
yielding a coarse-grained quantum master equation for the nuclear
spins. This reverses the standard perspective in which the nuclei
are considered as an environment for the electronic spins, but rather
views the nuclear spins as the quantum system coupled to an electronic
environment with an exceptional degree of tunability. Here, we have
focused on a regime favorable for the generation of entanglement in
the nuclear steady state, whereas the electrons are driven to an unpolarized,
classically correlated separable state. Therefore, in this setting,
electron dephasing turns out to be an asset rather than a liability.
Our central master equation directly incorporates nonlinear feedback
mechanisms resulting from the back-action of the Overhauser field
on the electron energy levels and thus explains the nuclear multi-stability
in a very transparent way. The associated instability of the nuclei
towards self-polarization can be used as a means for controlling the
nuclear spin distribution.\cite{rudner07} For example, as a prominent
application, we predict the deterministic generation of entanglement
between two (spatially separated) mesoscopic spin ensembles, induced
by electron transport and the common, collective coupling of the nuclei
to the electronic degrees of freedom of the DQD. The nuclear entangled
state is of EPR type, which is known to play a key role in continuous
variable quantum information processing\cite{hammerer10,furusawa07},
quantum sensing\cite{wasilewski10} and metrology\cite{appel09,schleier-smith10,gross10}.
Since the entanglement generation does not rely on coherent evolution,
but is rather stabilized by the dissipative dynamics, the proposed
scheme is inherently robust against weak random perturbations. Moreover,
as two large spin ensembles with $N\sim10^{6}$ get entangled, the
nuclear system has the potential to generate large amounts of entanglement,
i.e., many ebits. Lastly, the apparent relatively large robustness
of the nuclear steady state against charge noise shows that, when
viewed as (for example) a platform for spin-based quantum memories,
nuclear spin ensembles have certain, intrinsic advantages with respect
to their electronic cousins.

Our results provide a clear picture of the feedback-driven polarization
dynamics in a generic electron transport setting and, therefore, should
serve as a useful guideline for future experiments aiming at an enhanced,
dynamical control of the nuclear spins: While DNP experiments in double
quantum dots, for example, have revealed an instability towards large
Overhauser gradients, consistent with our results, the question of whether
or not this instability results from dot asymmetry or some other mechanism
is still unsettled.\cite{foletti09,petta08,barthel12} Here, we study
a generic DC setting, where the buildup of a large OH gradient straightforwardly
emerges even in the presence of a completely symmetric coherent hyperfine
interaction. From a more fundamental, conceptual point of view, our
theory gives valuable insights into the complex, non-equilibrium many-body
dynamics of localized electronic spins interacting with a mesoscopic
number of nuclear spins. Understanding the quantum dynamics of this
central spin model marks an important goal in the field of mesoscopic
physics, as a notable number of unexpected and intriguing phenomena
such as multi-stability, switching, hysteresis and long timescale
oscillations have been observed in this system.\cite{ono04,koppens05,danon09b,rudner13}

On the one hand, reversing again our approach, our scheme may lead
to a better quantum control over the nuclear spin bath and therefore
improved schemes to coherently control electron spin qubits, by reducing
the Overhauser field fluctuations and/or exploiting the gradient for
electron spin manipulation (as demonstrated experimentally already
for example in Ref.\cite{foletti09}). On the other hand, with nuclear
spin coherence times ranging from hundreds of microseconds to a millisecond,\cite{takahashi11,chekhovich13}
our work could be extended towards nuclear spin-based information
storage and manipulation protocols. The nuclear spin ensembles could
serve as a long-lived entanglement resource providing the basic building
block for an on-chip (solid-state) quantum network. The nodes of this
quantum network could be interconnected with electrons playing the
role of photons in more conventional atomic, molecular, and optical
(AMO) based approaches.\cite{kimble08} To wire up the system, coherent
transport of electron spins over \textit{long} distances (potentially
tens of microns in state-of-the-art experimental setups) could be
realized via QD arrays \cite{busl13,braakman13}, quantum Hall edge
channels\cite{fletcher13,feve07,bocquillon13,roulleau08,mahe10} or
surface acoustic waves\cite{hermelin11,mcneil11,sanada13,yamamoto12}.
Building upon this analogy to quantum optics, the localized nuclei
might also be used as a source to generate a current of \textit{many}
entangled electrons.\cite{christ08} Using the aforementioned tunability
of the electronic degrees of freedom, one could also engineer different
electronic quasisteady states, possibly resulting in nuclear stationary
states with on-demand properties. On a more fundamental level, our
work could also be extended towards deeper studies of dissipative
phase transitions in this rather generic transport setting. When combined
with driving---realized via, for example, a magnetic field $B_{x}$
perpendicular to the polarization direction---a variety of strong-correlation
effects, nonequilibrium, and dissipative phase transitions can be
expected\cite{kessler12,carmichael80,morrison08} and could now be
studied in a mesoscopic solid-state system, complementing other approaches
to dissipative phase transitions in quantum dots\cite{rudner10,chung09,borda06,leggett87}.
\begin{acknowledgments}
M.J.A.S., J.I.C. and G.G. acknowledge support by the DFG within SFB
631, the Cluster of Excellence NIM and the project MALICIA within
the 7th Framework Programme for Research of the European Commission,
under FET-Open Grant No. 265522. E.M.K. acknowledges support by the
Harvard Quantum Optics Center and the Institute for Theoretical Atomic
and Molecular Physics. L.M.K.V. acknowledges support by the Dutch
Foundation for Fundamental Research on Matter (FOM).
\end{acknowledgments}
\appendix

\section{Spin-Blockade Regime \label{sec:Two-Electron-Regime}}

In this appendix, for completeness we explicitly derive inequalities
involving the chemical potentials $\mu_{L\left(R\right)}$ of the
left and right lead, respectively, as well as the Coulomb energies
introduced in Eq.(\ref{eq:Anderson-Hamiltonian}) that need to be
satisfied in order to tune the DQD into the desired Pauli-blockade
regime in which at maximum two electrons reside on the DQD. For simplicity,
Zeeman splittings are neglected for the moment as they typically constitute
a much smaller energy scale compared to the Coulomb energies. Still,
an extension to include them is straight-forward. Then, the bare energies
$E_{\left(m,n\right)}$ for a state with $\left(m,n\right)$ charge
configuration can easily be read off from the Anderson Hamiltonian
$H_{S}$. In particular, we obtain 
\begin{eqnarray}
E_{\left(1,1\right)} & = & \epsilon_{L}+\epsilon_{R}+U_{LR},\\
E_{\left(2,1\right)} & = & 2\epsilon_{L}+\epsilon_{R}+U_{L}+2U_{LR},\\
E_{\left(1,2\right)} & = & \epsilon_{L}+2\epsilon_{R}+U_{R}+2U_{LR},\\
E_{\left(0,2\right)} & = & 2\epsilon_{R}+U_{R},\\
E_{\left(2,0\right)} & = & 2\epsilon_{L}+U_{L}.
\end{eqnarray}
In order to exclude the occupation of $\left(2,1\right)$ and $\left(1,2\right)$
states if the DQD is in a $\left(1,1\right)$ charge configuration
the left chemical potential must fulfill the inequality $\mu_{L}<E_{\left(2,1\right)}-E_{\left(1,1\right)}=\epsilon_{L}+U_{L}+U_{LR}$.
An analog condition needs to be satisfied for the right chemical potential
$\mu_{R}$ so that we can write in total 
\begin{equation}
\mu_{i}<\epsilon_{i}+U_{i}+U_{LR}.
\end{equation}
The same requirement should hold if the DQD is in a $\left(0,2\right)$
or $\left(2,0\right)$ charge configuration which leads to 
\begin{equation}
\mu_{i}<\epsilon_{i}+2U_{LR}.
\end{equation}
At the same time, the chemical potentials $\mu_{i}$ are tuned sufficiently
high so that an electron is added to the DQD from the leads whenever
only a single electron resides in the DQD. For example, this results
in $\mu_{L}>E_{\left(1,1\right)}-\epsilon_{R}=\epsilon_{L}+U_{LR}$.
An analog condition needs to hold for the right lead which gives 
\begin{equation}
\mu_{i}>\epsilon_{i}+U_{LR}.
\end{equation}
In particular this inequality guarantees that the right dot is always
occupied, since $\mu_{R}>\epsilon_{R}$. Moreover, localized singlet
states cannot populated directly if $\mu_{i}<\epsilon_{i}+U_{i}$
holds. Since $U_{LR}<U_{i}$, the conditions to realize the desired
two-electron regime can be summarized as 
\begin{equation}
\epsilon_{i}+U_{LR}<\mu_{i}<\epsilon_{i}+2U_{LR}.\label{eq:two-electron-regime-inequality}
\end{equation}
By applying a large bias that approximately compensates the charging
energy of the two electrons residing on the right dot, that is $\epsilon_{L}\approx\epsilon_{R}+U_{R}-U_{LR}$,
the occupation of a localized singlet with charge configuration $\left(2,0\right)$
can typically be neglected.\cite{stepaneko12,rudner10} In this regime,
only states with the charge configurations $\left(0,1\right)$, $\left(1,0\right)$,
$\left(1,1\right)$ and $\left(0,2\right)$ are relevant. Also, due
to the large bias, admixing within the one-electron manifold is strongly
suppressed---for typical parameters we estimate $t/\left(\epsilon_{L}-\epsilon_{R}\right)\approx10^{-2}$---such
that the relevant single electron states that participate in the transport
cycle in the spin-blockade regime are the two lowest ones $\left|0,\sigma\right\rangle =d_{R\sigma}^{\dagger}\left|0\right\rangle $
with $\left(0,1\right)$ charge configuration.\cite{giavaras13}

\section{Quantum Master Equation in Spin-Blockade Regime\label{sec:Quantum-Master-Equation}}

Following the essential steps presented in Ref.\cite{schuetz12},
we now derive an effective master equation for the DQD system which
experiences irreversible dynamics via the electron's coupling to the
reservoirs in the leads. We start out from the von Neumann equation
for the global density matrix given in Eq.(\ref{eq:von-Neumann}).
It turns out to be convenient to decompose $\mathcal{H}$ as 
\begin{equation}
\mathcal{H}=H_{0}+H_{1}+H_{T},
\end{equation}
with $H_{0}=H_{S}+H_{B}$ and $H_{1}=V_{\mathrm{HF}}+H_{t}$. We define
the superoperator $P$ as 
\begin{equation}
P\varrho=\mathsf{Tr}_{\mathsf{B}}\left[\varrho\right]\otimes\rho_{B}^{0}.
\end{equation}
It acts on the total system's density matrix $\varrho$ and projects
the environment onto their respective thermal equilibrium states,
labeled as $\rho_{B}^{0}$. The map $P$ satisfies $P^{2}=P$ and
is therefore called a projector. By deriving a closed equation for
the projection $P\varrho$ and tracing out the unobserved reservoir
degrees of freedom, we arrive at the Nakajima-Zwanzig master equation
for the system's density matrix 
\begin{eqnarray}
\dot{\rho} & = & \left[\mathcal{L}_{S}+\mathcal{L}_{1}\right]\rho\\
 &  & +\int_{0}^{t}d\tau\mathsf{Tr}_{\mathsf{B}}\left[\mathcal{L}_{T}e^{\left(\text{\ensuremath{\mathcal{L}}}_{0}+\mathcal{L}_{T}+\mathcal{L}_{1}\right)\tau}\mathcal{L}_{T}\rho\left(t-\tau\right)\otimes\rho_{B}^{0}\right].\nonumber 
\end{eqnarray}
where the Liouville superoperators are defined as usual via $\mathcal{L}_{\alpha}\cdot=-i\left[H_{\alpha},\cdot\right]$.
Next, we introduce two approximations: First, in the weak coupling
limit, we neglect all orders higher than two in $\mathcal{L}_{T}$.
This is well known as the Born approximation. Accordingly, we neglect
$\mathcal{L}_{T}$ in the exponential of the integrand. Second, we
apply the approximation of independent rates of variations \cite{cohen-tannoudji92}
which can be justified self-consistently, if the bath correlation
time $\tau_{c}$ is short compared to the typical timescales associated
with the system's internal interactions, that is $g_{\mathrm{hf}}\tau_{c}\ll1$
and $t\tau_{c}\ll1$, and if $H_{1}$ can be treated as a perturbation
with respect to $H_{0}$. In our system, the latter is justified as
$H_{0}$ incorporates the large Coulomb energy scales which energetically
separate the manifold with two electrons on the DQD from the lower
manifold with only one electron residing in the DQD, whereas $H_{1}$
induces couplings within these manifolds only. In this limit, the
master equation then reduces to 
\begin{eqnarray}
\dot{\rho} & = & \left[\mathcal{L}_{S}+\mathcal{L}_{1}\right]\rho\\
 &  & +\int_{0}^{t}d\tau\mathsf{Tr}_{\mathsf{B}}\left[\mathcal{L}_{T}e^{\text{\ensuremath{\mathcal{L}}}_{0}\tau}\mathcal{L}_{T}\rho\left(t-\tau\right)\otimes\rho_{B}^{0}\right].\nonumber 
\end{eqnarray}
In the next step, we write out the tunnel Hamiltonian $H_{T}$ in
terms of the relevant spin-eigenstates. Here, we single out one term
explicitly, but all others follow along the lines. We get 
\begin{eqnarray}
\dot{\rho} & = & \dots+\sum_{\sigma}\int_{0}^{t}d\tau\mathcal{C}\left(\tau\right)\left|0,\sigma\right\rangle \left\langle S_{02}\right|\\
 &  & \left[e^{-iH_{0}\tau}\rho\left(t-\tau\right)e^{iH_{0}\tau}\right]\left|S_{02}\right\rangle \left\langle 0,\sigma\right|,\nonumber 
\end{eqnarray}
where 
\begin{equation}
\mathcal{C}\left(\tau\right)=\int_{0}^{\infty}d\epsilon J\left(\epsilon\right)e^{i\left(\Delta E-\epsilon\right)\tau},
\end{equation}
and $J\left(\epsilon\right)=\left|T_{R}\right|^{2}n_{R}\left(\epsilon\right)\left[1-f_{R}\left(\epsilon\right)\right]$
is the spectral density of the right lead, with $n_{R}\left(\epsilon\right)$
being the density of states per spin of the right lead; $f_{\alpha}\left(\epsilon\right)$
denotes the Fermi function of lead $\alpha=L,R$ and $\Delta E$ is
the energy splitting between the two levels involved, i.e., for the
term explicitly shown above $\Delta E=\epsilon_{R}+U_{R}$. The correlation
time of the bath $\tau_{c}$ is determined by the decay of the memory-kernel
$\mathcal{C}\left(\tau\right)$. The Markov approximation is valid
if the spectral density $J\left(\epsilon\right)$ is flat on the scale
of all the effects that we have neglected in the previous steps. Typically,
the effective density of states $D\left(\epsilon\right)=\left|T_{R}\right|^{2}n_{R}\left(\epsilon\right)$
is weakly energy dependent so that this argument is mainly concerned
with the Fermi functions of the left (right) lead $f_{L\left(R\right)}\left(\epsilon\right)$,
respectively. Therefore, if $f_{i}\left(\epsilon\right)$ is flat
on the scale of $\sim t$, $\sim g_{\mathrm{hf}}$ and the dissipative
decay rates $\sim\Gamma$, it can be evaluated at $\Delta E$ and
a Markovian treatment is valid.\cite{schuetz12} In summary, this
results in 
\begin{equation}
\dot{\rho}=\dots+\Gamma_{R}\sum_{\sigma}\mathcal{D}\left[\left|0,\sigma\right\rangle \left\langle S_{02}\right|\right]\rho,
\end{equation}
where $\Gamma_{R}$ is the typical sequential tunneling rate $\Gamma_{R}=2\pi\left|T_{R}\right|^{2}n_{R}\left(\Delta E\right)\left[1-f_{R}\left(\Delta E\right)\right]$
describing direct hopping at leading order in the dot-lead coupling.\cite{qassemi09,schuetz12} 

\textit{Pauli blockade}.---The derivation above allows for a clear
understanding of the Pauli-spin blockade in which only the level $\left|S_{02}\right\rangle $
can decay into the right lead whereas all two electron states with
$\left(1,1\right)$ charge configuration are stable. If the $\left|S_{02}\right\rangle $
level decays, an energy of $\Delta E_{2}=E_{\left(0,2\right)}-\epsilon_{R}=\epsilon_{R}+U_{R}$
is released on the DQD which has to be absorbed by the right reservoir
due to energy conservation arguments. On the contrary, if one of the
$\left(1,1\right)$ levels were to decay to the right lead, an energy
of $\Delta E_{1}=E_{\left(1,1\right)}-\epsilon_{L}=\epsilon_{R}+U_{LR}$
would dissipate into the continuum. Therefore, the DQD is operated
in the Pauli blockade regime if $f_{R}\left(\Delta E_{2}\right)=0$
and $f_{R}\left(\Delta E_{1}\right)=1$ is satisfied. Experimentally,
this can be realized easily as $\Delta E_{2}$ scales with the on-site
Coulomb energy $\Delta E_{2}\sim U_{R}$, whereas $\Delta E_{1}$
scales only with the interdot Coulomb energy $\Delta E_{1}\sim U_{LR}$. 

Taking into account all relevant dissipative processes within the
Pauli-blockade regime and assuming the Fermi function of the left
lead $f_{L}\left(\epsilon\right)$ to be sufficiently flat, the full
quantum master equation for the DQD reads 
\begin{eqnarray}
\dot{\rho} & = & -i\left[H_{S}+H_{1},\rho\right]+\Gamma_{R}\sum_{\sigma}\mathcal{D}\left[\left|0,\sigma\right\rangle \left\langle S_{02}\right|\right]\rho\nonumber \\
 &  & +\Gamma_{L}\left\{ \mathcal{D}\left[\left|T_{+}\right\rangle \left\langle 0,\Uparrow\right|\right]\rho+\mathcal{D}\left[\left|\Downarrow\Uparrow\right\rangle \left\langle 0,\Uparrow\right|\right]\rho\right\} \nonumber \\
 &  & +\Gamma_{L}\left\{ \mathcal{D}\left[\left|T_{-}\right\rangle \left\langle 0,\Downarrow\right|\right]\rho+\mathcal{D}\left[\left|\Uparrow\Downarrow\right\rangle \left\langle 0,\Downarrow\right|\right]\rho\right\} ,\label{eq:QME-Markov-single-particle}
\end{eqnarray}
where the rate $\Gamma_{R}\sim\left[1-f_{R}\left(\Delta E_{2}\right)\right]$
describes the decay of the localized singlet $\left|S_{02}\right\rangle $
into the right lead, while the second and third line represent subsequent
recharging of the DQD with the corresponding rate $\Gamma_{L}\propto\left|T_{L}\right|^{2}$.\cite{secular-approximation} 

We can obtain a simplified description for the regime in which on
relevant timescales the DQD is always populated by two electrons.
This holds for sufficiently strong recharging of the DQD which can
be implemented experimentally by making the left tunnel barrier $T_{L}$
more transparent than the right one $T_{R}$.\cite{schuetz12,petersen13,giavaras13}
In this limit, we can eliminate the intermediate stage in the sequential
tunneling process $\left(0,2\right)\rightarrow\left(0,1\right)\rightarrow\left(1,1\right)$
and parametrize $H_{S}+H_{1}$ in the two-electron regime as $H_{\mathrm{el}}+H_{\mathrm{ff}}+H_{\mathrm{zz}}$.
Then, we arrive at the effective master equation 
\begin{equation}
\dot{\rho}=-i\left[H_{\mathrm{el}},\rho\right]+\mathcal{K}_{\Gamma}\rho+\mathcal{V}\rho,
\end{equation}
where the dissipator 
\begin{equation}
\mathcal{K}_{\Gamma}\rho=\Gamma\sum_{x\in\left(1,1\right)}\mathcal{D}\left[\left|x\right\rangle \left\langle S_{02}\right|\right]\rho\label{eq:transport-dissipator-two-electron-regime-bare-basis}
\end{equation}
models electron transport through the DQD; the sum runs over all four
electronic bare levels with $\left(1,1\right)$ charge configuration,
i.e., $\left|\sigma,\sigma'\right\rangle $ for $\sigma,\sigma'=\Uparrow,\Downarrow$:
Thus, in the limit of interest, the $\left(1,1\right)$ charge states
are reloaded with an effective rate $\Gamma=\Gamma_{R}/2$ via the
decay of the localized singlet $ $$\left|S_{02}\right\rangle $.\cite{petersen13,giavaras13} 

\begin{figure}
\includegraphics[width=1\columnwidth]{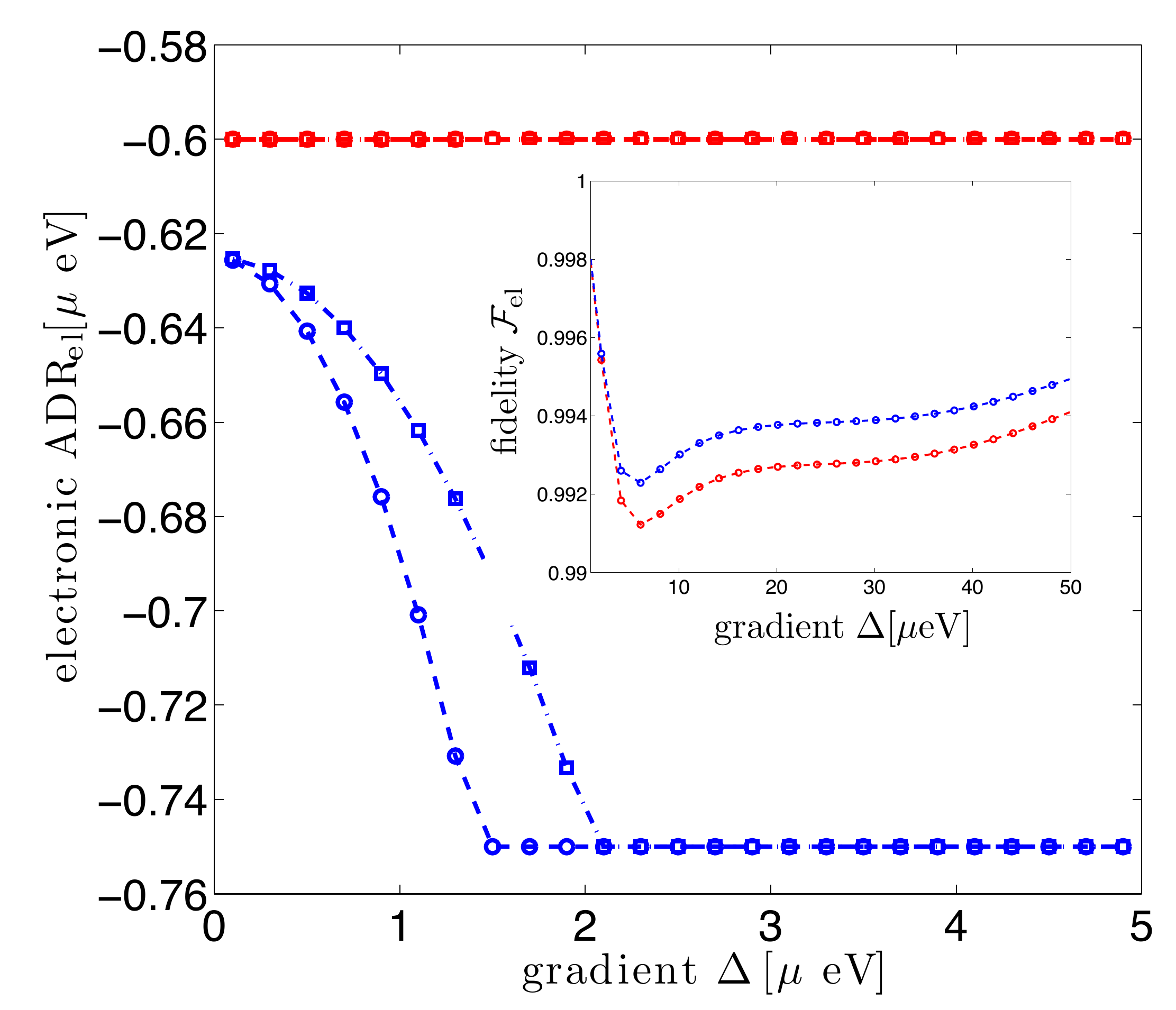}

\caption{\label{fig:ADR-simplified-transport-Liouvillian-check}(color online).
Electronic asymptotic decay rate $\mathrm{ADR}_{\mathrm{el}}$ and
fidelity $\mathcal{F}_{\mathrm{el}}$ for the purely electronic Lindblad
dynamics: The results obtained for the full dissipator given in Eq.(\ref{eq:electronic-Liouvilian0-full-spin-eigenstates})
(circles) are in good agreement with the results we get for the simplified
description as stated in Eq.(\ref{eq:electronic-Liouvilian0-simplified-energy-eigenstates})
(squares). The blue and red curves correspond to $\Gamma=25\mu\mathrm{eV}$,
$\Gamma_{\pm}=0.25\mu\mathrm{eV}$, $\Gamma_{\mathrm{deph}}=0.5\mu\mathrm{eV}$
and $\Gamma=25\mu\mathrm{eV}$, $\Gamma_{\pm}=0.3\mu\mathrm{eV}$,
$\Gamma_{\mathrm{deph}}=0$, respectively. Inset: The fidelity $\mathcal{F}_{\mathrm{el}}$
as a figure of merit for the similarity between the quasi-steady-state
solutions $\rho_{\mathrm{ss}}^{\mathrm{el}}$ and $\tilde{\rho}_{\mathrm{ss}}^{\mathrm{el}}$,
respectively. Other numerical parameters are: $t=20\mu\mathrm{eV}$,
$\epsilon=30\mu\mathrm{eV}$ and $\omega_{0}=0$. }
\end{figure}

\textit{Transport dissipator in eigenbasis of $H_{\mathrm{el}}$}.---The
electronic transport dissipator $\mathcal{K}_{\Gamma}$ as stated
in Eq.(\ref{eq:transport-dissipator-two-electron-regime-bare-basis})
describes electron transport in the bare basis of the two-orbital
Anderson Hamiltonian which does not correspond to the eigenbasis of
$H_{\mathrm{el}}$ due to the presence of the interdot tunnel coupling
$H_{t}$; in deriving Eq.(\ref{eq:transport-dissipator-two-electron-regime-bare-basis})
admixing due to $H_{t}$ has been neglected based on the approximation
of independent rates of variation.\cite{cohen-tannoudji92} It is
valid if $t\tau_{c}\ll1$ where $\tau_{c}\approx10^{-15}\text{\ensuremath{\mathrm{s}}}$
specifies the bath correlation time.\cite{schuetz12} Performing a
basis transformation $\tilde{\rho}=V^{\dagger}\rho V$ which diagonalizes
the electronic Hamiltonian $\tilde{H}_{\mathrm{el}}=V^{\dagger}H_{\mathrm{el}}V=\mathrm{diag}\left(\omega_{0},-\omega_{0},\epsilon_{1},\epsilon_{2},\epsilon_{3}\right)$
and neglecting terms rotating at a frequency of $\epsilon_{l}-\epsilon_{k}$
for $k\neq l$, the electronic transport dissipator takes on the form\cite{tilde}
\begin{eqnarray}
\mathcal{K}_{\Gamma}\tilde{\rho} & = & \sum_{k,\nu=\pm}\Gamma_{k}\mathcal{D}\left[\left|T_{\nu}\right\rangle \left\langle \lambda_{k}\right|\right]\tilde{\rho}\label{eq:electronic-transport-dissipator-after-RWA}\\
 &  & +\sum_{k,j}\Gamma_{k\rightarrow j}\mathcal{D}\left[\left|\lambda_{j}\right\rangle \left\langle \lambda_{k}\right|\right]\tilde{\rho},\nonumber 
\end{eqnarray}
where $\Gamma_{k}=\kappa_{k}^{2}\Gamma$ and $\Gamma_{k\rightarrow j}=\Gamma_{k}[1-\left|\kappa_{j}\right|^{2}]$.
Since only $\left(1,1\right)$ states can be refilled, the rate at
which the level $\left|\lambda_{j}\right\rangle $ is populated is
proportional to $\sim[1-\left|\kappa_{j}\right|^{2}]$; compare Ref.\cite{petersen13}.
While the first line in Eq.(\ref{eq:electronic-transport-dissipator-after-RWA})
models the decay from the dressed energy eigenstates $\left|\lambda_{k}\right\rangle $
back to the Pauli-blocked triplet subspace $\left|T_{\nu}\right\rangle \left(\text{\ensuremath{\nu}=\ensuremath{\pm}}\right)$
with an effective rate according to their overlap with the localized
singlet, the second line refers to decay and dephasing processes acting
entirely within the 'fast' subspace spanned by $\left\{ \left|\lambda_{k}\right\rangle \right\} $.
Intuitively, they should not affect the nuclear dynamics that take
place on a much longer timescale. This intuitive picture is corroborated
by exact diagonalization results: Leaving the HF interaction $\mathcal{V}$
aside for the moment, we compare the dynamics $\dot{\rho}=\mathcal{K}_{0}\rho$
generated by the full electronic Liouvillian
\begin{eqnarray}
\mathcal{K}_{0}\rho & = & -i\left[H_{\mathrm{el}},\rho\right]+\mathcal{K}_{\Gamma}\rho\label{eq:electronic-Liouvilian0-full-spin-eigenstates}\\
 &  & +\mathcal{K}_{\text{\ensuremath{\pm}}}\rho+\mathcal{L}_{\mathrm{deph}}\rho,\nonumber \\
\mathcal{K}_{\text{\ensuremath{\pm}}}\rho & = & \Gamma_{\pm}\sum_{\nu=\pm}\mathcal{D}\left[\left|T_{\bar{\nu}}\right\rangle \left\langle T_{\nu}\right|\right]\rho\\
 &  & +\Gamma_{\pm}\sum_{\nu=\pm}\left[\mathcal{D}\left[\left|T_{\nu}\right\rangle \left\langle T_{0}\right|\right]\rho+\mathcal{D}\left[\left|T_{0}\right\rangle \left\langle T_{\nu}\right|\right]\rho\right]\nonumber 
\end{eqnarray}
formulated in terms of the five undressed, bare levels $\left\{ \left|\sigma,\sigma'\right\rangle ,\left|S_{02}\right\rangle \right\} $
to the following Liouvillian 
\begin{eqnarray}
\mathcal{L}_{0}\tilde{\rho} & = & -i\left[\tilde{H}_{\mathrm{el}},\tilde{\rho}\right]+\mathcal{L}_{\Gamma}\tilde{\rho}\nonumber \\
 &  & +\mathcal{L}_{\pm}\tilde{\rho}+\mathcal{L}_{\mathrm{deph}}\tilde{\rho},\label{eq:electronic-Liouvilian0-simplified-energy-eigenstates}
\end{eqnarray}
which is based on the simplified form as stated in Eq.(\ref{eq:electronic-transport-dissipator-after-RWA}).\cite{cotunneling-dressed-vs-bare}
Here, we have also disregarded all dissipative processes acting entirely
within the fast subspace, that is all terms of the form $\mathcal{D}\left[\left|\lambda_{j}\right\rangle \left\langle \lambda_{k}\right|\right]$;
see the second line in Eq.(\ref{eq:electronic-transport-dissipator-after-RWA}).
First, as shown in Fig.~\ref{fig:ADR-simplified-transport-Liouvillian-check},
we have checked numerically that both $\mathcal{K}_{0}$ and $\mathcal{L}_{0}$
feature very similar electronic quasisteady states, fulfilling $ $$\mathcal{K}_{0}\left[\rho_{\mathrm{ss}}^{\mathrm{el}}\right]=0$
and $\mathcal{L}_{0}\left[\tilde{\rho}_{\mathrm{ss}}^{\mathrm{el}}\right]=0$,
respectively, with a Uhlmann fidelity\cite{uhlmann76} $\mathcal{F}_{\mathrm{el}}\left(\rho_{\mathrm{ss}}^{\mathrm{el}},\tilde{\rho}_{\mathrm{ss}}^{\mathrm{el}}\right)=\left\Vert \sqrt{\rho_{\mathrm{ss}}^{\mathrm{el}}}\sqrt{\tilde{\rho}_{\mathrm{ss}}^{\mathrm{el}}}\right\Vert _{\mathrm{tr}}$
exceeding $99\%$; here, $\left\Vert \cdot\right\Vert _{\mathrm{tr}}$
is the trace norm, the sum of the singular values. Second, we examine
the electronic asymptotic decay rate $\mathrm{ADR}_{\mathrm{el}}$,
corresponding to the eigenvalue with the largest real part different
from zero, which quantifies the typical timescale on which the electronic
subsystem reaches its quasi-steady state.\cite{kessler12} In other
words, the $\mathrm{ADR}_{\mathrm{el}}$ gives the spectral gap of
the electronic Liouvillian $\mathcal{K}_{0}\left(\mathcal{L}_{0}\right)$
setting the inverse relaxation time towards the steady state and therefore
characterizes the long-time behaviour of the electronic system. The
two models produce very similar results: Depending on the particular
choice of parameters, the electronic $\mathrm{ADR}_{\mathrm{el}}$
is set either by the eigenvectors $\left|\lambda_{2}\right\rangle \left\langle T_{\pm}\right|$,
$\left|T_{+}\right\rangle \left\langle T_{-}\right|$ and $\left|T_{+}\right\rangle \left\langle T_{+}\right|-\left|T_{-}\right\rangle \left\langle T_{-}\right|$
which explains the kinks observed in Fig.~\ref{fig:ADR-simplified-transport-Liouvillian-check}
as changes of the eigenvectors determining the $\mathrm{ADR}_{\mathrm{el}}$.
In summary, both the electronic quasisteady state $\left(\rho_{\mathrm{ss}}^{\mathrm{el}}\approx\tilde{\rho}_{\mathrm{ss}}^{\mathrm{el}}\right)$
and the electronic asymptotic decay rate $\mathrm{ADR}_{\mathrm{el}}$
are well captured by the approximative Liouvillian given in Eq.(\ref{eq:electronic-Liouvilian0-simplified-energy-eigenstates}).
Further arguments justifying this approximation are provided in Appendix
\ref{sec:Transport-Mediated-Transitions-Fast-Subspace}.

\section{Transport-Mediated Transitions In Fast Electronic Subspace\label{sec:Transport-Mediated-Transitions-Fast-Subspace}}

In this appendix, we provide analytical arguments why one can drop
the second line in Eq.(\ref{eq:electronic-transport-dissipator-after-RWA})
and keep only the first one to account for a description of electron
transport in the eigenbasis of $H_{\mathrm{el}}$. The second line,
given by 
\begin{equation}
\mathcal{L}_{\mathrm{fast}}\rho=\sum_{k,j}\Gamma_{k\rightarrow j}\mathcal{D}\left[\left|\lambda_{j}\right\rangle \left\langle \lambda_{k}\right|\right]\rho,
\end{equation}
describes transport-mediated transitions in the fast subspace $\left\{ \left|\lambda_{k}\right\rangle \right\} $.
The transition rate $\Gamma_{k\rightarrow j}=\kappa_{k}^{2}\left[1-\kappa_{j}^{2}\right]\Gamma$
refers to a transport-mediated decay process from $\left|\lambda_{k}\right\rangle $
to $\left|\lambda_{j}\right\rangle $. Here, we show that $\mathcal{L}_{\mathrm{fast}}$
simply amounts to an effective dephasing mechanism which can be absorbed
into a redefinition of the effective transport rate $\Gamma$. 

The only way our model is affected by $\mathcal{L}_{\mathrm{fast}}$
is that it adds another dephasing channel for the coherences $\left|\lambda_{k}\right\rangle \left\langle T_{\pm}\right|$
which are created by the hyperfine flip-flop dynamics; see Appendix
\ref{sec:Appendix-Effective-Nuclear-QME-High-Gradient-Regime}. In
fact, we have 
\begin{eqnarray}
\mathcal{L}_{\mathrm{fast}}\left[\left|\lambda_{k}\right\rangle \left\langle T_{\pm}\right|\right] & = & -\Gamma_{\mathrm{fast},k}\left|\lambda_{k}\right\rangle \left\langle T_{\pm}\right|,\\
\Gamma_{\mathrm{fast},k} & = & \frac{1}{2}\sum_{j}\Gamma_{k\rightarrow j}.
\end{eqnarray}
Due to the normalization condition $\sum_{j}\kappa_{j}^{2}=1$, the
new effective dephasing rate $\Gamma_{\mathrm{fast},k}$ is readily
found to coincide with the effective transport rate $\Gamma_{k}$,
that is $ $$\Gamma_{\mathrm{fast},k}=\Gamma_{k}=\kappa_{k}^{2}\Gamma$.
This equality is readily understood since all four $\left(1,1\right)$
levels are populated equally. While $\Gamma_{k}$ describes the decay
to the two Pauli-blocked triplet levels, $\Gamma_{\mathrm{fast},k}$
accounts for the remaining transitions within the $\left(1,1\right)$
sector. Therefore, when accounting for $\mathcal{L}_{\mathrm{fast}}$,
the total effective dephasing rates $\tilde{\Gamma}_{k}$ needs to
be modified as $\tilde{\Gamma}_{k}\rightarrow\tilde{\Gamma}_{k}+\Gamma_{k}=2\Gamma_{k}+3\Gamma_{\pm}+\Gamma_{\mathrm{deph}}/4$.
The factor of $2$ is readily absorbed into our model by a simple
redefinition of the overall transport rate $\Gamma\rightarrow2\Gamma$.

\section{Electronic Lifting of Pauli-Blockade \label{sec:Electronic-Lifting-of-Pauli-Blockade}}

This appendix provides a detailed analysis of purely electronic
mechanisms which can lift the Pauli-blockade without affecting directly
the nuclear spins. Apart from cotunneling processes discussed in the
main text, here we analyze virtual spin exchange processes and spin-orbital
effects.\cite{rudner07,rudner11b} It is shown, that these mechanisms,
though microscopically distinct, phenomenologically amount to effective
incoherent mixing and pure dephasing processes within the $(1,1)$
subspace which, for the sake of theoretical generality, are subsumed
under the term $\circled2$ in Eq.(\ref{eq:effective-QME-full-model}). 

Let us also note that electron spin resonance (ESR) techniques in
combination with dephasing could be treated on a similar footing.
\textcolor{black}{As recently shown in Ref.\cite{sanchez13}, in the
presence of a gradient $\Delta$, ESR techniques can be used to drive
the electronic system into the entangled steady state $\left|-\right\rangle =\left(\left|T_{+}\right\rangle -\left|T_{-}\right\rangle \right)/\sqrt{2}$.
Magnetic noise may then be employed to engineer the desired electronic
quasisteady state. }

\subsection{Spin Exchange with the Leads}

In the Pauli-blockade regime the $\left(1,1\right)$ triplet states
$\left|T_{\pm}\right\rangle $ do not decay directly, but---apart
from the cotunneling processes described in the main text---they may
exchange electrons with the reservoirs in the leads via higher-order
virtual processes.\cite{rudner07,rudner11b} We now turn to these
virtual, spin-exchange processes which can be analyzed along the lines
of the interdot cotunneling effects. Again, for concreteness we fix
the initial state of the DQD to be $\left|T_{+}\right\rangle $ and,
based on the approximation of independent rates of variation\cite{cohen-tannoudji92},
explain the physics in terms of the electronic bare states. The spin-blocked
level $\left|T_{+}\right\rangle $ can virtually exchange an electron
spin with the left lead yielding an incoherent coupling with the state
$\left|\Downarrow\Uparrow\right\rangle $; this process is mediated
by the intermediate singly occupied DQD level $\left|0,\Uparrow\right\rangle $
where no electron resides on the left dot. Then, from $\left|\Downarrow\Uparrow\right\rangle $
the system may decay back to the $\left(1,1\right)$ subspace via
the localized singlet $\left|S_{02}\right\rangle $. Therefore, for
this analysis, in Fig.~\ref{fig:scheme-cotunneling-spin-exchange}
we simply have to replace $\left|T_{+}(0,2)\right\rangle $ and $\Gamma_{\mathrm{ct}}$
by $\left|0,\Uparrow\right\rangle $ and $\Gamma_{\mathrm{se}}$,
respectively. Along the lines of our previous analysis of cotunneling
within the DQD, the bottleneck of the overall process is set by the
first step, labeled as $\Gamma_{\mathrm{se}}$. The main purpose of
this Appendix is an estimate for the rate $\Gamma_{\mathrm{se}}$.

The effective spin-exchange rate can be calculated in a ``golden
rule'' approach in which transitions for different initial and final
reservoir states are weighted according to the respective Fermi distribution
functions and added incoherently;\cite{kouwenhoven97} for more details,
see Refs.\cite{engel02,recher00}. Up to second order in $H_{T}$,
the cotunneling rate $\Gamma_{\mathrm{se}}$ for the process $\left|T_{+}\right\rangle \rightsquigarrow\left|\Downarrow\Uparrow\right\rangle $
is then found to be 
\begin{eqnarray}
\Gamma_{\mathrm{se}} & = & 2\pi n_{L}^{2}\left|T_{L}\right|^{4}\int_{\mu_{L}}^{\mu_{L}+\Delta}d\epsilon\frac{1}{\left(\epsilon-\delta_{+}\right)^{2}}\nonumber \\
 & \approx & \frac{\Gamma_{L}^{2}}{2\pi}\frac{\Delta}{\left(\mu_{L}-\delta_{+}\right)^{2}}.\label{eq:effective-spin-exchange-rate}
\end{eqnarray}
Here, $n_{L}$ is the left lead density of states at the Fermi energy,
$\mu_{L}$ is the chemical potential of the left lead, $\Delta=E_{T_{+}}-E_{\Downarrow\Uparrow}$
is the energy released on the DQD (which gets absorbed by the reservoir)
and $\delta_{+}=E_{T_{+}}-E_{0\Uparrow}=\text{\ensuremath{\epsilon}}_{L\uparrow}+U_{LR}$
refers to the energy difference between a doubly and singly occupied
DQD in the intermediate virtual state. Moreover, $\Gamma_{L}$ refers
to the first-order sequential tunneling rates $\Gamma_{L}=2\pi n_{L}\left|T_{L}\right|^{2}$
for the left $\left(L\right)$ lead. Note that in the limit $T\rightarrow0$
the DQD cannot be excited; accordingly, for $\Delta>0$, the transition
$\left|T_{\pm}\right\rangle \rightsquigarrow\left|\Uparrow\Downarrow\right\rangle $
is forbidden due to energy conservation.\cite{qassemi09} As expected,
$\Gamma_{\mathrm{se}}$ is proportional to $\sim\left|T_{L}\right|^{4}$,
but suppressed by the energy penalty $\Delta_{\mathrm{se}}^{+}=\mu_{L}-\delta_{+}$
which characterizes the violation of the two-electron condition in
Eq.(\ref{eq:two-electron-regime-inequality}) in the virtual intermediate
step. Notably, this can easily be tuned electrostatically via the
chemical potential $\mu_{L}$. Comparing the parameter dependence
$\Gamma_{\mathrm{se}}\sim\left|T_{L}\right|^{4}$ to $\Gamma_{\mathrm{ct}}\sim t^{2}\left|T_{L}\right|^{2}$
shows that, in contrast to the cotunneling processes $\Gamma_{\mathrm{ct}}$,
$\Gamma_{\mathrm{se}}$ is independent of the interdot tunneling parameter
$t$. Moreover, it can be made efficient by tuning properly the energy
penalty $\Delta_{\mathrm{se}}^{+}$ and the tunnel coupling to the
reservoir $T_{L}$. A similar analysis can be carried out for example
for the effective decay process $\left|T_{-}\right\rangle \rightsquigarrow\left|\Downarrow\Uparrow\right\rangle $
by spin-exchange with the right reservoir. The corresponding rates
are the same if $\Gamma_{L}/\Delta_{\mathrm{se}}^{+}=\Gamma_{R}/\Delta_{\mathrm{se}}^{-}$,
where $\Delta_{\text{\ensuremath{\mathrm{se}}}}^{-}=\mu_{R}-\left(\epsilon_{R\downarrow}+U_{LR}\right)$,
is satisfied. 
Taking the energy penalty as $\Delta_{\mathrm{se}}\approx\Delta_{\mathrm{st}}$, 
a comparison of $\Gamma_{\mathrm{se}}$ to interdot cotunneling transitions 
(as discussed in the main text) gives 
$\Gamma_{\mathrm{ct}}/\Gamma_{\mathrm{se}}\approx2\pi t^2/(\Gamma \Delta)$.
Thus, for $\Gamma\approx2\pi t$ and $t\approx \Delta$ (as considered in this work),
we get approximately $\Gamma_{\mathrm{ct}}\approx \Gamma_{\mathrm{se}}$.

The effective spin-exchange rate $\Gamma_{\mathrm{se}}$ can be made
very efficient in the high gradient regime. For example, to obtain
$\Gamma_{\mathrm{se}}\approx1\mu\mathrm{eV}$ when $\Delta\approx40\mu\mathrm{eV}$,
we estimate the required characteristic energy penalty to be $\Delta_{\mathrm{se}}\approx200\mu\mathrm{eV}$.
As stated in the main text, for an energy penalty of $\sim500\mu\mathrm{eV}$ 
and for $\Gamma_{L}\approx100\mu\mathrm{eV}$, 
we estimate $\Gamma_{\mathrm{se}}\approx0.25\mu\mathrm{eV}$, 
making $\Gamma_{\mathrm{se}}$ fast compared to typical nuclear timescales; 
note that for less transparent barriers with  $\Gamma_{L}\approx1\mu\mathrm{eV}$, 
 $\Gamma_{\mathrm{se}}$ is four orders of magnitude smaller, 
 in agreement with values given in Ref.\cite{rudner11b}.
Moreover, as apparent from Eq.(\ref{eq:effective-spin-exchange-rate}),
in the low gradient regime $\Gamma_{\mathrm{se}}\sim\Delta$ is suppressed
due to a vanishing phase space of reservoir electrons that can contribute
to this process without violating energy conservation. To remedy this,
one can lower the energy penalty $\Delta_{\mathrm{se}}$; however,
if $\Delta_{\mathrm{se}}$ becomes comparable to $\Gamma$, this leads
to a violation of the Markov approximation and tunes the system away
from the sequential tunneling regime. 
Note that the factor $\Delta$ appears in Eq.(\ref{eq:effective-spin-exchange-rate}) as we consider 
explicitly the inelastic transition $\left|T_{+}\right\rangle \rightsquigarrow\left|\Downarrow\Uparrow\right\rangle$.
In a more general analysis, $\Delta$ should be replaced by the energy separation $\Delta E$ (which is released 
by the DQD into the reservoir) 
for the particular transition at hand. \cite{qassemi09}

Here, we have considered spin-exchange via singly-occupied levels
in the virtual intermediate stage only; they are detuned by the characteristic
energy penalty $\delta=\left|\mu_{i}-\left(\epsilon_{i}+U_{LR}\right)\right|$
for $i=L,R$. In principle, spin exchange with the leads can also
occur via electronic levels with $\left(1,2\right)$ or $\left(2,1\right)$
charge configuration. However, here the characteristic energy penalty
can be estimated as $\delta=\left|\epsilon_{i}+U_{i}+U_{LR}-\mu_{i}\right|$
which can be significantly bigger due to the appearance of the on-site
Coulomb energies $U_{i}$ in this expression. Therefore, they have
been disregarded in the analysis above.

\subsection{Spin Orbit Interaction}

For the triplet states $\left|T_{\pm}\right\rangle $ interdot tunneling
is suppressed due to Pauli spin blockade, but---apart from HF interaction
with the nuclear spins---it can be mediated by spin-orbit interaction
which does not conserve the electronic spin. In contrast to hyperfine
mediated lifting of the spin blockade, spin-orbital effects provide
another purely electronic alternative to escape the spin blockade,
i.e., without affecting the nuclear spins. They describe interdot
hopping accompanied by a spin rotation thereby coupling the triplet
states $\left|T_{\pm}\right\rangle $ with single occupation of each
dot to the singlet state $\left|S_{02}\right\rangle $ with double
occupation of the right dot. Therefore, following Refs.\cite{danon09,rudner10,schreiber11,stepaneko12,giavaras13},
spin-orbital effects can be described phenomenologically in terms
of the Hamiltonian 
\begin{equation}
H_{\mathrm{so}}=t_{\mathrm{so}}\left(\left|T_{+}\right\rangle \left\langle S_{02}\right|+\left|T_{-}\right\rangle \left\langle S_{02}\right|+\mathrm{h.c.}\right),\label{eq:spin-orbit-Hamiltonian}
\end{equation}
where the coupling parameter $t_{\mathrm{so}}$ in general depends
on the orientation of the the DQD with respect to the crystallographic
axes. Typical values of $t_{\mathrm{so}}$ can be estimated as $t_{\mathrm{so}}\approx\left(d/l_{\mathrm{so}}\right)t$,
where $t$ is the usual spin-conserving tunnel coupling, $d$ the
interdot distance and $l_{\mathrm{so}}$ the material-specific spin-orbit
length ($l_{\mathrm{so}}\approx1-10\mu\mathrm{m}$ for GaAs); this
estimate is in good agreement with the exact equation given in Ref.\cite{stepaneko12}
and yields $t_{\mathrm{so}}\approx\left(0.01-0.1\right)t$. 

In Eq.(\ref{eq:spin-orbit-Hamiltonian}) we have disregarded the spin-orbit
coupling for the triplet $\left|T_{0}\right\rangle =\left(\left|\Uparrow\Downarrow\right\rangle +\left|\Downarrow\Uparrow\right\rangle \right)/\sqrt{2}$.
It may be taken into account by introducing the modified interdot
tunneling Hamiltonian $H_{t}\rightarrow H'_{t}$ with $H'_{t}=t_{\uparrow\downarrow}\left|\Uparrow\Downarrow\right\rangle \left\langle S_{02}\right|-t_{\downarrow\uparrow}\left|\Downarrow\Uparrow\right\rangle \left\langle S_{02}\right|+\mathrm{h.c.},$
where the tunneling parameters $t_{\uparrow\downarrow}$ and $t_{\downarrow\uparrow}$
are approximately given by $t_{\uparrow\downarrow\left(\downarrow\uparrow\right)}=t\pm t_{\mathrm{so}}/\sqrt{2}\approx t,$
since the second term marks only a small modification of the order
of 5\%. While $\left|T_{0}\right\rangle $ is dark under tunneling
in the singlet subspace, that is $H_{t}\left|T_{0}\right\rangle =0$,
similarly the slightly modified (unnormalized) state $\left|T_{0}'\right\rangle =t_{\downarrow\uparrow}\left|\Uparrow\Downarrow\right\rangle +t_{\uparrow\downarrow}\left|\Downarrow\Uparrow\right\rangle $
is dark under $H'_{t}$. Since this effect does not lead to any qualitative
changes, it is disregarded. 

\begin{figure}
\includegraphics[width=1\columnwidth]{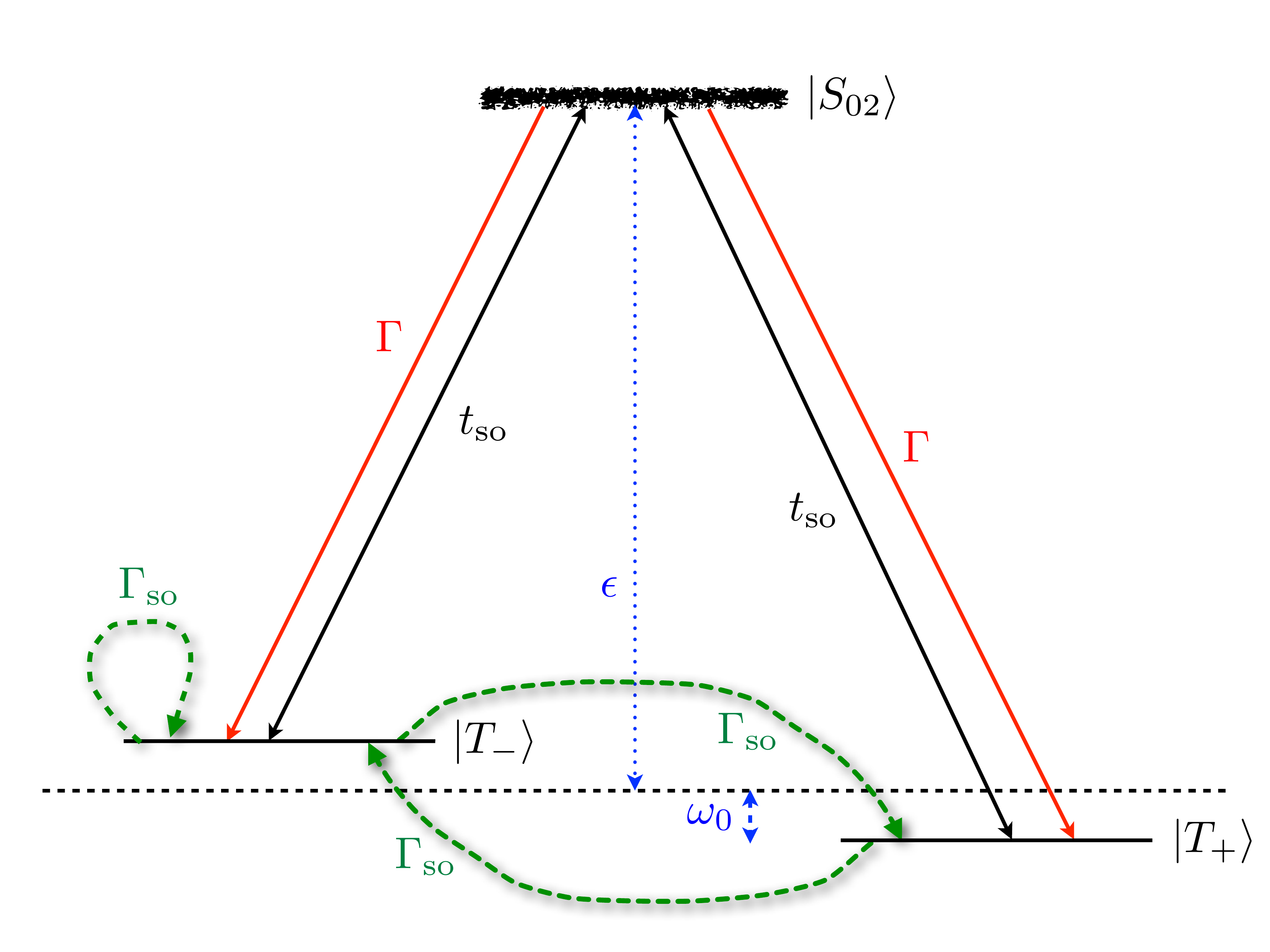}

\caption{\label{fig:schematic-SO-system}(color online). Phenomenological treatment
of spin-orbital effects in the spin-blockade regime. Scheme of the
simplified electronic system: The triplet states $\left|T_{\pm}\right\rangle $
are coherently coupled to the local singlet $\left|S_{02}\right\rangle $
by spin-orbit interaction. Via coupling to the leads, the DQD is discharged
and recharged again with an effective rate $\Gamma$. The triplet
states may experience a Zeeman splitting $\omega_{0}$. The parameter
$\epsilon$ specifies the interdot energy offset. Since $\Gamma,\epsilon\gg t_{\mathrm{so}}$,
the local singlet $\left|S_{02}\right\rangle $ can be eliminated
adiabatically yielding effective dissipative processes of strength
$\Gamma_{\mathrm{so}}$ (green dashed arrows). }
\end{figure}

\textit{Phenomenological treatment}.---In the following, we first
focus on the effects generated by $H_{\mathrm{so}}$ within the three-level
subspace $\left\{ \left|T_{\pm}\right\rangle ,\left|S_{02}\right\rangle \right\} $.
Within this reduced level scheme, the dynamics $\dot{\rho}=\mathcal{L}_{\mathrm{rd}}\rho$
are governed by the Liouvillian
\begin{eqnarray}
\mathcal{L}_{\mathrm{rd}}\rho & = & -i\left[\mathcal{H}_{\mathrm{rd}},\rho\right]+\Gamma\sum_{\nu=\pm}\mathcal{D}\left[\left|T_{\nu}\right\rangle \left\langle S_{02}\right|\right]\rho
\end{eqnarray}
where the relevant Hamiltonian within this subspace is 
\begin{equation}
\mathcal{H}_{\mathrm{rd}}=\omega_{0}\left(\left|T_{+}\right\rangle \left\langle T_{+}\right|-\left|T_{-}\right\rangle \left\langle T_{-}\right|\right)-\epsilon\left|S_{02}\right\rangle \left\langle S_{02}\right|+H_{\mathrm{so}}.
\end{equation}
This situation is schematized in Fig.~\ref{fig:schematic-SO-system}.
The external Zeeman splitting $\omega_{0}$ is assumed to be small
compared to the interdot detuning $\epsilon$ yielding approximately
equal detunings between the triplet states $\left|T_{\pm}\right\rangle $
and $\left|S_{02}\right\rangle $. In particular, we consider the
regime $t_{\mathrm{so}}\ll\epsilon,\Gamma$, with the corresponding
separation of timescales allowing for an alternative, effective description
of spin-orbital effects. Since the short-lived singlet state $\left|S_{02}\right\rangle $
is populated negligibly throughout the dynamics, it can be eliminated
adiabatically using standard techniques. The symmetric superposition
$\left|-\right\rangle =\left(\left|T_{+}\right\rangle -\left|T_{-}\right\rangle \right)/\sqrt{2}$
is a dark state with respect to the spin-orbit Hamiltonian $H_{\mathrm{so}}$.
Therefore, it is instructive to formulate the resulting effective
master equation in terms of the symmetric superposition states $\left|\pm\right\rangle =\left(\left|T_{+}\right\rangle \pm\left|T_{-}\right\rangle \right)/\sqrt{2}$.
Within the two-dimensional subspace spanned by the symmetric superpositions
$\left|\pm\right\rangle $, the effective dynamics is given by

\begin{eqnarray}
\dot{\rho} & = & +i\omega_{0}\left[\left|-\right\rangle \left\langle +\right|+\left|+\right\rangle \left\langle -\right|,\rho\right]\label{eq:effective-SO-QME}\\
 &  & -i\Omega_{\mathrm{so}}\left[\left|+\right\rangle \left\langle +\right|-\left|-\right\rangle \left\langle -\right|,\rho\right]\nonumber \\
 &  & +2\Gamma_{\mathrm{so}}\mathcal{D}\left[\left|-\right\rangle \left\langle +\right|\right]\rho\nonumber \\
 &  & +\frac{\Gamma_{\mathrm{so}}}{2}\mathcal{D}\left[\left|+\right\rangle \left\langle +\right|-\left|-\right\rangle \left\langle -\right|\right]\rho,\nonumber 
\end{eqnarray}
where the effective rate 
\begin{equation}
\Gamma_{\mathrm{so}}=\frac{t_{\mathrm{so}}^{2}}{\epsilon^{2}+\Gamma^{2}}\Gamma
\end{equation}
governs decay as well as pure dephasing processes within the triplet
subspace. We estimate $\Gamma_{\mathrm{so}}\approx(0.2-0.3)\mu\mathrm{eV}$
which is still fast compared to typical nuclear timescales. In Eq.(\ref{eq:effective-SO-QME})
we have also introduced the quantity $\Omega_{\mathrm{so}}=\left(\epsilon/\Gamma\right)\Gamma_{\mathrm{so}}$.
As we are particularly concerned with the nuclear dynamics in the
limit where one can eliminate the electronic degrees of freedom, Eq.(\ref{eq:effective-SO-QME})
provides an alternative way of accounting for spin-orbital effects:
In Eq.(\ref{eq:effective-SO-QME}) we encounter a decay term---see
the third line in Eq.(\ref{eq:effective-SO-QME})---which pumps the
electronic subsystem towards the dark state of the spin-orbit Hamiltonian,
namely the state $\left|-\right\rangle $. This state is also dark
under the Stark shift and pure dephasing terms in the second and last
line of Eq.(\ref{eq:effective-SO-QME}), respectively. However, by
applying an external magnetic field, the state $\left|-\right\rangle $
dephases due to the induced Zeeman splitting $\omega_{0}$. This becomes
apparent when examining the electronic quasisteady state corresponding
to the evolution given in Eq.(\ref{eq:effective-SO-QME}). In the
basis $\left\{ \left|T_{+}\right\rangle ,\mbox{\ensuremath{\left|T_{-}\right\rangle }}\right\} $,
it is found to be 
\begin{equation}
\rho_{\mathrm{ss}}^{\mathrm{el}}=\left(\begin{array}{cc}
\frac{1}{2}\left[1+\frac{\omega_{0}\Omega_{\mathrm{so}}}{\omega_{0}^{2}+\Gamma_{\mathrm{so}}^{2}+\Omega_{\mathrm{so}}^{2}}\right] & -\frac{\Gamma_{\mathrm{so}}^{2}+\Omega_{\mathrm{so}}^{2}+i\Gamma_{\mathrm{so}}\omega_{0}}{2\left(\omega_{0}^{2}+\Gamma_{\mathrm{so}}^{2}+\Omega_{\mathrm{so}}^{2}\right)}\\
-\frac{\Gamma_{\mathrm{so}}^{2}+\Omega_{\mathrm{so}}^{2}-i\Gamma_{\mathrm{so}}\omega_{0}}{2\left(\omega_{0}^{2}+\Gamma_{\mathrm{so}}^{2}+\Omega_{\mathrm{so}}^{2}\right)} & \frac{1}{2}\left[1-\frac{\omega_{0}\Omega_{\mathrm{so}}}{\omega_{0}^{2}+\Gamma_{\mathrm{so}}^{2}+\Omega_{\mathrm{so}}^{2}}\right]
\end{array}\right),
\end{equation}
which in leading orders of $\omega_{0}^{-1}$ reduces to 
\begin{equation}
\rho_{\mathrm{ss}}^{\mathrm{el}}\approx\left(\begin{array}{cc}
\frac{1}{2}+\frac{\Omega_{\mathrm{so}}}{2\omega_{0}} & -i\frac{\Gamma_{\mathrm{so}}}{2\omega_{0}}\\
i\frac{\Gamma_{\mathrm{so}}}{2\omega_{0}} & \frac{1}{2}-\frac{\Omega_{\mathrm{so}}}{2\omega_{0}}
\end{array}\right).
\end{equation}
Accordingly, for sufficiently large Zeeman splitting $\omega_{0}\gg\Omega_{\mathrm{so}},\Gamma_{\mathrm{so}}$,
the electronic subsystem is driven towards the desired equal mixture
of blocked triplet states $\left|T_{+}\right\rangle $ and $\left|T_{-}\right\rangle $.
Alternatively, the off-diagonal elements of $\left|-\right\rangle \left\langle -\right|$
are damped out in the presence of dephasing processes either mediated
intrinsically via cotunneling processes or extrinsically via engineered
magnetic noise yielding approximately the equal mixture $\rho_{\mathrm{target}}^{\mathrm{el}}=\left(\left|T_{+}\right\rangle \left\langle T_{+}\right|+\left|T_{-}\right\rangle \left\langle T_{-}\right|\right)/2$
in the quasisteady state. 

\begin{figure}
\includegraphics[width=1\columnwidth]{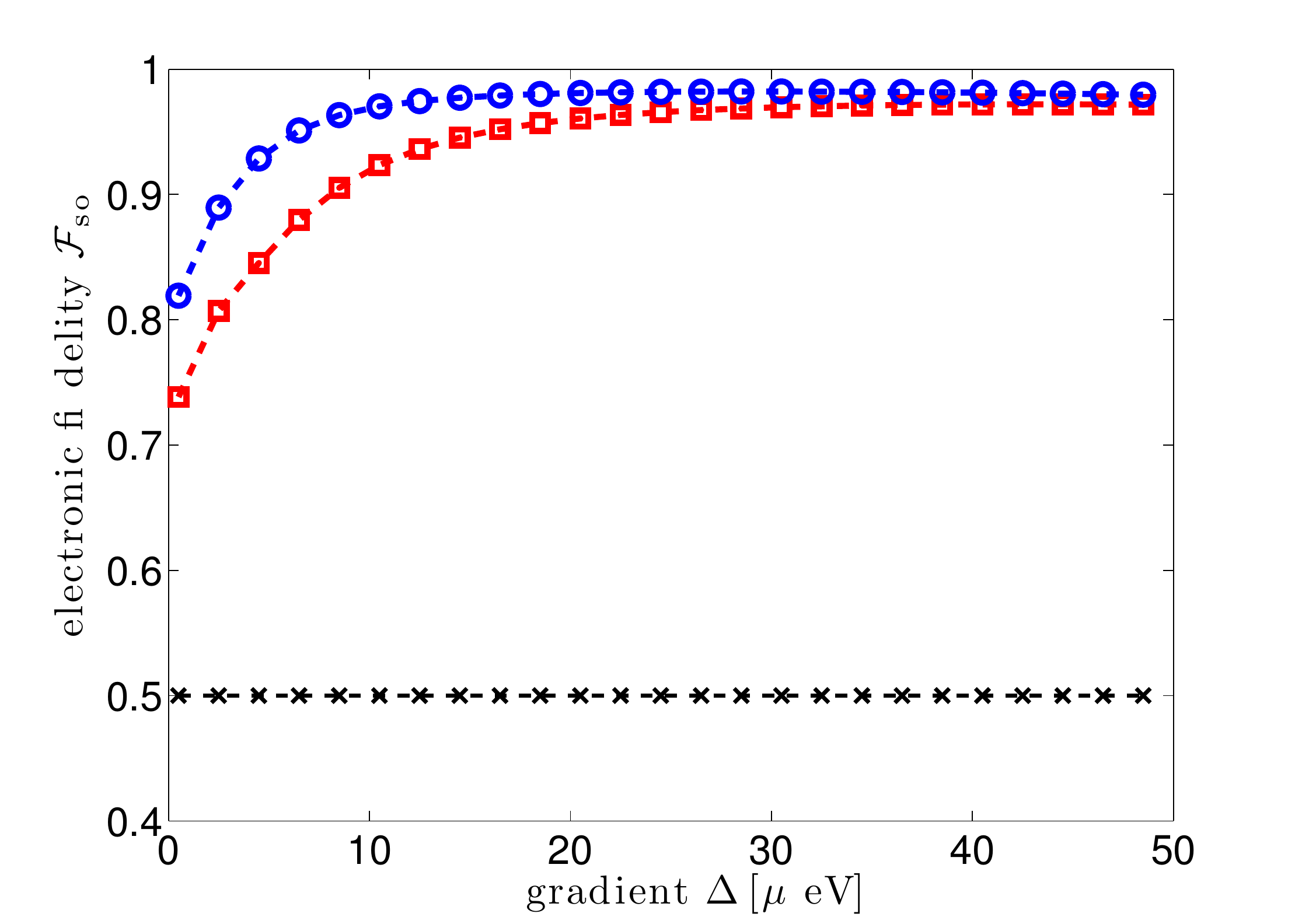}

\caption{\label{fig:SO-fidelity}(color online). Electronic quasi-steady-state
fidelities in the presence of spin-orbit coupling for the dynamics
generated by $\tilde{\mathcal{K}}_{0}$ as a function of the gradient
$\Delta$. As expected, in the absence of dephasing {[}$\Gamma_{\mathrm{deph}}=0$
(black curve){]}, the system settles into the dark state $\left|-\right\rangle $.
For $\Gamma_{\mathrm{deph}}=1\mu\mathrm{eV}$ (blue and red curves),
the off-diagonal elements of $\left|-\right\rangle $ are strongly
suppressed, leading to a high fidelity $\mathcal{F}_{\mathrm{so}}\gtrsim0.9$
with the desired mixed state $\rho_{\mathrm{target}}^{\mathrm{el}}$
in the high gradient regime: the blue and red curve refer to $t=20\mu\mathrm{eV}$
and $t=30\mu\mathrm{eV}$, respectively. Other numerical parameters
are: $t_{\mathrm{so}}=0.1t$, $\omega_{0}=0$, $\Gamma=25\mu\mathrm{eV}$
and $\epsilon=30\mu\mathrm{eV}$. }
\end{figure}

\textit{Numerical analysis}.---To complement the perturbative, analytical
study, we carry out a numerical evaluation of the electronic quasisteady
state in the presence of spin-orbit coupling. In the two-electron
subspace, the corresponding master equation (including spin-orbital
effects) under consideration reads 
\begin{equation}
\dot{\rho}=\tilde{\mathcal{K}}_{0}\rho=-i\left[H_{\mathrm{el}}+H_{\mathrm{so}},\rho\right]+\mathcal{K}_{\Gamma}\rho+\mathcal{L}_{\mathrm{deph}}\rho.\label{eq:Liouvillian-spin-orbit-Hamiltonian}
\end{equation}
We evaluate the exact electronic quasisteady state $\text{\ensuremath{\rho_{\mathrm{ss}}^{\mathrm{el}}}}$
fulfilling $\tilde{\mathcal{K}}_{0}\rho_{\mathrm{ss}}^{\mathrm{el}}=0$.
As a figure of merit, we compute the Uhlmann fidelity \cite{uhlmann76}
\begin{equation}
\mathcal{F}_{\text{\ensuremath{\mathrm{so}}}}=\mathrm{tr}\left[\left(\sqrt{\rho_{\mathrm{ss}}^{\mathrm{el}}}\rho_{\mathrm{target}}^{\mathrm{el}}\sqrt{\rho_{\mathrm{ss}}^{\mathrm{el}}}\right)^{1/2}\right]^{2}
\end{equation}
which measures how similar $\text{\ensuremath{\rho_{\mathrm{ss}}^{\mathrm{el}}}}$
and $\rho_{\mathrm{target}}^{\mathrm{el}}$ are. The results are illustrated
in Fig.~\ref{fig:SO-fidelity}: For $\Gamma_{\mathrm{deph}}=0$ the
electronic system settles into the pure dark state $\left|-\right\rangle \left\langle -\right|$.
However, in the presence of dephasing, the coherences are efficiently
damped out. In the low-gradient regime $\rho_{\mathrm{ss}}^{\mathrm{el}}$
has a significant overlap with the triplet $\left|T_{0}\right\rangle $,
whereas in the high-gradient regime it is indeed approximately given
by the desired mixed target state $\rho_{\mathrm{target}}^{\mathrm{el}}$.
Lastly, we have checked that in the high-gradient regime the corresponding
asymptotic decay rate can be approximated very well by $\mathrm{ADR}_{\mathrm{el}}\approx-2\Gamma_{\mathrm{so}}$.

\section{Effective Nuclear Master Equation\label{sec:Effective-Nuclear-Master-Equation-Adiabatic-Elimination}}

In this appendix, we present a detailed derivation of the effective
nuclear dynamics presented in Section \ref{sec:Effective-Nuclear-Dynamics}.
We use standard adiabatic elimination techniques to derive an effective
simplified description of the dynamics. To do so, we assume that electronic
coherences decay quickly on typical nuclear timescales. Conservatively,
i.e. not taking into account the detuning of the HF-mediated transitions,
this holds for $2\Gamma_{\pm}+\Gamma_{\mathrm{deph}}/4\gg g_{\mathrm{hf}}$,
where $g_{\mathrm{hf}}$ quantifies the typical HF interaction strength.
Alternatively, one may use a projection-operator based technique;\cite{schuetz12,kessler12}
this is done in detail in Appendix \ref{sec:Appendix-Effective-Nuclear-QME-High-Gradient-Regime}
for the high-gradient regime where $\rho_{\mathrm{ss}}^{\mathrm{el}}=\left(\left|T_{+}\right\rangle \left\langle T_{+}\right|+\left|T_{-}\right\rangle \left\langle T_{-}\right|\right)/2$,
but a generalization for the electronic quasisteady state in Eq.(\ref{eq:quasi-steady-state-general})
is straightforward. 

Throughout this appendix, for convenience we adopt the following notation:
$\left|a\right\rangle =\left|T_{+}\right\rangle $, $\left|b\right\rangle =\left|\lambda_{2}\right\rangle $,
$\left|c\right\rangle =\left|T_{-}\right\rangle $, $L=L_{2}$, $\mathbb{L}=\mathbb{L}_{2}$
and $\mathcal{D}\left[c\right]\rho=\mathcal{D}_{c}\rho$. Within this
simplified three-level model system, the flip-flop Hamiltonian $H_{\mathrm{ff}}$
reads 
\begin{equation}
H_{\mathrm{ff}}=\frac{a_{\mathrm{hf}}}{2}\left[L\left|b\right\rangle \left\langle a\right|+\mathbb{L}\left|b\right\rangle \left\langle c\right|+\mathrm{h.c.}\right].
\end{equation}
For simplicity, we assume $\omega_{0}=0$ and neglect nuclear fluctuations
arising from $H_{\mathrm{zz}}$. This approximation is in line with
the semiclassical approximation for studying the nuclear polarization
dynamics; for more details also see Appendix \ref{sec:Effective-Nuclear-Dynamics:OH-fluctuations}.
Within this reduced scheme, the dynamics are then described by the
Master equation 
\begin{eqnarray}
\dot{\rho} & = & -i\left[H_{\mathrm{ff}},\rho\right]-i\epsilon_{2}\left[\left|b\right\rangle \left\langle b\right|,\rho\right]+\frac{\Gamma_{\mathrm{deph}}}{2}\mathcal{D}_{\left|a\right\rangle \left\langle a\right|-\left|c\right\rangle \left\langle c\right|}\rho\nonumber \\
 &  & +\Gamma_{\pm}\left[\mathcal{D}_{\left|c\right\rangle \left\langle a\right|}\rho+\mathcal{D}_{\left|a\right\rangle \left\langle c\right|}\rho+\mathcal{D}_{\left|b\right\rangle \left\langle a\right|}\rho+\mathcal{D}_{\left|b\right\rangle \left\langle c\right|}\rho\right]\nonumber \\
 &  & +\left(\Gamma_{\pm}+\Gamma_{2}\right)\left[\mathcal{D}_{\left|a\right\rangle \left\langle b\right|}\rho+\mathcal{D}_{\left|c\right\rangle \left\langle b\right|}\rho\right].\label{eq:three-level-system-Liouvillian}
\end{eqnarray}

After adiabatic elimination of the electronic coherences $\rho_{ab}=\left<a|\rho|b\right>$,
$\rho_{cb}$ and $\rho_{ac}$ we obtain effective equations of motion
for the system's density matrix projected onto the electronic levels
$\left|a\right\rangle $, $\left|b\right\rangle $ and $\left|c\right\rangle $
as follows 
\begin{eqnarray}
\dot{\rho}_{aa} & = & \Gamma_{\pm}\left(\rho_{cc}-\rho_{aa}\right)+\Gamma_{\pm}\left(\rho_{bb}-\rho_{aa}\right)+\Gamma_{2}\rho_{bb}\label{eq:EOM-rho-aa}\\
 &  & +\gamma\left[L^{\dagger}\rho_{bb}L-\frac{1}{2}\left\{ L^{\dagger}L,\rho_{aa}\right\} \right]\nonumber \\
 &  & +i\delta\left[L^{\dagger}L,\rho_{aa}\right],\nonumber \\
\dot{\rho}_{cc} & = & \Gamma_{\pm}\left(\rho_{aa}-\rho_{cc}\right)+\Gamma_{\pm}\left(\rho_{bb}-\rho_{cc}\right)+\Gamma_{2}\rho_{bb}\label{eq:EOM-rho-cc}\\
 &  & +\gamma\left[\mathbb{L}^{\dagger}\rho_{bb}\mathbb{L}-\frac{1}{2}\left\{ \mathbb{L}^{\dagger}\mathbb{L},\rho_{cc}\right\} \right]\nonumber \\
 &  & +i\delta\left[\mathbb{L}^{\dagger}\mathbb{L},\rho_{cc}\right],\nonumber 
\end{eqnarray}
and
\begin{eqnarray}
\dot{\rho}_{bb} & = & -2\Gamma_{2}\rho_{bb}+\gamma\left[L\rho_{aa}L^{\dagger}-\frac{1}{2}\left\{ LL^{\dagger},\rho_{bb}\right\} \right]\label{eq:EOM-rho-bb}\\
 &  & -i\delta\left[LL^{\dagger},\rho_{bb}\right]\nonumber \\
 &  & +\gamma\left[\mathbb{L}\rho_{cc}\mathbb{L^{\dagger}}-\frac{1}{2}\left\{ \mathbb{L}\mathbb{L^{\dagger}},\rho_{bb}\right\} \right]-i\delta\left[\mathbb{L}\mathbb{L^{\dagger}},\rho_{bb}\right].\nonumber \\
 &  & +\Gamma_{\pm}\left(\rho_{aa}+\rho_{cc}-2\rho_{bb}\right).\nonumber 
\end{eqnarray}
Since this set of equations is entirely expressed in terms of $ $$\rho_{aa}$,
$\rho_{bb}$ and $\rho_{cc}$, the full density matrix of the system
obeys a simple block structure, given by 
\begin{equation}
\rho=\rho_{aa}\left|a\right\rangle \left\langle a\right|+\rho_{bb}\left|b\right\rangle \left\langle b\right|+\rho_{cc}\left|c\right\rangle \left\langle c\right|.
\end{equation}
Therefore, the electronic decoherence is fast enough to prevent the
entanglement between electronic and nuclear degrees of freedom and
the total density matrix of the system $\rho$ factorizes into a tensor
product for the electronic and nuclear subsystem,\cite{rudner07b}
respectively, that is $\rho=\rho_{\mathrm{el}}\otimes\sigma,$ where
$\sigma=\mathsf{Tr}_{\mathrm{el}}\left[\rho\right]$ refers to the
density matrix of the nuclear subsystem. This ansatz agrees with the
projection operator approach where $\mathcal{P}\rho=\sigma\otimes\rho_{\mathrm{el}}$
and readily yields $\rho_{aa}=p_{a}\sigma$, where we have introduced
the electronic populations 
\begin{equation}
p_{a}=\left<a|\rho_{\mathrm{el}}|a\right>=\mathsf{Tr}_{\mathrm{n}}\left[\rho_{aa}\right],
\end{equation}
and accordingly for $p_{b}$ and $p_{c}$; here, $\mathsf{Tr}_{\mathrm{n}}\left[\dots\right]$
denotes the trace over the nuclear degrees of freedom. With these
definitions, Eqs. (\ref{eq:EOM-rho-aa}), (\ref{eq:EOM-rho-cc}) and
(\ref{eq:EOM-rho-bb}) can be rewritten as 
\begin{eqnarray}
\dot{p}_{a} & = & \Gamma_{\pm}\left(p_{c}-p_{a}\right)+\Gamma_{2}p_{b}+\gamma\left[p_{b}\left\langle LL^{\dagger}\right\rangle -p_{a}\left\langle L^{\dagger}L\right\rangle \right]\nonumber \\
 &  & +\Gamma_{\pm}\left(p_{b}-p_{a}\right),\nonumber \\
\dot{p}_{c} & = & \Gamma_{\pm}\left(p_{a}-p_{c}\right)+\Gamma_{2}p_{b}+\gamma\left[p_{b}\left\langle \mathbb{L}\mathbb{L}^{\dagger}\right\rangle -p_{c}\left\langle \mathbb{L}^{\dagger}\mathbb{L}\right\rangle \right]\nonumber \\
 &  & +\Gamma_{\pm}\left(p_{b}-p_{c}\right),\nonumber \\
\dot{p}_{b} & = & -2\Gamma_{2}p_{b}+\Gamma_{\pm}\left(p_{a}+p_{c}-2p_{b}\right)\nonumber \\
 &  & +\gamma\left[p_{a}\left\langle L^{\dagger}L\right\rangle -p_{b}\left\langle LL^{\dagger}\right\rangle +p_{c}\left\langle \mathbb{L}^{\dagger}\mathbb{L}\right\rangle -p_{b}\left\langle \mathbb{L}\mathbb{L}^{\dagger}\right\rangle \right].\label{eq:electronic-rate-equations}
\end{eqnarray}
Similarly, the effective Master equation for the nuclear density matrix
$\sigma=\mathsf{Tr}_{\mathrm{el}}\left[\rho\right]$ is obtained from
$\dot{\sigma}=\mathsf{Tr}_{\mathrm{el}}\left[\dot{\rho}\right]=\dot{\rho}_{aa}+\dot{\rho}_{bb}+\dot{\rho}_{cc}$,
leading to 
\begin{eqnarray}
\dot{\sigma} & = & \gamma\left\{ p_{b}\mathcal{D}_{L^{\dagger}}\left[\text{\ensuremath{\sigma}}\right]+p_{b}\mathcal{D}_{\mathbb{L}^{\dagger}}\left[\sigma\right]+p_{a}\mathcal{D}_{L}\left[\sigma\right]+p_{c}\mathcal{D}_{\mathbb{L}}\left[\sigma\right]\right\} \nonumber \\
 &  & +i\delta\left\{ p_{a}\left[L^{\dagger}L,\sigma\right]+p_{c}\left[\mathbb{L}^{\dagger}\mathbb{L},\sigma\right]\right.\nonumber \\
 &  & \left.-p_{b}\left[LL^{\dagger},\sigma\right]-p_{b}\left[\mathbb{L}\mathbb{L}^{\dagger},\sigma\right]\right\} .\label{eq:QME-diffusion-nuclear-spins}
\end{eqnarray}
Equation (\ref{eq:QME-diffusion-nuclear-spins}) along with Eq.(\ref{eq:electronic-rate-equations})
describe the coupled electron-nuclear dynamics on a coarse-grained
timescale that is long compared to electronic coherence timescales.
Due to the normalization condition $p_{a}+p_{b}+p_{c}=1$, this set
of dynamical equations comprises three coupled equations. Differences
in the populations of the levels $\left|a\right\rangle $ and $\left|c\right\rangle $
decay very quickly on timescales relevant for the nuclear evolution;
that is, 
\begin{eqnarray}
\dot{p}_{a}-\dot{p}_{c} & = & -3\Gamma_{\pm}\left(p_{a}-p_{c}\right)+\gamma\left[p_{b}\left(\left\langle LL^{\dagger}\right\rangle -\left\langle \mathbb{L}\mathbb{L}^{\dagger}\right\rangle \right)\right.\nonumber \\
 &  & \left.-p_{a}\left\langle L^{\dagger}L\right\rangle +p_{c}\left\langle \mathbb{L}^{\dagger}\mathbb{L}\right\rangle \right]
\end{eqnarray}
Due to a separation of timescales, as $\Gamma_{\pm}\gg\gamma_{c}=N\gamma\approx10^{-4}\mu\mathrm{eV}$,
in a perturbative treatment the effect of the second term can be neglected
and the electronic subsystem approximately settles into $p_{a}=p_{c}$.
This leaves us with a single dynamical variable, namely $p_{a}$,
entirely describing the electronic subsystem on relevant timescales.
Thus, using $p_{c}=p_{a}$ and $p_{b}=1-2p_{a}$, the electronic quasi
steady state is uniquely defined by the parameter $p_{a}$ and the
nuclear evolution simplifies to 
\begin{eqnarray}
\dot{\sigma} & = & \gamma\left\{ p_{a}\left[\mathcal{D}_{L}\left[\sigma\right]+\mathcal{D}_{\mathbb{L}}\left[\sigma\right]\right]\right.\\
 &  & \left.+\left(1-2p_{a}\right)\left[\mathcal{D}_{L^{\dagger}}\left[\sigma\right]+\mathcal{D}_{\mathbb{L}^{\dagger}}\left[\sigma\right]\right]\right\} \nonumber \\
 &  & +i\delta\left\{ p_{a}\left(\left[L^{\dagger}L,\sigma\right]+\left[\mathbb{L}^{\dagger}\mathbb{L},\sigma\right]\right)\right.\nonumber \\
 &  & \left.-\left(1-2p_{a}\right)\left(\left[LL^{\dagger},\sigma\right]+\left[\mathbb{L}\mathbb{L}^{\dagger},\sigma\right]\right)\right\} ,\nonumber 
\end{eqnarray}
with $p_{a}$ obeying the dynamical equation 
\begin{eqnarray*}
\dot{p}_{a} & = & \Gamma_{\pm}\left(1-3p_{a}\right)+\Gamma_{2}\left(1-2p_{a}\right)\\
 &  & -\gamma\left[p_{a}\left\langle L^{\dagger}L\right\rangle +\left(1-2p_{a}\right)\left\langle LL^{\dagger}\right\rangle \right].
\end{eqnarray*}
Neglecting the HF terms in the second line, we recover the projection-operator-based
result for the quasisteady state, $p_{a}\approx\left(\Gamma_{\pm}+\Gamma_{2}\right)/\left(3\Gamma_{\pm}+2\Gamma_{2}\right)$
as stated in Eq.(\ref{eq:parameter-p}).

\section{Effective Nuclear Dynamics: Overhauser Fluctuations \label{sec:Effective-Nuclear-Dynamics:OH-fluctuations}}

In Sec. \ref{sec:Effective-Nuclear-Dynamics} we have disregarded
the effect of Overhauser fluctuations, described by $\dot{\rho}=-i\left[H_{\mathrm{zz}},\rho\right]=-ia_{\mathrm{hf}}\sum_{i}\left[S_{i}^{z}\delta A_{i}^{z},\rho\right].$
In the following analysis, this simplification is discussed in greater
detail. 

First of all, we note that this term cannot induce couplings within
the effective electronic three level system, $\left\{ \left|T_{\pm}\right\rangle ,\left|\lambda_{2}\right\rangle \right\} $,
since $\left|T_{\pm}\right\rangle $ are eigenstates of $S_{i}^{z}$,
that is explicitly $S_{i}^{z}\left|T_{\pm}\right\rangle =\pm\frac{1}{2}\left|T_{\pm}\right\rangle $,
which leads to 
\begin{equation}
\left<T_{\pm}|S_{i}^{z}|\lambda_{2}\right>=0.
\end{equation}
In other words, different $S_{\mathrm{tot}}^{z}$ subspaces are not
coupled by the action of $H_{\mathrm{zz}}$; this is in stark contrast
to the flip flop dynamics $H_{\mathrm{ff}}$. 

When also accounting for Overhauser fluctuations, the dynamical equations
for the coherences read
\begin{eqnarray}
\dot{\rho}_{ab} & = & \left(i\epsilon_{2}-\tilde{\Gamma}\right)\rho_{ab}-i\left[L^{\dagger}\rho_{bb}-\rho_{aa}L^{\dagger}\right]\\
 &  & -ia_{\mathrm{hf}}\sum_{i}\left[\left\langle S_{i}^{z}\right\rangle _{a}\delta A_{i}^{z}\rho_{ab}-\left\langle S_{i}^{z}\right\rangle _{b}\rho_{ab}\delta A_{i}^{z}\right],\nonumber 
\end{eqnarray}
where $\left\langle S_{i}^{z}\right\rangle _{a}=\left<a|S_{i}^{z}|a\right>$;
an analog equation holds for $\dot{\rho}_{cb}$. Typically, the second
line is small compared to the fast electronic quantities $\epsilon_{2},\tilde{\Gamma}$
in the first line. Therefore, it will be neglected. In Eqs.(\ref{eq:EOM-rho-aa}),
(\ref{eq:EOM-rho-cc}) and(\ref{eq:EOM-rho-bb}), the Overhauser fluctuations
lead to the following additional terms 
\begin{eqnarray}
\dot{\rho}_{aa} & = & \dots-\frac{i}{2}a_{\mathrm{hf}}\sum_{i}\left[\delta A_{i}^{z},\rho_{aa}\right],\\
\dot{\rho}_{cc} & = & \dots+\frac{i}{2}a_{\mathrm{hf}}\sum_{i}\left[\delta A_{i}^{z},\rho_{cc}\right],\\
\dot{\rho}_{bb} & = & \dots-ia_{\mathrm{hf}}\sum_{i}\left\langle S_{i}^{z}\right\rangle _{b}\left[\delta A_{i}^{z},\rho_{bb}\right].
\end{eqnarray}
First, this leaves the electronic populations $p_{a}=\mathsf{Tr}_{\mathrm{n}}\left[\rho_{aa}\right]$
untouched; $H_{\mathrm{zz}}$ does not induce any couplings between
them. Second, the dynamical equation for the nuclear density matrix
$\sigma=\mathsf{Tr}_{\mathrm{el}}\left[\rho\right]$ is modified as
\begin{eqnarray}
\dot{\sigma} & = & \dots-ia_{\mathrm{hf}}\sum_{i}\left[\frac{1}{2}\left(p_{a}-p_{c}\right)+p_{b}\left\langle S_{i}^{z}\right\rangle _{b}\right]\left[\delta A_{i}^{z},\sigma\right],\nonumber \\
 & \approx & \dots-i\left(1-2p_{a}\right)a_{\mathrm{hf}}\sum_{i}\left\langle S_{i}^{z}\right\rangle _{b}\left[\delta A_{i}^{z},\sigma\right].\label{eq:OH-fluctuations-nuclear-EOM}
\end{eqnarray}
In the second step, we have used again that differences in $p_{a}$
and $p_{c}$ are quickly damped to zero with a rate of $3\Gamma_{\pm}$.
Now, let us examine the effect of Eq.(\ref{eq:OH-fluctuations-nuclear-EOM})
for different important regimes: In the high gradient regime, where
$p_{b}$ is fully depleted, it does not give any contribution since
the electronic quasi steady state does not have any magnetization
$\left[\left\langle S_{i}^{z}\right\rangle _{b}=\left\langle S_{i}^{z}\right\rangle _{\mathrm{ss}}=0\right]$
and $p_{a}=1/2$. In the low gradient regime, $\left|b\right\rangle $
approaches the triplet $\left|T_{0}\right\rangle $ and again (since
$\left\langle S_{i}^{z}\right\rangle _{b}=0$) this term vanishes.
Finally, the intermediate regime has been studied within a semiclassical
approximation (see section \ref{sec:Polarization-Dynamics}): Note
that Eq.(\ref{eq:OH-fluctuations-nuclear-EOM}), however, leaves the
dynamical equation for the nuclear polarizations $I_{i}^{z}$ unchanged,
since they commute with $H_{\mathrm{zz}}$.

\section{Numerical Results for DNP\label{sec:Numerical-Results-for-DNP}}

\begin{figure}
\includegraphics[width=1\columnwidth]{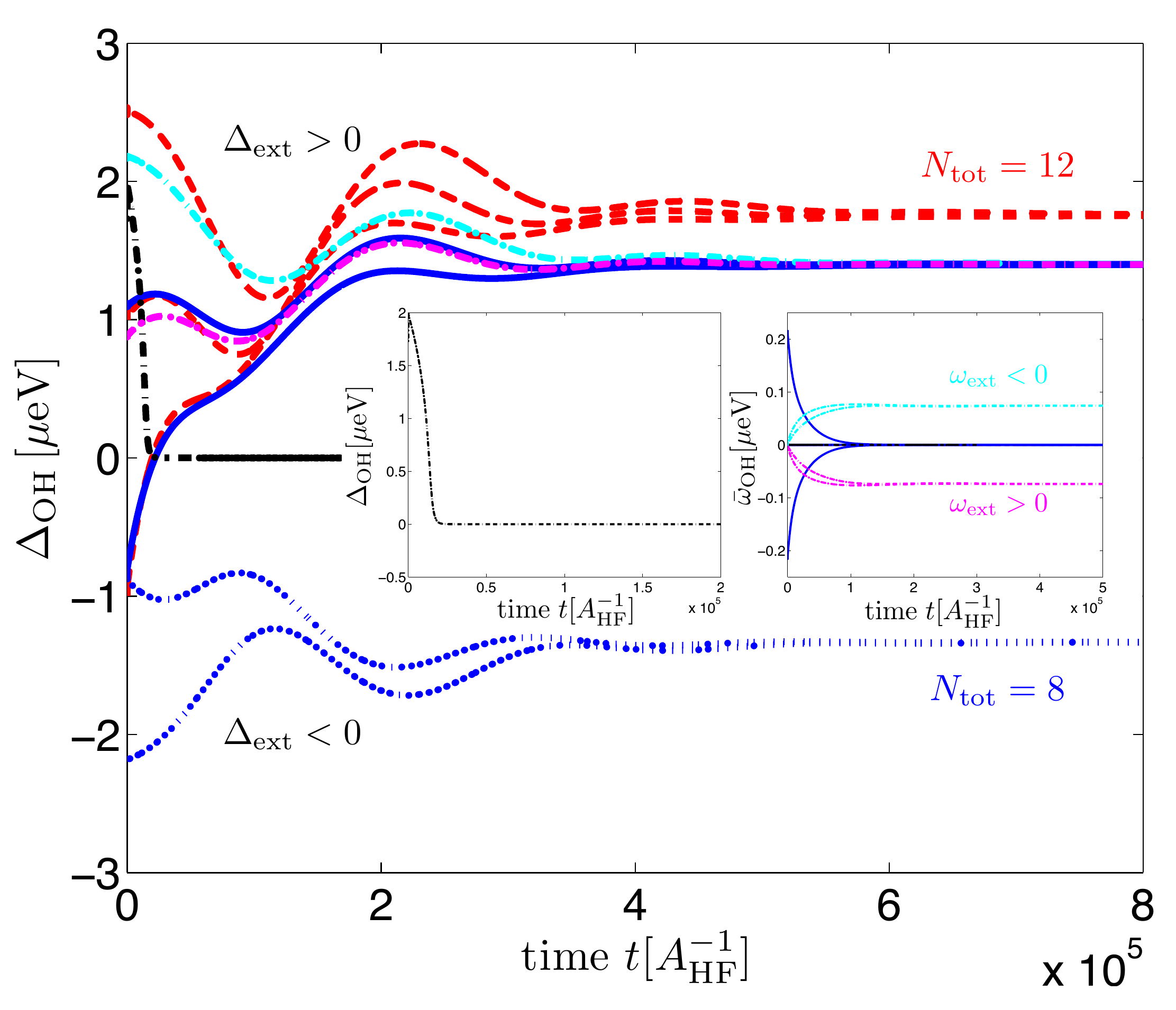}

\caption{\label{fig:time-evolution-N8}(color online). Exact time evolution
for $N=8$ and $N=12$ (red dashed curves) nuclear spins, four and
six in each quantum dot, respectively. Depending on the initial value
of the gradient, the nuclear system either runs into the trivial,
unpolarized state or into the highly polarized one, if the initial
gradient exceeds the critical value; the blue dotted, black dash-dotted
and all other refer to $\Delta_{\mathrm{ext}}=-5\mu\mathrm{eV}$,
$\Delta_{\mathrm{ext}}=0$ and $\Delta_{\mathrm{ext}}=5\mu\mathrm{eV}$,
respectively. For $\omega_{\mathrm{ext}}\neq0$, also a homogeneous
OH field $\omega_{\mathrm{OH}}$ builds up which partially compensates
$\omega_{\mathrm{ext}}$: here, $\omega_{\mathrm{ext}}=0.1\mu\mathrm{eV}$
(magenta dash-dotted) and $\omega_{\mathrm{ext}}=-0.1\mu\mathrm{eV}$
(cyan dash-dotted). Other numerical parameters: $t=10\mu\mathrm{eV}$,
$\epsilon=30\mu\mathrm{eV}$, $\Gamma=25\mu\mathrm{eV}$, $\Gamma_{\pm}=\Gamma_{\mathrm{deph}}=0.1\mu\mathrm{eV}$. }
\end{figure}

In this appendix the analytical findings of the semiclassical model
are corroborated by exact numerical simulations for small sets of
nuclear spins. This treatment complements our analytical DNP analysis
in several aspects: First, we do not restrict ourselves to the effective
three level system $\left\{ \left|T_{\pm}\right\rangle ,\left|\lambda_{2}\right\rangle \right\} $.
Second, the electronic degrees of freedom are not eliminated adiabatically
from the dynamics. Lastly, this approach does not involve the semiclassical
decorrelation approximation stated in Eq.(\ref{eq:semiclassical-factorization-1}). 

\textit{Technical details}.---We consider the idealized case of homogeneous
hyperfine coupling for which an exact numerical treatment is feasible
even for a relatively large number of coupled nuclei as the system
evolves within the totally symmetric low-dimensional subspace $\left\{ \left|J,m\right\rangle ,m=-J,\dots,J\right\} $,
referred to as \textit{Dicke ladder}. We restrict ourselves to the
fully symmetric subspace where $J_{i}=N_{i}/2\approx3$. Moreover,
to mimic the separation of timescales in experiments where $N\approx10^{6}$,
the HF coupling is scaled down appropriately to the constant value
$g_{\mathrm{hf}}\approx0.1\mu\mathrm{eV}$; also compare the numerical
results presented in Fig.~\ref{fig:external-gradient-N10}.

Our first numerical approach is based on simulations of the time evolution.
Starting out from nuclear states with different initial Overhauser
gradient $\Delta_{\mathrm{OH}}\left(t=0\right)$, we make the following
observations, depicted in Fig.~\ref{fig:time-evolution-N8}: First
of all, the tri-stability of the Overhauser gradient with respect
to the initial nuclear polarization is confirmed. If the initial gradient
$\Delta_{\mathrm{OH}}\left(t=0\right)+\Delta_{\mathrm{ext}}$ exceeds
a certain threshold value, the nuclear system runs into the highly-polarized
steady state, otherwise it gets stuck in the trivial, zero-polarization
solution. There are two symmetric high-polarization solutions that
depend on the sign of $\Delta_{\mathrm{OH}}\left(t=0\right)+\Delta_{\mathrm{ext}}$;
also note that the Overhauser gradient $\Delta_{\mathrm{OH}}$ may
flip the sign as determined by the total initial gradient $\Delta_{\mathrm{OH}}\left(t=0\right)+\Delta_{\mathrm{ext}}$.
Second, in the absence of an external Zeeman splitting $\omega_{\mathrm{ext}}$,
a potential initial homogeneous Overhauser polarizations $\bar{\omega}_{\mathrm{OH}}$
is damped to zero in the steady state. For finite $\omega_{\mathrm{ext}}\neq0$,
a homogeneous Overhauser polarization $\bar{\omega}_{\mathrm{OH}}$
builds up which partially compensates $\omega_{\mathrm{ext}}$. Lastly,
the high-polarization solutions $ $$\Delta_{\mathrm{OH}}^{\mathrm{ss}}\approx2\mu\mathrm{eV}$
are far away from full polarization. This is an artifact of the small
system sizes $J_{i}\approx3$: As we deal with very short Dicke ladders,
even the ideal, nuclear two-mode squeezedlike target state $\left|\xi\right\rangle _{\mathrm{ss}}$
given in Eq.(\ref{fig:ideal-target-state-uniform-HF-coupling}) does
not feature a very high polarization. Pictorially, it leaks with a
non-vanishing factor $\sim\xi^{m}$ into the low-polarization Dicke
states. This argument is supported by the fact that (for the same
set of parameters) we observe tendency towards higher polarization
for an increasing number of nuclei $N$ (which features a larger Dicke
ladder) and confirmed by our second numerical approach to be discussed
below. 

\begin{figure}
\includegraphics[width=1\columnwidth]{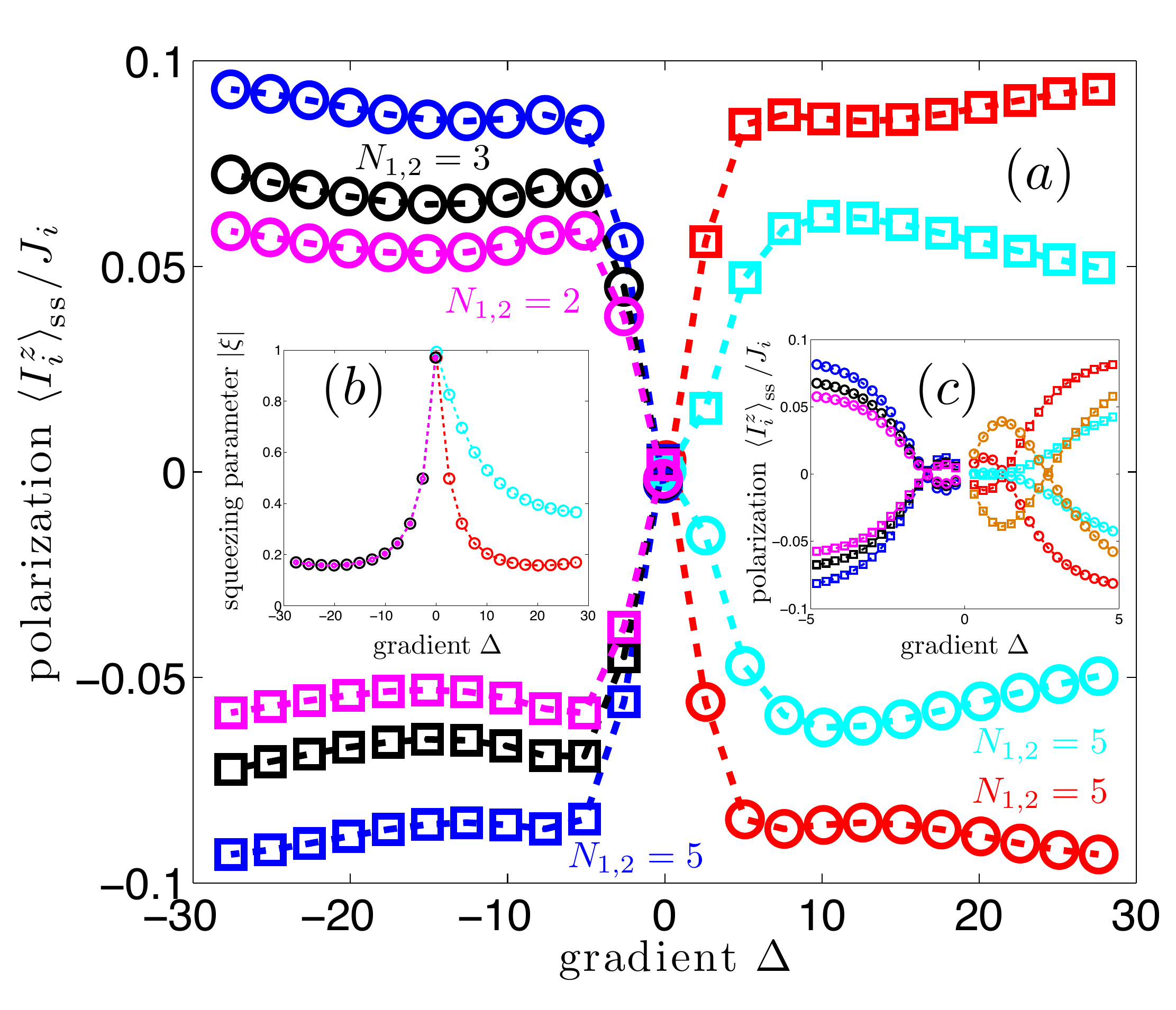}

\caption{\label{fig:instability_Ohgradient_exact}(color online). Instability
towards nuclear self-polarization: Exact numerical results for small
system sizes $J_{i}=N_{i}/2$. The exact steady state of the coupled
electron-nuclear dynamics is computed as a function of the gradient
$\Delta$. The circles (squares) refer to the polarization in the
left (right) dot, respectively. (a) For $\Delta>\left|\Delta_{\mathrm{OH}}^{\mathrm{crt}}\right|$,
we find $\Delta_{\mathrm{OH}}>0$, whereas for $\Delta<-\left|\Delta_{\mathrm{OH}}^{\mathrm{crt}}\right|$
we get $\Delta_{\mathrm{OH}}<0$, i.e., outside of the small-gradient
regime {[}see inset (c){]} the nuclear system is seen to be unstable
towards the buildup of a OH gradient with opposite polarizations in
the two dots. The nuclear polarization depends on the system size
$J_{i}$ and the parameter $\left|\xi\right|$; compare inset (b).
(c) The critical value of $\Delta_{\mathrm{OH}}^{\mathrm{crt}}\approx3\mu\mathrm{eV}$
agrees with the semiclassical estimate; it becomes smaller for smaller
values of $\Gamma_{\pm}$. Numerical parameters in $\mu\mathrm{eV}$:
$\epsilon=30$, $\Gamma=10$, $\Gamma_{\pm}=\Gamma_{\mathrm{deph}}=0.3$,
$\omega_{\mathrm{ext}}=0$ and $t=10$ except for the cyan curve where
$t=20$ and $\Gamma_{\pm}=\Gamma_{\mathrm{deph}}=0.6$ for the orange
curve in (c). }
\end{figure}

Our second numerical approach is based on exact diagonalization: As
we tune the parameter $\Delta$, we compute the steady state for the
full electronic-nuclear system directly giving the corresponding steady-state
nuclear polarizations $\left\langle I_{i}^{z}\right\rangle _{\mathrm{ss}}$.
We see a clear instability towards the buildup of an Overhauser gradient
$\Delta_{\mathrm{OH}}^{\mathrm{ss}}$ (Fig.~\ref{fig:instability_Ohgradient_exact}): Inside the small-gradient region
$\left(\left|\Delta\right|<\left|\Delta_{\mathrm{OH}}^{\mathrm{crt}}\right|\right)$
we observe negative feedback $\mathrm{sgn}\left(\Delta_{\mathrm{OH}}^{\mathrm{ss}}\right)=-\mathrm{sgn}\left(\Delta\right)$,
whereas outside of it $\left(\left|\Delta\right|>\left|\Delta_{\mathrm{OH}}^{\mathrm{crt}}\right|\right)$
the nuclear system experiences positive feedback $\mathrm{sgn}\left(\Delta_{\mathrm{OH}}^{\mathrm{ss}}\right)=\mathrm{sgn}\left(\Delta\right)$.
The latter leads to the build-up of large OH gradients, in agreement
with our semiclassical analysis.

\section{Effective Nuclear Master Equation in High-Gradient Regime\label{sec:Appendix-Effective-Nuclear-QME-High-Gradient-Regime}}

This Appendix provides background material for the derivation of the
effective nuclear master equation as stated in Eq.(\ref{eq:effective-QME-uniqueSS-explicit})
using projection-operator techniques\cite{kessler12,schuetz12}. We
start with 
\begin{equation}
\mathsf{Tr}_{\mathrm{el}}\left[\mathcal{P}\mathcal{V}\mathcal{P}\rho\right]=\mathsf{Tr}_{\mathrm{el}}\left[\mathcal{P}\mathcal{L}_{\mathrm{ff}}\mathcal{P}\rho\right]+\mathsf{Tr}_{\mathrm{el}}\left[\mathcal{P}\mathcal{L}_{\mathrm{zz}}\mathcal{P}\rho\right]
\end{equation}
The first term is readily found to be
\begin{equation}
\mathsf{Tr}_{\mathrm{el}}\left[\mathcal{P}\mathcal{L}_{\mathrm{ff}}\mathcal{P}\rho\right]=-i\frac{a_{\mathrm{hf}}}{2}\sum_{i,\alpha=\pm}\left\langle S_{i}^{\alpha}\right\rangle _{\mathrm{ss}}\left[A_{i}^{\bar{\alpha}},\sigma\right],
\end{equation}
where $\left\langle \cdot\right\rangle _{\mathrm{ss}}=\mathsf{Tr}_{\mathrm{el}}\left[\cdot\rho_{\mathrm{ss}}^{\mathrm{el}}\right]$
denotes the steady-state expectation value. An analog calculation
yields 
\begin{equation}
\mathsf{Tr}_{\mathrm{el}}\left[\mathcal{P}\mathcal{L}_{\mathrm{zz}}\mathcal{P}\rho\right]=-ia_{\mathrm{hf}}\sum_{i}\left\langle S_{i}^{z}\right\rangle _{\mathrm{ss}}\left[\delta A_{i}^{z},\sigma\right].
\end{equation}
Using that $\left\langle S_{i}^{\alpha}\right\rangle _{\mathrm{ss}}=0$
and $\left\langle S_{i}^{z}\right\rangle _{\mathrm{ss}}=0$ {[}the
Knight shift seen by the nuclear spins is zero since the electronic
quasi steady-state carries no net magnetization{]}, the first two
Hamiltonian terms vanish.

The second-order term of interest 
\begin{equation}
\mathcal{K}\sigma=\mathsf{Tr}_{\mathrm{el}}\left[\mathcal{P}\mathcal{V}\mathcal{Q}\left(-\mathcal{L}_{0}^{-1}\right)\mathcal{Q}\mathcal{V}\mathcal{P}\rho\right]
\end{equation}
can be decomposed as $\mathcal{K}\sigma=\mathcal{K}_{\mathrm{ff}}\sigma+\mathcal{K}_{\mathrm{zz}}\sigma$,
where 
\begin{eqnarray}
\mathcal{K}_{\mathrm{ff}}\sigma & = & \mathsf{Tr}_{\mathrm{el}}\left[\mathcal{P}\mathcal{L}_{\mathrm{ff}}\mathcal{Q}\left(-\mathcal{L}_{0}^{-1}\right)\mathcal{Q}\mathcal{L}_{\mathrm{ff}}\mathcal{P}\rho\right],\\
\mathcal{K}_{\mathrm{zz}}\sigma & = & \mathsf{Tr}_{\mathrm{el}}\left[\mathcal{P}\mathcal{L}_{\mathrm{zz}}\mathcal{Q}\left(-\mathcal{L}_{0}^{-1}\right)\mathcal{Q}\mathcal{L}_{\mathrm{zz}}\mathcal{P}\rho\right].
\end{eqnarray}
All other second order terms containing combinations of the superoperators
$\mathcal{L}_{\mathrm{ff}}$ and $\mathcal{L}_{\mathrm{zz}}$ can
be shown to vanish. In the following, we will evaluate the two terms
separately. 

\textit{Hyperfine flip-flop dynamics}.---First, we will evaluate $\mathcal{K}_{\mathrm{ff}}$
which can be rewritten as 
\begin{eqnarray}
\mathcal{K}_{\mathrm{ff}}\sigma & = & \underset{\circled a}{\underbrace{\int_{0}^{\infty}d\tau\mathsf{Tr}_{\mathrm{el}}\left[\mathcal{P}\mathcal{L}_{\mathrm{ff}}e^{\mathcal{L}_{0}\tau}\mathcal{L}_{\mathrm{ff}}\mathcal{P}\rho\right]}}\nonumber \\
 &  & \underset{\circled b}{\underbrace{-\int_{0}^{\infty}d\tau\mathsf{Tr}_{\mathrm{el}}\left[\mathcal{P}\mathcal{L}_{\mathrm{ff}}\mathcal{P}\mathcal{L}_{\mathrm{ff}}\mathcal{P}\rho\right]}}.
\end{eqnarray}
Here, we used the Laplace transform $-\mathcal{L}_{0}^{-1}=\int_{0}^{\infty}d\tau e^{\mathcal{L}_{0}\tau}$
and the property $e^{\mathcal{L}_{0}\tau}\mathcal{P}=\mathcal{P}e^{\mathcal{L}_{0}\tau}=\mathcal{P}$.\cite{kessler12}
The first term labeled as $\circled a$ is given by 
\begin{equation}
\circled a=-\int_{0}^{\infty}d\tau\mathsf{Tr}_{\mathrm{el}}\left(\left[H_{\mathrm{ff}},e^{\mathcal{L}_{0}\tau}\left[H_{\mathrm{ff}},\sigma\otimes\rho_{\mathrm{ss}}^{\mathrm{el}}\right]\right]\right).\label{eq:flip-flop-2nd-order-derivation1}
\end{equation}
Then, using the relations 
\begin{eqnarray}
\mathcal{L}_{0}\left[\left|\lambda_{k}\right\rangle \left\langle T_{\pm}\right|\right] & = & -i\left(\delta_{k}^{\pm}-i\tilde{\Gamma}_{k}\right)\left|\lambda_{k}\right\rangle \left\langle T_{\pm}\right|,\\
\mathcal{L}_{0}\left[\left|T_{\pm}\right\rangle \left\langle \lambda_{k}\right|\right] & = & +i\left(\delta_{k}^{\pm}+i\tilde{\Gamma}_{k}\right)\left|T_{\pm}\right\rangle \left\langle \lambda_{k}\right|,
\end{eqnarray}
where (to shorten the notation) $\delta_{k}^{+}=\Delta_{k}$ and $\delta_{k}^{-}=\delta_{k}$,
respectively, we find 
\begin{eqnarray}
e^{\mathcal{L}_{0}\tau}\left(H_{\mathrm{ff}}\sigma\rho_{\mathrm{ss}}^{\mathrm{el}}\right) & = & \frac{a_{\mathrm{hf}}}{4}\sum_{k}\left[e^{-i\left(\delta_{k}^{+}-i\tilde{\Gamma}_{k}\right)\tau}\left|\lambda_{k}\right\rangle \left\langle T_{+}\right|L_{k}\sigma\right.\nonumber \\
 &  & \left.+e^{-i\left(\delta_{k}^{-}-i\tilde{\Gamma}_{k}\right)\tau}\left|\lambda_{k}\right\rangle \left\langle T_{-}\right|\mathbb{L}_{k}\sigma\right],\label{eq:flip-flop-2nd-order-aux1}
\end{eqnarray}
and along the same lines 
\begin{eqnarray}
e^{\mathcal{L}_{0}\tau}\left(\sigma\rho_{\mathrm{ss}}^{\mathrm{el}}H_{\mathrm{ff}}\right) & = & \frac{a_{\mathrm{hf}}}{4}\sum_{k}\left[e^{+i\left(\delta_{k}^{+}+i\tilde{\Gamma}_{k}\right)\tau}\left|T_{+}\right\rangle \left\langle \lambda_{k}\right|\sigma L_{k}^{\dagger}\right.\nonumber \\
 &  & \left.+e^{+i\left(\delta_{k}^{-}+i\tilde{\Gamma}_{k}\right)\tau}\left|T_{-}\right\rangle \left\langle \lambda_{k}\right|\sigma\mathbb{L}_{k}^{\dagger}\right].\label{eq:flip-flop-2nd-order-aux2}
\end{eqnarray}
Plugging Eq.(\ref{eq:flip-flop-2nd-order-aux1}) and Eq.(\ref{eq:flip-flop-2nd-order-aux2})
into Eq.(\ref{eq:flip-flop-2nd-order-derivation1}), tracing out the
electronic degrees of freedom, performing the integration in $\tau$
and separating real and imaginary parts of the complex eigenvalues
leads to
\begin{eqnarray}
\circled a & = & \sum_{k}\left[\frac{\gamma_{k}^{+}}{2}\mathcal{D}\left[L_{k}\right]\sigma+i\frac{\Delta_{k}^{+}}{2}\left[L_{k}^{\dagger}L_{k},\sigma\right]\right.\\
 &  & \left.+\frac{\gamma_{k}^{-}}{2}\mathcal{D}\left[\mathbb{L}_{k}\right]\sigma+i\frac{\Delta_{k}^{-}}{2}\left[\mathbb{L}_{k}^{\dagger}\mathbb{L}_{k},\sigma\right]\right].
\end{eqnarray}
This corresponds to the flip-flop mediated terms given in Eq.(\ref{eq:effective-QME-uniqueSS-explicit})
in the main text. The second term labeled as $\circled b$ can be
computed along the lines: due to the additional appearance of the
projector $\mathcal{P}$, it contains factors of $\left\langle S_{i}^{\alpha}\right\rangle _{\mathrm{ss}}$
and is therefore found to be zero. 

\textit{Overhauser fluctuations}.---In the next step, we investigate
the second-order effect of Overhauser fluctuations with respect to
the effective QME for the nuclear dynamics. Our analysis starts out
from the second-order expression $\mathcal{K}_{\mathrm{zz}}$ which,
as above, can be rewritten as 
\begin{eqnarray}
\mathcal{K}_{\mathrm{zz}}\sigma & = & \underset{\circled1}{\underbrace{\int_{0}^{\infty}d\tau\mathsf{Tr}_{\mathrm{el}}\left[\mathcal{P}\mathcal{L}_{\mathrm{zz}}e^{\mathcal{L}_{0}\tau}\mathcal{L}_{\mathrm{zz}}\mathcal{P}\rho\right]}}\nonumber \\
 &  & \underset{\circled2}{\underbrace{-\int_{0}^{\infty}d\tau\mathsf{Tr}_{\mathrm{el}}\left[\mathcal{P}\mathcal{L}_{\mathrm{zz}}\mathcal{P}\mathcal{L}_{\mathrm{zz}}\mathcal{P}\rho\right]}}.
\end{eqnarray}
First, we evaluate the terms labeled by $\circled1$ and $\circled2$
separately. We find 
\begin{eqnarray}
\circled1 & = & -a_{\mathrm{hf}}^{2}\sum_{i,j}\int_{0}^{\infty}d\tau\left[\left\langle S_{i}^{z}\left(\tau\right)S_{j}^{z}\right\rangle _{\mathrm{ss}}\left[\delta A_{i}^{z},\delta A_{j}^{z}\sigma\right]\right.\nonumber \\
 &  & \left.-\left\langle S_{j}^{z}S_{i}^{z}\left(\tau\right)\right\rangle _{\mathrm{ss}}\left[\delta A_{i}^{z},\sigma\delta A_{j}^{z}\right]\right]
\end{eqnarray}
where we used the Quantum Regression theorem yielding the electronic
auto-correlation functions 
\begin{eqnarray}
\left\langle S_{i}^{z}\left(\tau\right)S_{j}^{z}\right\rangle _{\mathrm{ss}} & = & \mathsf{Tr}_{\mathrm{el}}\left[S_{i}^{z}e^{\mathcal{L}_{0}\tau}\left(S_{j}^{z}\rho_{\mathrm{ss}}^{\mathrm{el}}\right)\right],\\
\left\langle S_{j}^{z}S_{i}^{z}\left(\tau\right)\right\rangle _{\mathrm{ss}} & = & \mathsf{Tr}_{\mathrm{el}}\left[S_{i}^{z}e^{\mathcal{L}_{0}\tau}\left(\rho_{\mathrm{ss}}^{\mathrm{el}}S_{j}^{z}\right)\right].
\end{eqnarray}
In a similar fashion, the term labeled by $\circled2$ is found to
be 
\begin{equation}
\circled2=a_{\mathrm{hf}}^{2}\sum_{i,j}\int_{0}^{\infty}d\tau\left\langle S_{i}^{z}\right\rangle _{\mathrm{ss}}\left\langle S_{j}^{z}\right\rangle _{\mathrm{ss}}\left[\delta A_{i}^{z},\left[\delta A_{j}^{z},\sigma\right]\right].
\end{equation}
Putting together the results for $\circled1$ and $\circled2$, we
obtain 
\begin{eqnarray}
\mathcal{K}_{\mathrm{zz}}\sigma & = & \sum_{i,j}\Pi_{ij}\left[\delta A_{j}^{z}\sigma\delta A_{i}^{z}-\delta A_{i}^{z}\delta A_{j}^{z}\sigma\right]\nonumber \\
 &  & +\Upsilon_{ij}\left[\delta A_{j}^{z}\sigma\delta A_{i}^{z}-\sigma\delta A_{i}^{z}\delta A_{j}^{z}\right],
\end{eqnarray}
which can be rewritten as 
\begin{eqnarray}
\mathcal{K}_{\mathrm{zz}}\sigma & = & \sum_{i,j}\left(\Pi_{ij}+\Upsilon_{ij}\right)\left[\delta A_{j}^{z}\sigma\delta A_{i}^{z}-\frac{1}{2}\left\{ \delta A_{i}^{z}\delta A_{j}^{z},\sigma\right\} \right]\nonumber \\
 &  & -\frac{i}{2}\left[\frac{1}{i}\left(\Pi_{ij}-\Upsilon_{ij}\right)\delta A_{i}^{z}\delta A_{j}^{z},\sigma\right].
\end{eqnarray}
Here, we have introduced the integrated electronic auto-correlation
functions\cite{kessler12}
\begin{eqnarray*}
\Pi_{ij} & = & a_{\mathrm{hf}}^{2}\int_{0}^{\infty}d\tau\left(\left\langle S_{i}^{z}\left(\tau\right)S_{j}^{z}\right\rangle _{\mathrm{ss}}-\left\langle S_{i}^{z}\right\rangle _{\mathrm{ss}}\left\langle S_{j}^{z}\right\rangle _{\mathrm{ss}}\right),\\
\Upsilon_{ij} & = & a_{\mathrm{hf}}^{2}\int_{0}^{\infty}d\tau\left(\left\langle S_{i}^{z}S_{j}^{z}\left(\tau\right)\right\rangle _{\mathrm{ss}}-\left\langle S_{i}^{z}\right\rangle _{\mathrm{ss}}\left\langle S_{j}^{z}\right\rangle _{\mathrm{ss}}\right).
\end{eqnarray*}
For an explicit calculation, we use the relation 
\begin{eqnarray}
S_{j}^{z}\rho_{\mathrm{ss}}^{\mathrm{el}} & = & \rho_{\mathrm{ss}}^{\mathrm{el}}S_{j}^{z}=\frac{1}{4}\left(\left|T_{+}\right\rangle \left\langle T_{+}\right|-\left|T_{-}\right\rangle \left\langle T_{-}\right|\right),
\end{eqnarray}
and $ $the fact that $\left|T_{+}\right\rangle \left\langle T_{+}\right|-\left|T_{-}\right\rangle \left\langle T_{-}\right|$
is an eigenvector of $\mathcal{L}_{0}$ with eigenvalue $-5\Gamma_{\pm},$
which readily yield $\Pi_{ij}=\Upsilon_{ij}=\gamma_{\mathrm{zz}}/2$.
From this, we immediately obtain the corresponding term appearing
in the effective nuclear dynamics as 
\begin{equation}
\mathcal{K}_{\mathrm{zz}}\sigma=\gamma_{\mathrm{zz}}\sum_{i,j}\left[\delta A_{j}^{z}\sigma\delta A_{i}^{z}-\frac{1}{2}\left\{ \delta A_{i}^{z}\delta A_{j}^{z},\sigma\right\} \right].
\end{equation}

\section{\textcolor{black}{Diagonalization of Nuclear Dissipator\label{sec:Diagonalization-of-Nuclear-Disspator}}}

\textcolor{black}{The flip-flop mediated terms $\mathcal{K}_{\mathrm{ff}}$
in Eq.(\ref{eq:effective-QME-uniqueSS-explicit}) can be recast into
the following form 
\begin{equation}
\dot{\sigma}=\sum_{i,j}\frac{\gamma_{ij}}{2}\left[A_{i}\sigma A_{j}^{\dagger}-\frac{1}{2}\left\{ A_{j}^{\dagger}A_{i},\sigma\right\} \right]+i\frac{\Delta_{ij}}{2}\left[A_{j}^{\dagger}A_{i},\sigma\right],\label{eq:effective-QME-collective-spins-local}
\end{equation}
where we have introduced the vector $\mathbf{A}$ containing the }\textit{\textcolor{black}{local}}\textcolor{black}{{}
nuclear jump operators as $\mathbf{A}=\left(A_{1}^{+},A_{2}^{+},A_{2}^{-},A_{1}^{-}\right)$.
The matrices $\gamma$ and $\Delta$ obey a simple block-structure
according to 
\begin{eqnarray}
\gamma & = & \gamma^{+}\oplus\gamma^{-},\\
\Delta & = & \Delta^{+}\oplus\Delta^{-},
\end{eqnarray}
where the 2-by-2 block entries are given by 
\begin{equation}
\gamma^{\pm}=\left(\begin{array}{cc}
\gamma_{11}^{\pm} & \gamma_{12}^{\pm}\\
\gamma_{21}^{\pm} & \gamma_{22}^{\pm}
\end{array}\right)=\left(\begin{array}{cc}
\sum_{k}\gamma_{k}^{\pm}\nu_{k}^{2} & \sum_{k}\gamma_{k}^{\pm}\mu_{k}\nu_{k}\\
\sum_{k}\gamma_{k}^{\pm}\mu_{k}\nu_{k} & \sum_{k}\gamma_{k}^{\pm}\mu_{k}^{2}
\end{array}\right),
\end{equation}
and similarly 
\begin{eqnarray}
\Delta^{\pm} & = & \left(\begin{array}{cc}
\Delta_{11}^{\pm} & \Delta_{12}^{\pm}\\
\Delta_{21}^{\pm} & \Delta_{22}^{\pm}
\end{array}\right)\\
 & = & \left(\begin{array}{cc}
\sum_{k}\Delta_{k}^{\pm}\nu_{k}^{2} & \sum_{k}\Delta_{k}^{\pm}\mu_{k}\nu_{k}\\
\sum_{k}\Delta_{k}^{\pm}\mu_{k}\nu_{k} & \sum_{k}\Delta_{k}^{\pm}\mu_{k}^{2}
\end{array}\right).\nonumber 
\end{eqnarray}
The nuclear dissipator can be brought into diagonal form
\begin{equation}
\tilde{\gamma}=U^{\dagger}\gamma U=\mathrm{diag}\left(\tilde{\gamma}_{1}^{+},\tilde{\gamma}_{2}^{+},\tilde{\gamma}_{1}^{-},\tilde{\gamma}_{2}^{-}\right),
\end{equation}
where 
\begin{eqnarray}
\tilde{\gamma}_{1}^{\pm} & = & \frac{1}{2}\left[\gamma_{11}^{\pm}+\gamma_{22}^{\pm}+\sqrt{\left(\gamma_{11}^{\pm}-\gamma_{22}^{\pm}\right)^{2}+4\left(\gamma_{12}^{\pm}\right)^{2}}\right],\\
\tilde{\gamma}_{2}^{\pm} & = & \frac{1}{2}\left[\gamma_{11}^{\pm}+\gamma_{22}^{\pm}-\sqrt{\left(\gamma_{11}^{\pm}-\gamma_{22}^{\pm}\right)^{2}+4\left(\gamma_{12}^{\pm}\right)^{2}}\right],
\end{eqnarray}
and $U=U^{+}\oplus U^{-}$ with 
\begin{equation}
U^{\pm}=\left(\begin{array}{cc}
\cos\left(\theta_{\pm}/2\right) & -\sin\left(\theta_{\pm}/2\right)\\
\sin\left(\theta_{\pm}/2\right) & \cos\left(\theta_{\pm}/2\right)
\end{array}\right).
\end{equation}
Here, we have defined $\theta_{\pm}$ via the relation $\tan\left(\theta_{\pm}\right)=2\gamma_{12}^{\pm}/\left(\gamma_{11}^{\pm}-\gamma_{22}^{\pm}\right)$,
$0\leq\theta_{\pm}<\pi$. Introducing a new set of operators $\tilde{\mathbf{A}}=\left(\tilde{A}_{1},\tilde{A}_{2},\tilde{B}_{1},\tilde{B}_{2}\right)$
according to 
\begin{equation}
\tilde{\mathbf{A}}_{k}=\sum_{j}U_{jk}A_{j},
\end{equation}
that is explicitly 
\begin{eqnarray}
\tilde{A}_{1} & = & \cos\left(\theta_{+}/2\right)A_{1}^{+}+\sin\left(\theta_{+}/2\right)A_{2}^{+},\\
\tilde{A}_{2} & = & -\sin\left(\theta_{+}/2\right)A_{1}^{+}+\cos\left(\theta_{+}/2\right)A_{2}^{+},\\
\tilde{B}_{1} & = & \sin\left(\theta_{-}/2\right)A_{1}^{-}+\cos\left(\theta_{-}/2\right)A_{2}^{-},\\
\tilde{B}_{2} & = & \cos\left(\theta_{-}/2\right)A_{1}^{-}-\sin\left(\theta_{-}/2\right)A_{2}^{-},
\end{eqnarray}
the effective nuclear flip-flop mediated dynamics simplifies to 
\begin{eqnarray}
\dot{\sigma} & = & \sum_{l}\frac{\tilde{\gamma}_{l}}{2}\left[\tilde{\mathbf{A}}_{l}\sigma\tilde{\mathbf{A}}_{l}^{\dagger}-\frac{1}{2}\left\{ \tilde{\mathbf{A}}_{l}^{\dagger}\tilde{\mathbf{A}}_{l},\sigma\right\} \right]\nonumber \\
 &  & +i\sum_{k,l}\frac{\tilde{\Delta}_{kl}}{2}\left[\tilde{\mathbf{A}}_{l}^{\dagger}\tilde{\mathbf{A}}_{k},\sigma\right],
\end{eqnarray}
where the matrix $\tilde{\Delta}_{kl}=\sum_{ij}U_{ki}^{\dagger}\Delta_{ij}U_{jl}$
associated with second-order Stark shifts is in general not diagonal.
This gives rise to the Stark term mediated criticality in the nuclear
spin dynamics. }

\textcolor{black}{In general, the matrices $\gamma^{\pm}$ have $\mathrm{rank}\left(\gamma^{\pm}\right)=2$,
yielding four non-zero decay rates $\tilde{\gamma}_{1,2}^{\pm}$ and
four linear independent Lindblad operators $\tilde{\mathbf{A}}_{l}$;
therefore, in general, no pure, nuclear dark state $\left|\Psi_{\mathrm{dark}}\right\rangle $
fulfilling $\tilde{\mathbf{A}}_{l}\left|\Psi_{\mathrm{dark}}\right\rangle =0$
$\forall l$ exists. In contrast, when keeping only the supposedly
dominant coupling to the electronic eigenstate $\left|\lambda_{2}\right\rangle $,
they simplify to 
\begin{equation}
\gamma_{\mathrm{ideal}}^{\pm}=\gamma_{2}^{\pm}\left(\begin{array}{cc}
\nu_{2}^{2} & \mu_{2}\nu_{2}\\
\mu_{2}\nu_{2} & \mu_{2}^{2}
\end{array}\right),
\end{equation}
which fulfills $\mathrm{rank}\left(\gamma_{\mathrm{ideal}}^{\pm}\right)=1$.
Still, also in the non-ideal setting, for realistic experimental parameters
we observe a clear hierarchy in the eigenvalues, namely $\tilde{\gamma}_{2}^{\pm}/\tilde{\gamma}_{1}^{\pm}\lesssim0.1$.}

\section{Nonidealities In Electronic Quasisteady State \label{sec:Nonidealities-In-Electronic-Quasisteady-State}}

\begin{figure}
\includegraphics[width=1\columnwidth]{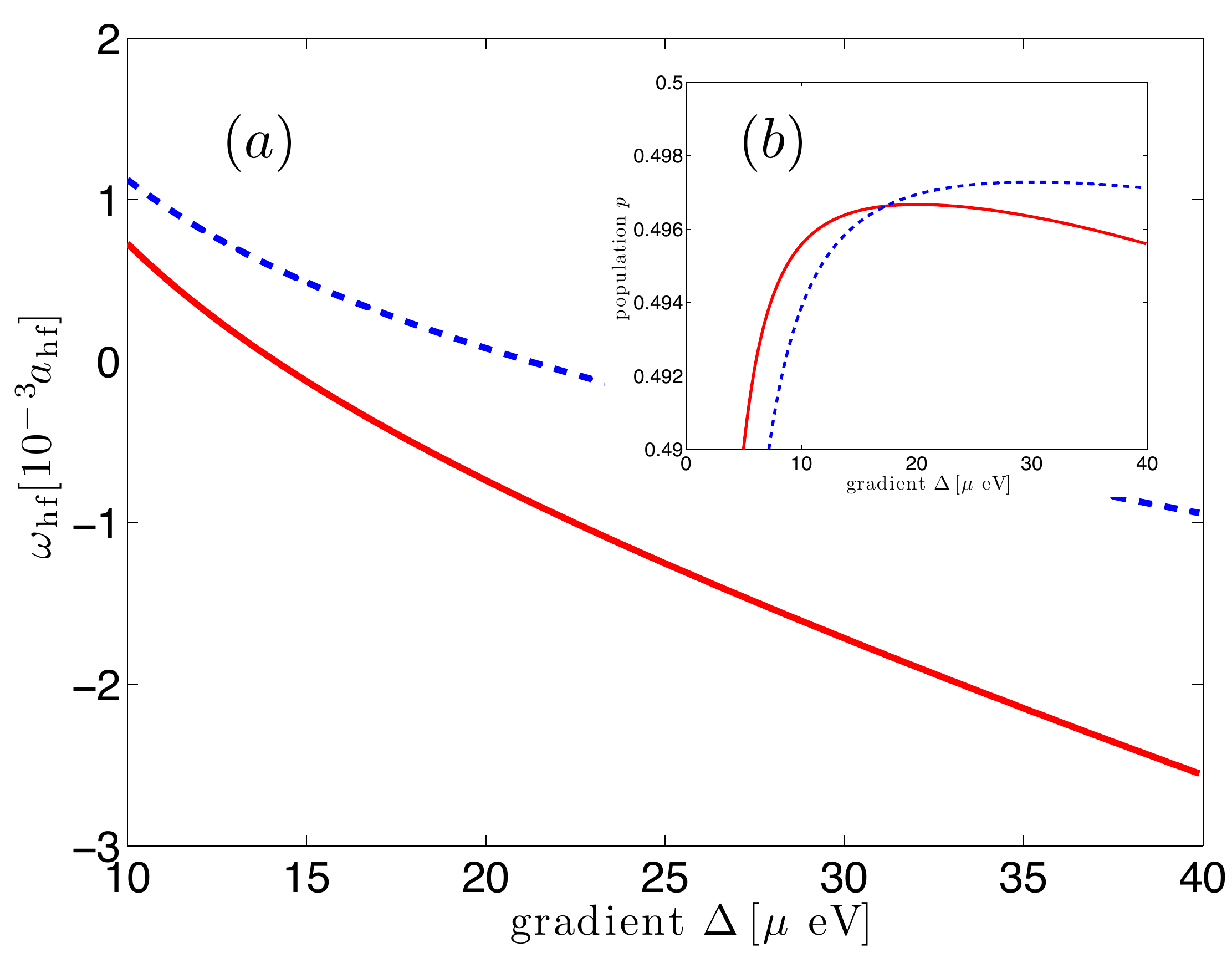}

\caption{\label{fig:finite-population_lambda-levels}(color online). (a) Knight
shift $\omega_{\mathrm{hf}}$ due to nonzero populations $p_{k}$
in the electronic quasisteady state for $t=20\mu\mathrm{eV}$ (solid)
and $t=30\mu\mathrm{eV}$ (dashed). (b) In the high gradient regime,
for $\Gamma\gg\Gamma_{\pm}$ the levels $\left|\lambda_{k}\right\rangle $
get depleted efficiently, such that $p_{k}<1\%\ll p$. Other numerical
parameters: $\epsilon=30\mu\mathrm{eV}$ and $x_{\pm}=10^{-3}$. }
\end{figure}

In Section \ref{sec:Steady-State-Entanglement} we have analyzed the
nuclear spin dynamics in the submanifold of the electronic quasisteady
state $\rho_{\mathrm{ss}}^{\mathrm{el}}=\left(\left|T_{+}\right\rangle \left\langle T_{+}\right|+\left|T_{-}\right\rangle \left\langle T_{-}\right|\right)/2$.
In this Appendix we consider (small) deviations from this ideal electronic
quasisteady state due to populations of the levels $\left|\lambda_{k}\right\rangle \left(k=1,2,3\right)$,
labeled as $p_{k}$. Since all coherences are damped out on electronic
timescales, the generalized electronic quasisteady state under consideration
is
\begin{equation}
\rho_{\mathrm{ss}}^{\mathrm{el}}=p\left(\left|T_{+}\right\rangle \left\langle T_{+}\right|+\left|T_{-}\right\rangle \left\langle T_{-}\right|\right)+\sum_{k}p_{k}\left|\lambda_{k}\right\rangle \left\langle \lambda_{k}\right|.
\end{equation}
Using detailed balance, $p_{k}$ can be calculated via the equations
$p_{k}\left(\kappa_{k}^{2}+x_{\pm}\right)=px_{\pm}$, where $x_{\pm}=\Gamma_{\pm}/\Gamma$
and $p=\left(1-\sum p_{k}\right)/2$ gives the population in $\left|T_{\pm}\right\rangle $,
respectively. The electronic levels $\left|\lambda_{k}\right\rangle $
get depleted efficiently for $\Gamma_{k}\gg\Gamma_{\pm}$: In contrast
to the low-gradient regime where $p_{2}\approx1/3$, in the high-gradient
regime, we obtain $p_{k}<1\%\ll p$ such that the electronic system
settles to a quasisteady state very close to the ideal limit where
$p=1/2$; compare Fig.~\ref{fig:finite-population_lambda-levels}.
In describing the effective nuclear dynamics, nonzero populations
$p_{k}$ lead to additional terms which are second order in $\varepsilon$,
but strongly suppressed further as $p_{k}\ll1$. 

\textit{Knight shift}.---For nonzero populations $p_{k}$, the Knight
shift seen by the nuclear spins does not vanish, leading to the following
(undesired) additional term for the effective nuclear spin dynamics
\begin{equation}
\dot{\sigma}=-i\omega_{\mathrm{hf}}\left[\delta A_{1}^{z}-\delta A_{2}^{z},\sigma\right],
\end{equation}
where 
\begin{equation}
\omega_{\mathrm{hf}}=\frac{a_{\mathrm{hf}}}{2}\sum_{k}p_{k}\left(\mu_{k}^{2}-\nu_{k}^{2}\right).
\end{equation}
with $a_{\mathrm{hf}}\approx10^{-4}\mu\mathrm{eV}$. As shown in Fig.~\ref{fig:finite-population_lambda-levels},
however, $\omega_{\mathrm{hf}}\approx10^{-7}\mu\text{\ensuremath{\mathrm{eV}}}$
is further suppressed by approximately three orders of magnitude;
in particular, $\omega_{\mathrm{hf}}$ is small compared to the dissipative
gap of the nuclear dynamics $\mathrm{ADR}\approx2\times10^{-5}\mu\mathrm{eV}$
and can thus be neglected. 

\textit{Hyperfine flip-flop dynamics}.---Moreover, nonzero populations
$p_{k}$ lead to additional Lindblad terms of the form $\dot{\sigma}=\dots+p_{k}\gamma_{k}^{+}\mathcal{D}[L_{k}^{\dagger}]\sigma$.
They contain terms which are incommensurate with the ideal two-mode
squeezedlike target state. Since $p_{k}\ll p$, however, they are
strongly suppressed compared to the ones absorbed into $\mathcal{L}_{\mathrm{nid}}$
and thus do not lead to any significant changes in our analysis.


\begin{thebibliography}{100}
\bibitem{amplitudes}The analytic expressions for the amplitudes $\mu_{k},\nu_{k}$
and $\kappa_{k}$ are not instructive and therefore not explicitly
given.
\bibitem{cotunneling-basis}Note that, in the spirit of the approximation
of independent rates of variation \cite{cohen-tannoudji92}, the relevant
(in-)coherent couplings are shown in Fig.~\ref{fig:scheme-cotunneling-spin-exchange}
in terms of the bare basis $\left\{ \left|\sigma,\sigma'\right\rangle ,\left|S_{02}\right\rangle \right\} $
for $\sigma=\Uparrow,\Downarrow$. After a basis transformation to
the dressed basis $\left\{ \left|T_{\pm}\right\rangle ,\left|\lambda_{k}\right\rangle \right\} $
one obtains the dissipative terms given in Eq.(\ref{eq:Liouvillian-mixing}).
\bibitem{cotunneling-asymmetry}If the nuclear spins in the second
dot are highly polarized, the doubly occupied triplet levels get Zeeman
split by the corresponding local Overhauser field. In principle, this
could lead to a small asymmetry in the incoherent rates describing
decay from $\left|T_{+}\right\rangle $ to $\left|T_{-}\right\rangle $
and vice versa. However, in the Appendix it is shown that in this
regime a similar incoherent decay process within the Pauli-blocked
triplet space $\left\{ \left|T_{+}\right\rangle ,\left|T_{-}\right\rangle \right\} $
can be made much more efficient by working in a regime of fast spin-exchange
with the reservoirs in the leads.
\bibitem{simiplification-delta=00003D0}Here, we have set $\delta=0$
for simplicity. However, similar results (not shown) have been obtained
for finite values of $\delta$.
\bibitem{small-gradient-entanglement}We have numerically checked
that the entanglement generation persists for a small external gradient
$\Delta_{\mathrm{ext}}$ provided that the corresponding induced nuclear
Zeeman splitting is small compared to the dissipative gap $\mathrm{ADR}=\tau_{\mathrm{gap}}^{-1}$.
Thus, to support the generation of steady-state entnaglement, we estimate
that the magnetic gradient should be smaller than $\sim1\mathrm{mT}$.
\bibitem{secular-approximation}In deriving Eq.(\ref{eq:QME-Markov-single-particle}),
we have ignored level shifts arising from the coupling to the environment;
as usual, they can be absorbed into renormalized energy levels. Moreover,
we have applied the so-called secular approximation which is mathematically
correct in the weak coupling limit and ensures the positivity of the
dynamics \cite{rivas12}.
\bibitem{tilde}Here, the tilde symbol explicitly refers to the dressed
electronic basis $\left\{ \left|T_{\nu}\right\rangle ,\left|\lambda_{k}\right\rangle \right\} $.
For notational convenience, it is dropped in the main text.
\bibitem{cotunneling-dressed-vs-bare}For simplicity, in the definition
of $\mathcal{K}_{\pm}$ we have included dissipative cotunneling-mediated
transitions in the bare triplet subspace only $\left\{ \left|T_{\pm}\right\rangle ,\left|T_{0}\right\rangle \right\} $.
To make the comparison with the Liouvillian in the dressed basis $\mathcal{L}_{0}\tilde{\rho}$,
the corresponding mixing terms $\mathcal{D}\left[\left|T_{\nu}\right\rangle \left\langle \lambda_{k}\right|\right]$
and $\mathcal{D}\left[\left|\lambda_{k}\right\rangle \left\langle T_{\nu}\right|\right]$
in $\mathcal{L}_{\pm}\tilde{\rho}$ appear with a rate $\Gamma_{\pm}\left|\left<\lambda_{k}|T_{0}\right>\right|^{2}$.
In particular, in the low-gradient regime where $\left|\lambda_{2}\right\rangle \approx\left|T_{0}\right\rangle $
this captures well the dissipative mixing between $\left|T_{\pm}\right\rangle $
and $\left|\lambda_{2}\right\rangle $ which is the most adverse process
to our scheme. We have also verified that simply replacing $\Gamma_{\pm}\left|\left<\lambda_{k}|T_{0}\right>\right|^{2}\rightarrow\Gamma_{\pm}$
as it is stated in Eq.(\ref{eq:Liouvillian-mixing}) does not change
$\mathcal{F}_{\mathrm{el}}$ nor $\mathrm{ADR}_{\mathrm{el}}$ severely.
\end{thebibliography}
\end{document}